\documentclass[conference]{IEEEtran}

\usepackage{graphicx}

\newif\ifnotes\notesfalse
\usepackage[utf8]{inputenc}
\usepackage{amsmath,amssymb,pifont}
\usepackage{algorithm,algorithmic}
\usepackage{amstext,amssymb,amsmath}
\usepackage{amsthm}
\usepackage{multirow}
\usepackage{booktabs}
\usepackage{subcaption}
\usepackage{lipsum}
\usepackage[shortlabels]{enumitem}
\usepackage{wrapfig}
\usepackage{array}
\usepackage{siunitx}
\usepackage{csvsimple}
\usepackage[multidot]{grffile}
\usepackage{bbm}
\usepackage{dblfloatfix}
\usepackage{makecell}

\renewcommand{\epsilon}{\varepsilon}
\newcommand{\mypar}[1]{\smallskip
	\noindent{\textbf{{#1}:}}}

\newcommand{\anoteinline}[1]{\todo[color=blue!30,inline]{AT: #1}}
\newcommand{\unote}[1]{\todo[color=green!30]{{\'U}E: #1}}

\newcommand{\vnoteinline}[1]{\todo[color=cyan!30,inline]{VF: #1}}

\newcommand{\zo}{\ensuremath{\{0,1\}}}
\newcommand{\epsl}{\epsilon_\ell}
\newcommand{\epsle}{\epsilon_{\ell e}}
\newcommand{\epslt}[1]{\epsilon_{\ell^{#1}}}

\newcommand{\epsc}{\epsilon_c}
\newcommand{\deltac}{\delta_c}

\newcommand{\eps}{\ensuremath{\varepsilon}}

\newcommand{\afrag}{\ensuremath{\mathsf{att}\textsf{-}\mathsf{frag}}}
\newcommand{\rfrag}{\ensuremath{\mathsf{r}\textsf{-}\mathsf{frag}}}

\newcommand{\boldg}{\ensuremath{\boldsymbol{g}}}
\newcommand{\boldh}{\ensuremath{\boldsymbol{h}}}

\newcommand{\boldv}{\ensuremath{\boldsymbol{v}}}

\newcommand{\boldx}{\ensuremath{\boldsymbol{x}}}

\newcommand{\boldz}{\ensuremath{\boldsymbol{z}}}
\newcommand{\boldzero}{\ensuremath{\boldsymbol{0}}}

\newcommand{\bfS}{\ensuremath{\mathbf{S}}}

\newcommand{\calC}{\ensuremath{\mathcal{C}}}
\newcommand{\calD}{\ensuremath{\mathcal{D}}}

\newcommand{\calF}{\ensuremath{\mathcal{F}}}

\newcommand{\calI}{\ensuremath{\mathcal{I}}}

\newcommand{\calM}{\ensuremath{\mathcal{M}}}

\newcommand{\calQ}{\ensuremath{\mathcal{Q}}}
\newcommand{\calR}{\ensuremath{\mathcal{R}}}
\newcommand{\calS}{\ensuremath{\mathcal{S}}}

\renewcommand{\Pr}{\mathop{\mathbf{Pr}}}

\newcommand{\E}{\mathop{\mathbf{E}}}

\newtheorem{lem}{Lemma}[section]
\newtheorem{thm}[lem]{Theorem}
\newtheorem{cor}[lem]{Corollary}

\newtheorem{definition}[lem]{Definition}

\newcommand{\rmse}{\textsc{RMSE}}

\DeclareMathOperator*{\argmax}{arg\,max}
\DeclareMathOperator*{\argmin}{arg\,min}

\newcommand{\nbr}[1]{\num[group-separator={,}]{#1}}
\makeatletter
\newcommand{\vast}{\bBigg@{4}}
\newcommand{\Vast}{\bBigg@{5}}
\makeatother

\newcommand{\indi}[1]{\mathbbm{1}\left(#1\right)}

\newcommand{\sks}[0]{\kappa}

\newcommand{\horseRev}{\text{Horse}}
\newcommand{\girl}{\text{Child}}
\newcommand{\map}{\text{Map}}
\newcommand{\real}{\text{Heavy-hitter}}

\providecommand{\alequn}[1]{\begin{align*} #1 \end{align*}}

\ifnotes
\usepackage[textsize=scriptsize,textwidth=1cm]{todonotes}
\else
\usepackage[disable]{todonotes}
\fi

\newif\ifbigpdf\bigpdffalse

\IEEEoverridecommandlockouts

\begin{document}

\title{Encode, Shuffle, Analyze Privacy Revisited: \\
Formalizations and Empirical Evaluation}

\author{\'Ulfar Erlingsson, Vitaly Feldman, Ilya Mironov,$^\dagger$\thanks{$\dagger$~~Work done while at Google Brain.} Ananth Raghunathan \\ Shuang Song, Kunal Talwar, Abhradeep Thakurta \\ \textit{Google Research -- Brain}}

\date{}
\maketitle
\thispagestyle{plain}
\pagestyle{plain}

\newif\ifsketch\sketchtrue

\begin{abstract}
Recently, a number of approaches and techniques
have been introduced 
for reporting software statistics
with strong privacy guarantees,
spurred by the large-scale deployment of mechanisms
such as Google's RAPPOR~\cite{rappor}.

Ranging from abstract algorithms
to comprehensive systems,
and varying in their
assumptions and applicability,
this work has built upon
\emph{local differential privacy} mechanisms,
sometimes augmented by anonymity.
Most recently,
based on 
the \emph{Encode, Shuffle, Analyze} (ESA) framework~\cite{prochlo},
a notable set of results
has formally clarified how
making reports anonymous
can greatly improve
privacy guarantees 
without loss of utility~\cite{soda-shuffling,dpmixnets}.
Unfortunately,
these results have comprised
either 
systems with seemingly incomparable mechanisms and attack models,
or formal statements that
have given little guidance to practitioners.

To address this, 
in this work we provide a formal treatment and offer prescriptive guidelines
for 
privacy-preserving reporting with anonymity,
i.e.,
for deployments of 
``privacy amplification by shuffling.''
To achieve this, 
we revisit the ESA framework
and craft 
a simple, abstract model of attackers and assumptions
covering it and other proposed systems of anonymity.
\ifsketch
In light of the new formal privacy bounds,
we examine the limitations of
sketch-based encodings and
ESA mechanisms
such as data-dependent crowds.
\else
In light of the new formal privacy bounds,
we examine the limitations of
ESA mechanisms
such as data-dependent crowds.
\fi
However, we also 
demonstrate how the ESA notion of 
\emph{fragmentation}---i.e.,
reporting different data aspects in separate, unlinkable messages---is
essential for improving 
the privacy/utility tradeoff
both in terms of local and central differential-privacy guarantees.

Finally, 
to help practitioners 
understand the applicability and limitations
of privacy-preserving reporting,
we report on a large number of empirical experiments.
In these, we mostly use real-world datasets with
heavy-tailed or near-flat distributions,
since these pose the greatest difficulty for our techniques;
in particular, we focus on
data drawn from images,
since it can be easily visualized 
in a way that highlights
errors in its reconstruction.
Showing the promise of the approach,
and of independent interest, 
we also report on experiments
using
anonymous, privacy-preserving reporting
to
train high-accuracy
deep neural networks
on standard tasks, such as MNIST and CIFAR-10.
  
\end{abstract}

\section{Introduction}
\label{sec:intro}

\begin{figure}[t]
    \centering
    \includegraphics[width=0.95\columnwidth]{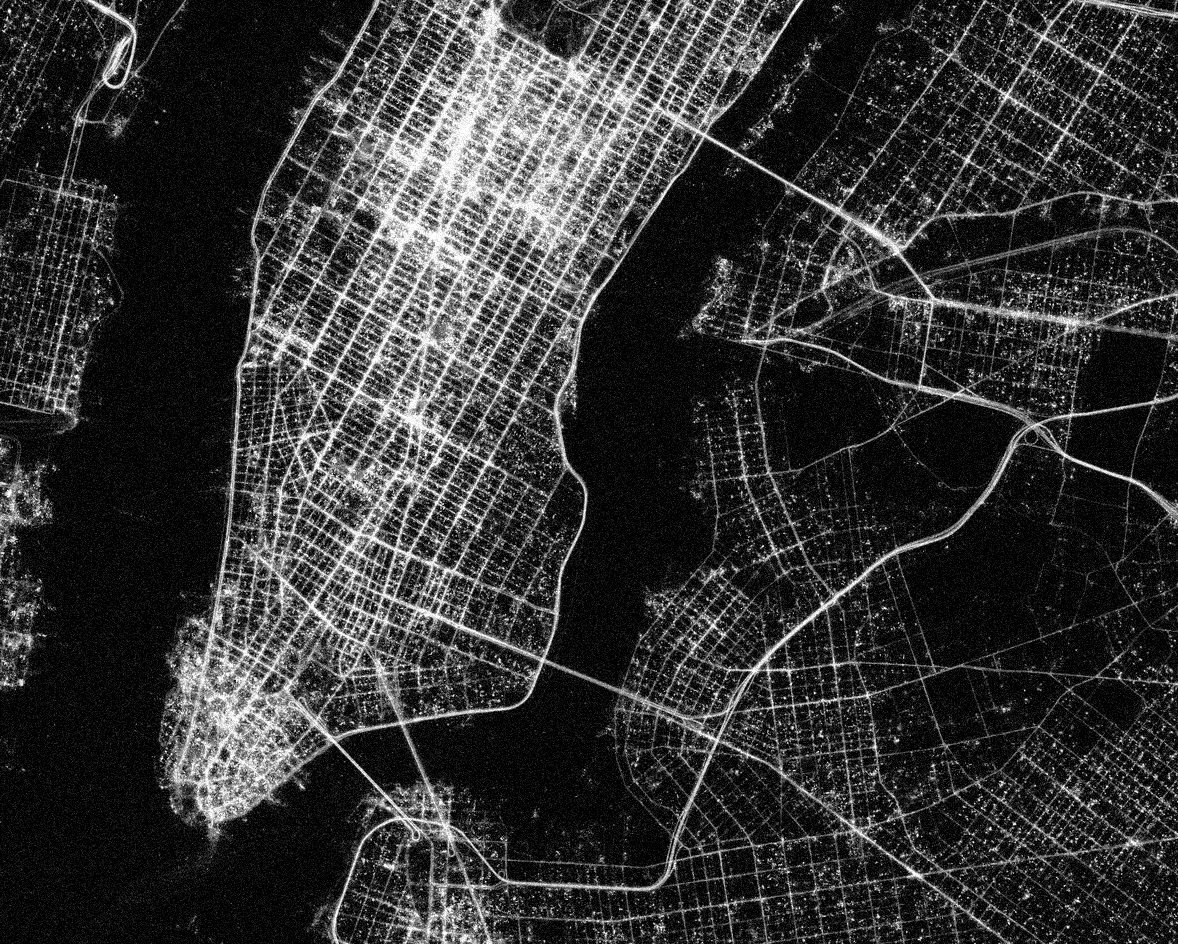}
    \caption{A differentially-private view of the NYC smartphone-location data published by the New York Times in 2018~\cite{nytimesdata}.
      Anonymous, randomized location reports
      allow high accuracy
      with a strong central differential privacy guarantee
      ($\epsc=0.5$)
      and a weaker local guarantee
      ($\epslt{\infty}\approx 12$)
      that still provides
      uncertainty 
      even if all parties collude 
      and break 
      report anonymity.\vspace*{-2ex}}
    \label{fig:nychist}
\end{figure}

To guide their efforts,
public health officials must sometimes
gather statistics
based on sensitive, private information
(e.g., to survey the prevalence of vaping
among middle-school children).
Due to privacy concerns---or simple
reluctance to admit the truth---respondents 
may fail to answer 
such surveys, or purposefully answer incorrectly,
despite the societal benefits of
improved public-health measures.
To remove such discouragement,
and still compute accurate statistics,
epidemiologists can turn to \emph{randomized response}
and have respondents 
\emph{not} report their true answer,
but instead report the results of random coin flips
that are just biased by that true answer~\cite{Warner}.

In computing,
such randomized-response mechanisms
that guarantee \emph{local differential privacy} (LDP)
have become a widely-deployed, best-practice
means of 
gathering potentially-sensitive information
about software and its users
in a responsible manner~\cite{rappor,appledp}.
Simultaneously,
many systems have been developed
for anonymous communication and messaging~\cite{tor,karaoke},
many of which 
are designed to 
gather aggregate statistics
with privacy~\cite{prochlo,prio,secureagg,honeycrisp}.
As shown in Figure~\ref{fig:nychist},
when combined with anonymity,
LDP reports can permit high-accuracy
central visibility into distributed, sensitive data
(e.g., different users' private attributes)
with strong worst-case privacy guarantees that
hold 
for even the most unlucky respondents---even
when fate and other parties conspire against them.
Thereby, a key dilemma can be resolved: how to 
usefully learn about a population's data distribution
without collecting 
distinct, identifiable population data 
into a database
whose very existence forms an unbounded privacy risk,
especially as it may be abused for surveillance.

\subsection{Statistical Reporting with Privacy, in Practice}
\label{sec:general_motivation}
Unfortunately, in practice, %
there remains little clarity
on how statistical reporting
should be implemented and deployed with strong privacy guarantees---especially
if LDP reports are to be made anonymous~\cite{prochlo,soda-shuffling,cheu2019distributed,BalleBG18}.
A daunting number 
of LDP reporting protocols
have been recently proposed and formally analyzed,
each using slightly different assumptions and techniques,
such as strategies for randomization 
and encoding of binary, categorical,
and other types of data~\cite{rappor,BS15,BST17,duchi2018minimax}.
However,
these protocols may not be suitable
to the specifics of any given application domain,
due to their different assumptions,
(e.g., about adaptivity~\cite{soda-shuffling,BalleBG18},
sketching~\cite{rappor,rapporunknowns,BS15,BST17},
or succinctness of communication~\cite{BS15,BST17,ghazi2019private}).
Thus, these protocols may exhibit lackluster performance
on real-world data distributions
of limited size,
even when
accompanied by a formal proof of
asymptotically-optimal privacy/utility tradeoffs.
\ifsketch
In particular,
many of these protocols
perform dimensionality-reduction using sketches
whose added noise
may greatly thwart visibility
into the tail of distributions
(as shown in the experiments of Section~\ref{sec:expt}).
\else
In particular,
many of these protocols
perform dimensionality-reduction using sketches
whose added noise
greatly thwarts visibility
into the tail of distributions.
\fi
Finally,
the option
of simply replicating the details of
prominent LDP-reporting deployments
is not very attractive,
since these have been criticized 
both for a lack of privacy and a lack of utility~\cite{rappor,prochlo,korolova-apple}.

Similarly,
multiple, disparate approaches
have been developed for ensuring anonymity,
including some comprehensive systems 
that have seen wide deployment~\cite{tor}.
However, 
most of these are not well suited
to gathering statistics with strong privacy,
as they are focused on low-latency communication
or point-to-point messaging~\cite{tor,karaoke,atom}.
The few that are
well-suited to ensuring the anonymity
of long-term, high-latency statistical reporting
are somewhat incomparable,
due to their
different technical mechanisms
and their varying
assumptions and threat models.
Whether they rely on Tor-like mixnets or trusted hardware,
some proposed systems
output sets of reports unlinkable to their origin~\cite{prochlo,cheu2019distributed},
while others
output only
a summary of the reports
made anonymous by the use of 
a commutative, associative
aggregation operation~\cite{prio,secureagg}.
Also,
these systems' abilities 
are constrained by
the specifics of their construction
and 
mechanisms (e.g., built-in sampling rates
and means of multi-party cryptographic computation, as in~\cite{secureagg});
some systems are more specific still,
and focus only on certain applications,
such as the maintenance of (statistical) models~\cite{honeycrisp,sage}.
Finally,
all of these systems have slightly different threat models,
e.g., with some assuming an honest-but-curious central coordinator~\cite{secureagg}
and other assuming a non-colluding, trusted set of parties~\cite{prochlo,prio}.
(Interestingly,
these threat models
typically exclude the risk of statistical inference,
even though
limiting this risk
is often a primary privacy goal, as it is in this paper.)
All of this tends to obscure
how these anonymity systems
can be best applied to
learning statistics with strong privacy guarantees.

This lack of clarity is especially
concerning
because of 
recent formal results---known colloquially as
``privacy amplification by shuffling''~\cite{soda-shuffling,cheu2019distributed,BalleBG18,ghazi2019private}---which
have fundamentally changed privacy/utility tradeoffs
and forced
a reconsideration of previous approaches,
like those described above.
These amplification results
prove how central privacy guarantees
can be strengthened by orders of magnitude
when LDP reports
can be made anonymous---i.e., unlinkable to their source---in particular,
by having them get ``lost in the crowd''
through their
shuffling or aggregate summarization
with a sufficiently-large set of other reports.

The source of these
privacy amplification results
are efforts to formalize
how LDP reporting mechanisms benefit from anonymity in 
the \emph{Encode, Shuffle, Analyze} (ESA) framework~\cite{prochlo}.
The ESA architecture is rather abstract---placing
few restrictions on 
specifics such as randomization schemes, report encoding,
or the means of establishing  anonymity---and,
not surprisingly,
can be a suitable foundation 
for implementations that aim to benefit from privacy amplification by anonymity.

\subsection{Practical Experiments, Primitives, and Attack Models}
\label{sec:ESA_motivation}
\label{sec:high_epsilon_motivation}
In this work, we revisit the specifics of the ESA framework
and explore
statistical reporting with strong privacy guarantees augmented by anonymity,
with the goal of providing clear, practical implementation guidelines.

At the center of this paper
are a set of empirical experiments,
modeled on real-world monitoring tasks,
that achieve different levels of privacy
on a representative set of data distributions.
For most of our experiments
we use data distributions derived from images,
which we choose because
they are both representative of
certain sensitive data---such as user-location data, as in Figure~\ref{fig:nychist}---and
their reconstruction accuracy 
can be easily estimated, visually.
Reconstructing images with strong privacy
is  particularly challenging 
since images are a
naturally high-dimensional dataset
with
a low maximum amplitude
(e.g., 
the per-pixel distribution
of an 8-bit gray-scale image 
will have a luminescence bound of 255),
and
which can be either dense or sparse.
In addition, following most previous work,
we also include experiments that use
a real-world, Zipfian dataset
with high-amplitude heavy hitters. 

The overall conclusion of this paper is
that
high-accuracy statistical reporting with
strong, anonymity-amplified privacy guarantees
can be implemented 
using a small set of simple primitives:
(i) a new ``removal'' basis for the analysis of LDP reporting, %
(ii) one-hot encoding of categorical data,
(iii) fragmenting of data and reports into multiple messages,
and (iv) anonymous shuffling or aggregate sums.
Although novel in combination,
most of these individual primitives have
been explored in previous work;
the exception
is our ``removal LDP'' report definition
which can strengthen
the local privacy guarantees by a factor of two.
For several common statistical reporting tasks,
we argue 
that these four primitives are difficult to improve upon,
and we verify this in experiments.

Interestingly,
we find that some of the more advanced primitives
from the related work
may offer
little benefits
and can, in some cases, be detrimental to privacy and utility.
\ifsketch
These include
ESA's Crowd IDs
and the
heterogeneous privacy levels
they induce, by identifying subsets of reports,
as well as---most surprisingly---the
use of the sketch-based encodings
like those popularized by RAPPOR~\cite{rappor,BST17}.
As we shown in experiments,
while sketching will
always reduce the number of sent reports,
sketching may add
noise that greatly exceeds that required for privacy,
unless the sketch construction
is fine-tuned to the target data-distribution specifics.
\else
These include
ESA's Crowd IDs
and the
heterogeneous privacy levels
they induce, by identifying subsets of reports.
\fi

However,
we find great benefits in
the ESA concept of \emph{fragments}:
breaking up the information to be reported
and
leveraging anonymity to
send multiple unlinkable reports,
instead of
sending the same information in
just one report.
As an example,
using \emph{attribute fragmentation},
a respondent with different attributes
encoded into a long, sparse Boolean bitvector
can send multiple, separately-anonymous reports
for the index of each bit set in an LDP version of the bitvector.
In particular, we show how
privacy/utility tradeoffs
can be greatly improved by
applying such attribute fragmentation to
LDP reports based on one-hot encodings of categorical data.
Another useful form is
\emph{report fragmentation},
where  respondents send multiple, re-randomized reports
based on an LDP backstop (e.g., an underlying, permanent LDP report, like the PRR of~\cite{rappor});
this
can allow for a more refined attack model
and lower the per-report privacy risk,
while maintaining a strict cap on the overall, long-term privacy loss.

Finally,
we propose a simple, abstract model
of threats and assumptions
that
abstracts away from the how shuffling is performed
and assumes only that LDP report anonymization satisfy
a few, clear requirements;
thereby, we hope to 
help practitioners 
reason about and choose from
the disparate set of anonymization systems,
both current and future.
The requirements of our attack model
can be met using
a variety of mechanisms,
such as mixnets, anonymous messaging,
or a variety of cryptographic multi-party mechanisms
including ESA's ``blinded shuffling''~\cite{prochlo}.
Furthermore,
while simple,
our attack model
still allows for refinements---such as
efficient in-system aggregation of summaries, and
gradual loss of privacy due to partial compromise or collusion---which
may be necessary for practical, real-world deployments.

\subsection{Summary of Contributions}
This paper gives clear guidelines
for how practitioners can
implement 
high-accuracy
statistical reporting with strong privacy guarantees---even
for difficult, high-dimensional data distributions,
and as little as a few dozen respondents---and
best leverage recent privacy-amplification results based on  anonymity.
In particular, this paper contributes the following:

\begin{enumerate}
\item[\S\ref{Sec:prelims}] %
  We explain how
  the reports in anonymous statistical monitoring 
  are well suited to
  a ``removal LDP'' definition of local differential privacy
  and how this can strengthen
  respondents' local privacy guarantees by a factor of two, without compromise.

\item[\S\ref{sec:expt}]
  We give the results of numerous experiments
  that are representative of real-world tasks and data distributions
  and
  show that strong central privacy guarantees are compatible with high utility---even
  for low-amplitude and long-tail distributions---but
  that this requires high-epsilon LDP reports
  and, correspondingly, great trust in how reports are anonymized.

\item[\S\ref{sec:crowd}]
  We clarify how---given the strong central privacy guarantees allowed by anonymity---the
  use of higher-epsilon LDP reports
  is almost always preferable to 
  mechanisms, like ESA Crowd IDs,
  which perform data-dependent grouping of reports during anonymization.

\item[\S\ref{sec:attribFrag}]
  We outline how privacy and utility can be maximized
  by having respondents use \emph{attribute fragmentation}
  to break up their data
  (such as the different bits of their reports)
  and send as separate, unlinkable LDP reports.

\item[\S\ref{sec:record_fragmenting}]
  We formally analyze 
  how---along the lines of RAPPOR's permanent randomized response~\cite{rappor}---\emph{report fragmentation}
  can reduce the per-report privacy risk,
  while strictly bounding the overall, long-term privacy loss.

\item[\S\ref{sec:expt}]
  \ifsketch
  We empirically
  show the advantages of
  simple one-hot LDP report encodings
  and---as
  a warning to practitioners---empirically highlight the need to
  fine-tune the parameters of sketch-based encodings.
  \else
  We show the benefits of
  LDP reporting of categorical data
  based on simple one-hot encodings.
  \fi

\item[\S\ref{Sec:prelims}] %
  We provide a simple, abstract attack model
  that makes it easier to reason about
  the assumptions and specifics
  of anonymity mechanisms and LDP reporting schemes,
  and compose them into practical systems.

\item[\S\ref{sec:expt}]
  Finally, we demonstrate  how anonymous LDP reports
  can be usefully applied
  to the training of benchmark deep learning models with high accuracy,
  with clear central privacy guarantees
  and minimal empirical loss of privacy.

\end{enumerate}

\section{Definitions and Assumptions}
\label{Sec:prelims}
We first
lay a foundation for the remainder of this paper
by defining notation, terms, and stating clear assumptions.
In particular, we clarify what we mean by LDP reports, their encoding and fragmentation,
as well as our model of attackers and anonymization.

\subsection{Local Differential Privacy and Removal vs.\ Replacement}
\label{sec:dp}

Differential privacy (DP), introduced by Dwork et al.~\cite{DMNS,Dwork-ICALP}, is a privacy definition
that captures how randomized algorithms that operate on a dataset
can be bounded in their sensitivity
to the presence or absence of any particular data item.
Differential privacy is measured as the maximum possible divergence between the output distributions
of such algorithms
when applied to two datasets that differ by any one record.
The most common definition of this metric
is based on the worst-case replacement of any dataset record:
\begin{definition}[Replacement $(\eps,\delta)$-DP~\cite{odo}]
\label{def:dp-replace}
A randomized algorithm $\calM\colon \calD^n \rightarrow \calS$ satisfies {\em replacement} $(\eps,\delta)$-differential privacy if for all $S \subset \calS$ and for all $i\in [n]$ and datasets $D=(x_1,\ldots,x_n), D'=(x'_1,\ldots,x'_n) \in \calD^n$ such that $x_j = x'_j$ for all $j \neq i$ we have:
\[
\Pr[\calM(D) \in S] \leq e^\eps \Pr[\calM(D') \in S] + \delta.
\]
\end{definition}

\label{sec:notation} \noindent
Above, as in the rest of this paper,
we let $[n]$ denote the set of integers $\{1, \ldots, n\}$, $[a,b]$ denote $\{v \colon a \leq v \leq b\}$, and $(a \wedge b)$ denote $\text{max}(a, b)$. Symbols such as $x$ typically represent scalars, symbols such as $\boldx$ represent vectors of appropriate length. Elements of $\boldx$ are represented by $x_i$.
Respectively, $\|\boldx\|_1$ and $\|\boldx\|_2$ represent $\sum |x_i|$ and $\sqrt{\sum x_i^2}$. Additionally, all logarithms in this paper are natural logarithms, unless the base is explicitly mentioned.

\emph{Local} differential privacy (LDP) considers a distributed dataset or data collection task
where an attacker is assumed to see and control the reports or records for all-but-one respondent,
and where
the entire transcript of all communication must satisfy differential privacy
for each respondent.
Commonly, LDP guarantees are achieved by having
respondents
communicate only randomized reports
that result from applying a differentially private algorithm $\calR$ to their data.

For any given level of privacy,
there are strict limits to the utility of
datasets gathered via LDP reporting.
The uncertainty in each LDP report
creates a ``noise floor''
below which no signal can be detected.
This noise floor
grows %
with the dimensionality of the reported data;
therefore,
compared to a Boolean question  (``Do you vape?''),
a high-dimensional question about location (``Where in the world are you?'')
can be expected to have dramatically more noise
and a correspondingly worse signal.
This noise floor
also
grows in proportion to the square root of the number of reports;
therefore,
somewhat counter-intuitively,
as more data is collected
it will become harder to detect
any fixed-magnitude signal
(e.g., the global distribution of the limited, fixed set of people named Sandiego).

The
algorithms used to create per-respondent LDP reports---referred to as \emph{local randomizers}---must
satisfy the definition of differential privacy
for a dataset of size one;
in particular, they may satisfy the following definition based on replacement:
\begin{definition}[Replacement LDP]
\label{def:ldp-replace}
An algorithm $\calR\colon \calD  \rightarrow \calS$ is a \emph{replacement $(\eps,\delta)$-differentially private local randomizer} if for all $S \subseteq \calS$ and for all $x,x' \in \calD$:
\[
\Pr[\calR(x) \in S] \leq e^{\eps} \Pr[\calR(x) \in S] + \delta.
\]
\end{definition}

\noindent
However, this replacement-based LDP definition is
unnecessarily conservative---at least
for finding
good privacy/utility tradeoffs
in statistical reporting---although it
has often been used in prior work, because it simplifies certain analyses.

Replacement LDP compares
the presence of any respondent's report
against the counterfactual
of being replaced with its worst-case alternative.
For distributed monitoring,
a more suitable
counterfactual is one
where the respondent has decided not to send any report,
and thereby has removed themselves from the dataset.
It is well known that
replacement LDP has a
differential-privacy $\epsilon$ upper bound
that for some mechanisms can be twice
that of an $\epsilon$ based on
the removal of a respondent's report.
For the $\eps > 1$ regime that is typical in LDP applications,
this factor-of-two change makes a major difference
because the probability of $S$ depends exponentially on $\epsilon$.
Thus, a removal-based definition is more appropriate for our practical privacy/utility tradeoffs.
Unfortunately,
a removal-based LDP definition
cannot be directly adopted in the local model
due to a technicality:
removing any report will change the support of the output distribution
because the attacker is assumed to observe all communication.
To avoid this,
we can define removal-based differential privacy generally
with respect to algorithms defined only on inputs of fixed length $n$,
and from this define a corresponding local randomizer:
\begin{definition}[Generalized removal $(\eps,\delta)$-DP]
\label{def:dp-remove}
A randomized algorithm $\calM\colon \calD^n \rightarrow \calS$ satisfies {\em removal} $(\eps,\delta)$-differential privacy if there exists an algorithm $\calM'\colon \calD^n \times 2^{[n]} \rightarrow \calS$ with the following properties:
\begin{enumerate}
\item for all $D\in \calD^n$, $\calM'(D,[n])$ is identical to $\calM(D)$;
\item for all $D\in \calD^n$ and $I\subseteq [n]$, $\calM'(D,I)$ depends only on the elements of $D$ with indices in $I$;
\item for all $S \subset \calS$, $D\in \calD^n$ and $I, I'\subseteq [n]$ where we have that $|I \bigtriangleup I'|=1$:
\[
\Pr[\calM'(D,I) \in S] \leq e^\eps \Pr[\calM'(D,I') \in S] + \delta.
\]
\end{enumerate}
\end{definition}

\noindent
(Notably,
this definition
generalizes the more standard definition of removal-based differential privacy
where $\calM$ is defined for datasets of all sizes,
by setting
$\calM'(D,I):=\calM((x_i)_{i\in I})$---i.e., by defining $\calM'(D,I)$ to be $\calM$
applied to the elements of $D$ with indices in $I$.)

In the distributed setting it suffices to
define removal-based LDP---as follows---by combining the above definition
with the use of a local randomizer whose properties satisfy
Definition~\ref{def:dp-remove} when restricted to datasets of size 1.
(For convenience, we state this only for $\delta=0$,
since extensions to $\delta>0$ and other notions of DP are straightforward.)
\begin{definition}[Removal LDP]
\label{def:ldp-remove}
An algorithm $\calR\colon \calD  \rightarrow \calS$ is a \emph{removal $\eps$-differentially private local randomizer} if there exists a random variable $\calR_0$ such that for all $S \subseteq \calS$ and for all $x \in \calD$:
\[
e^{-\eps} \Pr[\calR_0 \in S] \leq \Pr[\calR(x) \in S] \leq e^{\eps} \Pr[\calR_0 \in S].
\]
\end{definition}
Here $\calR_0$ should be thought of as the output of the randomizer when a respondent's data is absent. This definition is equivalent, up to a factor of two, to the replacement version of the definitions. To distinguish between these two notions we will always explicitly state ``removal differential privacy" but often omit ``replacement" to refer to the more common notion.

\subsection{Attributes, Encodings, and Fragments of Reports}
\label{sec:encodefragment}

There are various means by which
LDP reports can be crafted from
a respondent's data record, $\boldx\in\calD$ in a domain $\calD$,
using a local randomizer $\calR$.
This paper
considers three specific
LDP report constructions,
that stem from the ESA framework~\cite{prochlo}---report
encoding,
attribute fragmentation, and report fragmentation---each of which
provides a lever for controlling different aspects of the utility/privacy tradeoffs.

\ifsketch
\mypar{Encodings} Given a data record $\boldx$, depending on its domain $\calD$, the type of encoding can have a strong impact on the utility of a differentially private algorithm. Concretely, consider a setting where the domain $\calD$ is a dictionary of elements (e.g., words in a language), and one wants to estimate the frequency of elements in this domain, with each data record $\boldx$ holding an element. One natural way to encode $\boldx$ is via \emph{one-hot} encoding if the cardinality of $\calD$ is \emph{not too large}. 
For large domains, in order to reduce communication/storage one can use a sketching algorithm (e.g., count-mean-sketch~\cite{cormode2005improved}) to establish a compact encoding.
(For any given dataset and task, and at any given level of privacy, the choice of such an encoding will impact the empirical utility;
we explore this empirical tradeoff in the evaluations of Section \ref{sec:expt}.)
\else
\textcolor{red}{Here we used to have a paragraph about encoding. I guess we didn't need to talk about encoding if we don't talk about sketching. Is that correct?}
\fi

\mypar{Attribute fragments} Respondents' data records may hold multiple independent or dependent aspects.
We can, without restriction, consider the setting where each such data record $\boldx$ is encoded as a binary vector
with $k$ or fewer bits set (i.e., no more than $k$ non-zero coordinates).
We can refer to each of those $k$ vector coordinates as attributes
and write $\boldx=\sum\limits_{i=1}^k \boldx_i$, where each $\boldx_i$ is a \emph{one-hot} vector.
Given any bounded LDP budget, there are two distinct choices for satisfying privacy by  randomizing $\boldx$:
either send each $\boldx_i$ independently through the randomizer $\calR$,
splitting the privacy budget accordingly, or
sample one of the $\boldx_i$'s at random and spend all of privacy budget to send it through $\calR$.
As demonstrated empirically in Section~\ref{sec:expt},
we find
that sampling is always better for the privacy/utility tradeoff
(thereby, we verify what has been shown analytically~\cite{BST17,bun2018heavy}).
\vnoteinline{I don't see a discussion in Sec III that supports this. Also not in experiments.}
\anoteinline{@Shuang: Can you verify the reference is correct, and hence the comment by Vitaly is addressed?}
Once a one-hot vector $\boldsymbol{z}$ is sampled from $\{\boldx_i\colon i\in[k]\}$,
we establish analytically and empirically that for both local and central differential-privacy tradeoffs
it is advantageous to send each attribute of $\boldsymbol{z}$ independently to LDP randomizers that produce anonymous reports.
(There are other natural variants of attributes based on this encoding scheme e.g., in the context of learning algorithms~\cite{smith2017interaction},
but these are not considered in this paper.)

\mypar{Report fragments} Given an $\eps$ LDP budget and an encoded data record $\boldx$,
a sequence of LDP reports may be generated by multiple independent applications of the randomizer $\calR$ to $\boldx$,
while still ensuring an overall $\eps$ bound on the privacy loss.
Each such report is a \emph{report fragment},
containing less information than the entire LDP report sequence.
Anonymous report
fragments allow improved privacy guarantees in more refined threat models,
as we show in Section~\ref{sec:record_fragmenting}.

\ifsketch
\mypar{Sketch-based reports}
Locally-differentially-private variants of sketching~\cite{BST17,appledp,ghazi2019private} have been used for optimizing communication, computation, and storage tradeoffs w.r.t. privacy/utility in the context of estimating distributions. Given a domain $\{0,1\}^k$, the main idea is to reduce the domain to $\{0,1\}^{\sks}$, with $\sks \ll k$, via hashing and then use locally private protocols to operate over a domain of size $\sks$. To avoid significant loss of information due to hashing, and in turn boost the accuracy, the above procedure is performed with multiple independent hash functions.
Sketching techniques can be used in conjunction with all of the fragmentation schemes explored in this paper,
with the benefits of sketching extending seamlessly, as we corroborate in experiments.

As a warning to practitioners,
we note that sketching must be deployed carefully,
and only in conjunction with tuning of its parameters.
Sketching will add additional estimation error---on top of the error introduced by differential privacy---and
this error can easily exceed the error introduced by differential privacy,
unless the sketching parameters are tuned to a specific, known target dataset,

We also observe that
sketching is not a requirement
for practical deployments
in regimes with high local-differential privacy, such as those explored in this paper.
A primary reason for using sketching is to reduce communication cost, by reducing the domain size from $k$ to $\sks\ll k$,
but 
for high-epsilon LDP reports
only a small number of bit may need to be sent, even without sketching.
If the probability of flipping a bit is $p$ for one-hot encodings of a domain size $d$,
then only
the indices of $p (d-1) + (1-p)$ bits need be sent---the non-zero bits---and
each such index can be sent in $\log_2 d$ bits or less.
For high-epsilon one-hot-encoded LDP reports, which apply small~$p$ to domains of modest size~$d$,
the resulting communication cost
may well be acceptable, in practice.

Table~\ref{fig:real-1-crowd} shows some examples of applying one-hot and sketch-based LDP report encodings
to a real-world dataset, with sketching configured as in a practical deployment~\cite{appledp}.
As the table shows,
for a central privacy guarantee of $\epsc=1$, only the indices of one or two bits must be sent in sketch-based LDP reports;
on the other hand, five or six bit indices must be sent using one-hot encodings
(because the attribute-fragmented LDP reports must have $\epslt{\infty}=12.99$,
which corresponds to $p=2.28\times 10^{-6}$).
However, this sixfold increase in communication cost is coupled
with greatly increased utility:
the top $10\mathrm{,}000$ items can be recovered quite accurately using the one-hot encoding, while only the top $100$ or so can be recovered using the count sketch.
Such a balance of utility/privacy and communication-cost tradeoffs
arises naturally in high-epsilon one-hot encodings, while with sketching
it can be achieved
only by hand-tuning the configuration of
sketching parameters to the target data distribution.
\fi

\subsection{Anonymity and Attack Models}
\label{sec:attackmodel}
\label{sec:esa-prelim}

The basis of our
attack model
are the guarantees
of local differential privacy,
which are quantified by  $\epsl$
and place an
$e^{\epsl}$
worst-case upper bound on the information loss
of each \emph{respondent}
that contributes reports to the statistical monitoring.
These guarantees are consistent with a particularly simple attack model
for any one respondent,
because the~$\epsl$ privacy guarantees
hold true
even when all other parties (including other respondents)
conspire to attack them---as
long as that one respondent
constructs reports correctly
using good randomness.
We write
$\epslt{\infty}$
when this guarantee
holds even if
the respondent invokes the protocol multiple (possibly unbounded) number of times, without changing its private input.

Statistical reporting with strong privacy
is also quantified
by $\epsc$,
as its goal is to ensure
that a central \emph{analyzer}
can never reduce by more than $e^{\epsc}$
its uncertainty about any respondent's data---even
in the worst case,
for the most vulnerable and unlucky respondent.
The analyzer is assumed to be a potential attacker
which may adversarially
compromise or collude with
anyone involved in the statistical reporting;
if successful in such attacks,
the analyzer may be able to reduce
their uncertainty
from $e^{\epsc}$ to $e^{\epsl}$
for at least some respondents.
Unless the analyzer is successful in such collusion,
our attack model
assumes that
its $\epsc$ privacy guarantee will hold.

In addition to the above,
as in the ESA~\cite{prochlo} architecture,
an intermediary
termed the \emph{shuffler}
can be used to ensure the anonymity of reports
without having visibility into report contents (thanks to cryptography).
Our attack model includes such a middleman
even though it adds complexity,
because anonymization
can greatly strengthen
the $\epsc$ guarantee that guards privacy against the prying eyes of the analyzer,
as established
in recent amplification results~\cite{soda-shuffling,cheu2019distributed,privacy-blanket}.
However,
our attack model
requires that the shuffler
can learn nothing about the content of reports
unless it colludes with the analyzer
(this entails assumptions,
e.g., about traffic analysis,
which are discussed below).

\mypar{Anonymization Intermediary}
In our attack model,
the shuffler is assumed to be an intermediary layer
that consists of $K$ independent \emph{shuffler instances}
that can transport multiple \emph{reporting channels}.
The shuffler
must be a well-authenticated, networked system
that can securely receive and collect reports from identifiable
respondents---simultaneously, on separate reporting channels, to efficiently use resources---and
forward those reports to the analyzer after their anonymization,
without ever having visibility into report contents (due to encryption).
Each shuffler instance
must separately collect reports on each channel
into a sufficiently large set, or crowd,
from enough distinct respondents,
and must output that crowd
only to the analyzer destination that is appropriate for the channel,
and only in a manner that guarantees anonymity: i.e., that origin, order, and timing of report reception is hidden.
In particular,
this anonymity can be achieved by
outputting the crowd's records in a randomly-shuffled order, stripped of any metadata.

Our attack model
abstracts away from the
specifics of disparate anonymity techniques
and is \emph{not}
limited to shuffler instances that
output reports in a randomly shuffled order.
Depending on the primitives used to encrypt the reports,
shuffler instances may output
an aggregate summary of the reports
by using a commutative, associative operator
that can compute such a summary without decryption.
Such anonymous summaries are less general than shuffled reports
(from which they can be constructed by post-processing),
but they can be practically
computed using cryptographic means~\cite{prio,secureagg,ishai2006cryptography}
and have seen formal analysis~\cite{ghazi2019private,balle2019summation}.
However,
if the output is only an aggregate summary, the shuffler instance
must provide quantified means of guaranteeing the integrity of that summary;
in particular,
summaries must be robust
in the face of corruption
or malicious construction
of any single respondent's report,
e.g., via techniques like those in~\cite{prio}.

By utilizing $K$ separate shuffler instances, each in a different trust domain,
our attack model
captures the possibility of partial compromise.
The $K$ instances
should be appropriately isolated
to represent a variety of different trust assumptions,
e.g., by being resident in separate administrative domains (physical, legislative, etc.);
thereby,
by choosing to which instance they sent their reports,
respondents can limit their potential privacy risk
(e.g., by choosing randomly, or in a manner that represents their trust beliefs).
Thereby, respondents may retain some privacy guarantees
even when certain shuffler instances
collude with attackers
or are compromised.
The effects of any compromise
may be further limited, temporally,
in realizations
that regularly reset to a known good state;
when a respondent
uses fragmentation techniques to
send multiple reports,
simultaneously, or over time,
we quantify
as
$\epslt{1}$
the worst-case privacy loss
due to attacker capture
of a single report,
noting that
$\epslt{1} \leq \epslt{\infty}$
will always hold.

Our attack model assumes
a binary state for each
shuffler instance, in which it is either
fully compromised, or fully trustworthy
and, further, that the
compromise of one instance
does not affect the others.
However, notably,
in many realizations---such as those based on Prio~\cite{prio},
mixnets~\cite{tor},
or ESA's blinding~\cite{prochlo}---a
single shuffler instance
can be constructed
from $M$ independent entities,
such that attackers
must compromise all $M$ entities,
to be successful.
Thereby,
by using a large $M$ number of entities,
and placing them
in different, separately-trusted protection domains,
each shuffler instance
can be made arbitrarily trustworthy---albeit at the cost of reduced efficiency.

Our attack model
assumes that an adversary (colluding with the analyzer)
is able to monitor the network without breaking cryptography.
As a result,
attackers must not benefit from learning
the identity of shufflers or reporting channels
to which respondents are reporting;
this may entail that respondents
must send more reports, and send to more destinations than strictly necessary,
e.g., creating cover traffic
using incorrectly-encrypted ``chaff''
that will be discarded by the analyzer.
Our attack model also
abstracts away from most other concerns
relating to how information may leak
due to the manner in which respondents send reports,
such as via timing- or traffic-analysis,
via mistakes like
report encodings that accidentally include an identifier,
or include insufficient randomization such that reports can be linked
(see the PRR discussion in~\cite{rappor}),
or via respondents' participation in multiple
reporting systems that convey overlapping information.

Much like in~\cite{prochlo}, our attack model abstracts away from
the choice of cryptographic mechanisms
or how respondents acquire trusted software or keys,
and how those are updated.
Finally, our attack model also abstracts away from policy decisions
such as
which of their attributes respondents should report upon,
what privacy guarantees should be considered acceptable,
the manner or frequency by which respondents' self-select for reporting,
how they sample what attributes to report upon,
when or whether they should send empty chaff reports,
and what an adequate size of a crowd should be.

\subsection{Central Differential Privacy and Amplification by Shuffling}
\label{Sec:amplif-prelim}
To state the differential privacy guarantees that hold for the view of the analyzer (to which we often refer as central privacy) we rely on privacy amplification properties of shuffling. First results of this type were established by Erlingsson et al.~\cite{soda-shuffling} who showed that shuffling amplifies privacy of arbitrary local randomizers and Cheu et al.~\cite{cheu2019distributed} who gave a tighter analysis for the shuffled binary randomized response. Balle et al.~\cite{privacy-blanket} showed tighter bounds for non-interactive local randomizers via an elegant analysis. We state here two results we use in the rest of the paper. The first \cite[Corollary 5.3.1]{privacy-blanket} is for general non-interactive mechanisms, and the second for a binary mechanism~\cite[Corollary 17]{cheu2019distributed}.

\begin{lem}
\label{lem:gen-amplif} For $\delta \in [0,1]$ and $\epsl \leq \log(n/\log(1/\delta))/2$, the output of a shuffler that shuffles $n$ reports that are outputs of a $\epsl$-DP local randomizers satisfy $(\eps, \delta)$-DP where $\eps = O\left((e^{\epsl}-1)\sqrt{\log(1/\delta)/n}\right)$.
\label{lem:borja}
\end{lem}

\begin{lem}
Let $\delta\in[0,1]$, $n\in\mathbb{N}$, and $\lambda\in\left[14\log(4/\delta),n\right]$. Consider a dataset $X=(x_1,\ldots,x_n)\in\{0,1\}^n$. For each bit $x_i$ consider the following randomization: $\hat{x}_i\leftarrow x_i$ w.p. $\left(1-\frac{\lambda}{2n}\right)$, and $1-x_i$ otherwise. The algorithm computing an estimation of the sum $S^{\sf priv}=\frac{1}{n-\lambda}\left(\sum\limits_{i=1}^n \hat{x}_i-\frac{\lambda}{2}\right)$ satisfies $(\epsilon,\delta)$-central differential privacy where
\begin{equation}\epsilon=\sqrt{\frac{32\log(4/\delta)}{\lambda-\sqrt{2\lambda\log(2/\delta)}}}\left(1-\frac{\lambda-\sqrt{2\lambda\log(2/\delta)}}{n}\right).\label{eq:mixnet-bound}
\end{equation}
\label{lem:abcfd}
\end{lem}

We will also use the advanced composition results for differential privacy by Dwork, Rothblum and Vadhan~\cite{dwork2010boosting}
and sharpened by Bun and Steinke~\cite[Corollary 8.11]{bun2016concentrated}.
\begin{thm}[Advanced Composition Theorem~\cite{bun2016concentrated}]
Let $\calM_1, \ldots, \calM_k\colon \calD^n \times \calS \to \calS$ be algorithms such that for all $z \in \calS$, $i\in [k]$, $\calM_i(\cdot,z)$ satisfies $(\eps,\delta)$-DP.
The {\em adaptive composition} of these algorithms is the algorithm that given $D\in \calD^n$ and $z_0\in \calS$, outputs $(z_1,\ldots,z_k)$, where $z_i$ is the output of $\calM_i(D,z_{i-1})$ for $i\in [k]$. Then $\forall \,\delta'>0$ and $z_0\in \calS$, the adaptive composition satisfies $\left(k\eps^2/2 + \sqrt{k}\eps\cdot \sqrt{2\log(\sqrt{k\pi/2}\eps/\delta')}, \delta' + k\delta\right)$-DP.
\label{thm:advc}
\end{thm}

When these amplification and composition results
are used to derive
central privacy guarantees 
for %
collections of LDP reports,
the details matter.
Depending on
how information
is encoded and fragmented into the LDP reports that are sent by each respondent,
the resulting central privacy guarantee
that can be derived may vary greatly.
For some types of LDP reports,
new amplification results may be required
to precisely account for the balance of utility and privacy.
Specifically---as
described in the next section and further detailed in our experiments---for
sketch-based LDP reports,
more precise analysis have yet to be developed;
as a result, the central privacy guarantees
that are known to hold for anonymous, sketch-based reporting
are quite unfavorable compared to those
known to hold
for one-hot-encoded LDP reports.

\begin{algorithm}[t]
	\caption{$\afrag(\calR_{k\text{-RAPPOR}})$: Attribute fragmented $k$-RAPPOR.}
	\begin{algorithmic}[1]
	    \REQUIRE Respondent data $x \in \calD$, LDP parameter $\epsl$.
	    \STATE Compute $\boldx \in \zo^k$, a one-hot encoding of $x$.
	    \STATE For each $j \in [k]$, define \[\calR_j(b, \eps) := \left\{ \begin{array}{cr}
	    b & \text{w.p.~}\; e^{\eps}/\left(1+e^{\eps}\right) \\
	    1-b & \text{w.p.~}\; 1/\left(1+e^{\eps}\right)\end{array}\right.
	    \]
	    \STATE \textbf{send} $\calR_j(x^{(j)}, \epsl)$ to shuffler $\calS_j$ for $j \in [k]$
	\end{algorithmic}
	\label{Alg:Frag}
\end{algorithm}

\section{Histograms via Attribute Fragmenting}
\label{sec:attribFrag}
In this section we revisit and formalize the idea of \emph{attribute fragmenting} \cite{prochlo}.
We demonstrate its applicability in estimating high-dimensional histograms\footnote{Following a tradition in the differential-privacy literature~\cite{dwork2014algorithmic},
  this paper uses the term \emph{histogram} 
  for a count of the frequency of each distinct element in a multiset drawn from a finite domain of elements.}
with strong privacy/utility tradeoffs. By applying recent results on \emph{privacy amplification by shuffling} \cite{soda-shuffling,cheu2019distributed,BalleBG18}, we show that attribute fragmenting helps achieve \emph{nearly optimal} privacy/utility tradeoffs both in the central and local differential privacy models w.r.t~the $\ell_\infty$-error in the estimated distribution. Through an extensive set of experiments  with data sets having \emph{long-tail distributions} we show that attribute fragmenting help recover much larger fraction of the tail for the same central privacy guarantee (as compared to generically applying privacy amplification by shuffling for locally private algorithms \cite{soda-shuffling,privacy-blanket}). In the rest of this section, we formally state the idea of attribute fragmenting and provide the theoretical guarantees. We defer the experimental evaluation to Section~\ref{sec:fthist}.

Consider a local randomizer $\calR$ taking inputs with $k$ attributes, i.e., inputs are of the form $\boldx_i = (x_i^{(1)}, \ldots, x_i^{(k)})$.
\emph{Attribute fragmenting} comprises two ideas:
First, decompose the local randomizer $\calR$ into $\afrag(\calR) := (\calR_1, \ldots, \calR_k)$, a tuple of independent randomizers each acting on a single attribute.
Second, have each respondent report $\calR_j(x_i^{(j)})$ to $\calS_j$,  one of $k$ \emph{independent} shuffler instances $\calS_1, \ldots, \calS_k$ that separately anonymize all reports of a single attribute.
Attribute fragmenting is applicable whenever LDP reports about individual attributes are sufficient for the task at hand, such as when estimating marginals.

Attribute fragmenting can also be applied to scenarios where the respondent's data is not naturally in the form of fragmented tuples. Thus, we can consider two broad scenarios when applying attribute fragments: (1) \emph{Natural} attributes such as when reporting demographic information about age, gender, etc., which constitute the attributes. Another example would be app usage statistics across different apps with disjoint information about load times, screen usage etc. (2) \emph{Synthetic} fragments where a single piece of respondent data can be \emph{cast} into a form that comprises several attributes to apply this fragmenting technique.

An immediate application of (synthetic) fragments is to the problem of learning histograms over a domain $\calD$ of size $k$ where each input $x_i \in \calD$ can be represented as a ``one-hot vector'' in $\zo^k$. Algorithm~\ref{Alg:Frag} shows how to (naturally) apply attribute fragmenting when the local randomizer $\calR$ is what is referred to as the $k$-RAPPOR randomizer~\cite{Kairouz-extremal}. Theorems~\ref{thm:att-frag-privacy} and \ref{thm:att-frag-accuracy} demonstrate the near optimal utility/privacy tradeoff of this scheme. We remark that Algorithm \ref{Alg:Frag} is briefly described and analyzed in~\cite{cheu2019distributed} (for replacement LDP).

To estimate the histogram of reports from $n$ respondents, the server receives and sums up bits from each shuffler instance $\calS_j$ to construct attribute-wise sums. The estimate for element $j \in \calD$ is computed as:
\[
\hat{h}_j = \frac{1}{n} \cdot \frac{e^{\epsl}+1}{e^{\epsl}-1} \cdot \underbrace{\sum_{i=1}^n\calR_j(x_i^{(j)}, \epsl)}_{\text{from shuffler $\calS_j$}} - \frac{1}{e^{\epsl}-1}.
\]

We show that $\afrag(\calR_{k\text{-RAPPOR}})$ achieves nearly optimal utility/privacy tradeoffs both for local and central privacy guarantees. Accuracy is defined via the $\ell_\infty$ error: $\alpha := \max\limits_{j\in[k]}\left|\hat{h}_j-\frac{1}{n}\sum_{i=1}^n x_i^{(j)}\right|$.

Informally, the following theorems state that in the high-epsilon regime, $\afrag(\calR_{k\text{-RAPPOR}})$ achieves privacy amplification satisfying $\left(O(e^{\epsl/2}/\sqrt{n}), \delta\right)$-central DP, and achieves error bounded by $\Theta\left(\sqrt{\frac{\log k}{ne^{\epsl}}}\right)$ and $\Theta\left(\frac{\sqrt{\log k}}{n\epsc}\right)$ in terms of its local ($\epsl$) and central ($\epsc$) privacy  respectively. Proofs are deferred to Appendix \ref{app:attribFrag}.

Standard lower bounds for central differential privacy imply that the dependence of $\alpha$ on $k$, $n$, and $\epsc$ are within logarithmic factors of \emph{optimal}. To the best of our knowledge, the analogous dependence for $\epsl$ in the local DP model is the best known.

\begin{thm}[Privacy guarantee]
\label{thm:att-frag-privacy}
Algorithm $\afrag(\calR_{k\text{-RAPPOR}})$ satisfies removal $\epsl$-local differential privacy and for $\epsl\in \left[1,\log n-\log\left(14\log\left(\frac{4}{\delta}\right)\right)\right]$ and $\delta\geq n^{-\log n}$, $\afrag(\calR_{k\text{-RAPPOR}})$ satisfies removal $(\epsc, \delta)$-central differential privacy in the Shuffle model where:
\[
\epsc = \sqrt{\frac{64 \cdot e^{\epsl}\cdot\log(4/\delta)}{n}}.
\]\label{thm:ahd}
\end{thm}
\begin{thm}[Utility/privacy tradeoff]
\label{thm:att-frag-accuracy}
Algorithm $\afrag(\calR_{k\text{-RAPPOR}})$ simultaneously satisfies $\epsl$-local differential privacy, $(\epsc, \delta)$-central differential privacy (in the Shuffle model), and has $\ell_\infty$-error at most $\alpha$ with probability at least $1-\beta$, where
\begin{small}
\begin{equation*}
\alpha=\Theta\left(\sqrt{\frac{\log(k/\beta)}{ne^{\epsl/2}}}\right);\text{ equiv.\ } \alpha=\Theta\left(\frac{\sqrt{\log(k/\beta)\log(1/\delta)}}{n\epsc}\right).
\end{equation*}
\end{small}
\label{thm:ahds}
\end{thm}

Unlike
one-hot-encoded LDP reports,
for deployed
sketch-based LDP reporting schemes---such
as the count sketch of~\cite{BST17,appledp}---there
are no analyses 
that are known to 
derive
precise central privacy guarantees,
while both leveraging amplification-by-shuffling
and being able to account for attribute fragmentation.
One known approach to analyzing sketch-based LDP reports
is to ignore all fragmentation
and apply a generic privacy amplification-by-shuffling result, such as Lemma~\ref{lem:borja};
since it ignores attribute fragments
its $\epsl$ dependence is $e^{\epsl}$, instead of $e^{\epsl/2}$,
and its central privacy bound
suffers compared to that of $k$-RAPPOR.
A second known approach
observes that the randomizer for each individual hash function is an instance of $k$-RAPPOR,
for which the lower $e^{\epsl/2}$-type dependence holds.
However, for this second analysis,
the effective size of the crowd $n$ is reduced by the number of hash functions used---making
anonymity less effective in amplifying privacy---and
a large number of hash functions is often required to achieve good utility.
Thus,
for sketch-based LDP reports,
the best known privacy/utility tradeoffs may not be favorable,
in the eyes of practitioners,
compared to those of one-hot-encoded LDP reports.
\unote{comment on communication cost?}

In real-world applications---unlike what is proposed above---the
number of attributes 
may be far too large
for it to be practical
to use a separate shuffler instance for each attribute.
For example,
this can be seen in
the datasets of Table~\ref{tab:dataset_statistics},
which we use in our experiments.

However,
in our attack model,
efficient realizations of shuffling
are possible
for high-epsilon LDP reports with attribute fragmenting.
For this, there need only be
$K$ shuffler instances
with each instance having a separate reporting channel
for every single attribute,
for a number $K$ that is sufficiently large
for the dataset and task at hand.
For high-epsilon LDP reports,
the report encoding can be constructed
such that
each respondent will send
only a few LDP reports for a few attributes---and
if this number is small enough,
those reports can still be arranged to be sent to independent shuffler instances, 
e.g., in expectation, by randomly selecting the destination shuffler instance.
In particular,
for the experiments of Table~\ref{fig:unified_attr},
our assumption of independence will hold
as long as the number of $K$ shuffler instances
is large enough for each bit
to be sent to a separate instance,
with high confidence, in expectation.

\section{Report Fragmenting}
\label{sec:record_fragmenting}

While the shuffle model enables respondents to send randomized reports of local data with large local differential privacy values and still enjoy the benefits of privacy amplification, it might be desirable to further reduce the risk to respondents' privacy by reducing the privacy cost of each individual report. As an example, consider randomizing a single bit with the randomizer defined in Section \ref{sec:attribFrag}. For $\epsl=10$, the probability of sending a flipped bit is $\approx e^{-10}$. Therefore, given a report from a respondent, there is a roughly $99.996\%$ chance of the report being identical to the respondent's data. This probability drops to $63.21\%$ with $\epsl=1$.

Extending the ideas of fragmenting from Section~\ref{sec:attribFrag}, one might be tempted to consider the following different way to fragment the reports: given an LDP budget of $\epsl$, send several reports (specifically, $\epsl/\epslt{f}$ reports) each with LDP $\epslt{f} \ll \epsl$. While this certainly reduces the privacy cost of each report, it has an impact on the utility. To replace one report of $\epsl=4$, with several reports of $\epslt{f}=2$ while achieving the same utility one would need roughly $\exp(\epsl/\epslt{f})=\exp(2)\sim 7$ reports, which blows up the local privacy loss. Equivalently (see Corollary~\ref{cor:recFrag} in Appendix~\ref{app:record_fragmenting}), for a given local privacy budget $\epsl$, the $\ell_\infty$ error increases by a factor of roughly $\sqrt{\exp(\epsl/2)/\epsl}$. %

\begin{algorithm}[t]
	\caption{$\rfrag(\calR_b, \calR_f, \epslt{b}, \epslt{f}, \tau)$: Report fragmenting.}
	\begin{algorithmic}[1]
	    \REQUIRE Respondent data $x$, LDP $\epslt{b}$, fragment LDP $\epslt{f}$, number of fragments~$\tau$
	    \STATE $x' \leftarrow \calR_b(x;\epslt{b})$
	    \FOR{$i \in [\tau]$}
	        \STATE $y_i = \calR_f(x'; \epslt{f})$
	    \ENDFOR
	    \STATE \textbf{send} $(i, y_i)$ to shuffler $\calS_i$ for $i \in [\tau]$
	\end{algorithmic}
	\label{Alg:ReportFrag}
\end{algorithm}

\mypar{Report fragments with privacy backstops} Inspired by the concept of a \emph{permanent} randomized response~\cite{rappor}, we propose a simple fix to the unfavorable tradeoff described above. Instead of working with reports of local privacy $\epslt{f}$ on the original respondent data, the respondent first constructs a randomized response of the original data with a higher epsilon $\epslt{b}$ (for backstop) and only outputs lower epsilon reports on this randomized data. More precisely, given $\eps$-DP local randomizers $\calR_b(\cdot;\eps)$ and $\calR_f(\cdot;\eps)$, on input data $d$, a backstop randomized report $d' \leftarrow \calR_b(d;\epslt{b})$ is first computed. Then, we fragment the report into several reports $r_i \leftarrow \calR_f(d'; \epslt{f})$ for several independent applications of $\calR_f$.

We claim to get the best of both worlds with this construction. With sufficiently many reports, we get utility/privacy results that are essentially what we can achieve with local privacy budget of $\epslt{b}$ while ensuring that each report continues to have small LDP. The backstop ensures that even with sufficiently many reports sent to the same shuffler, the privacy guarantee does not degrade linearly with the number of reports, but stops degrading beyond the backstop $\epslt{b}$. The only price we pay is in additional communication overhead. The number of fragments is only constrained by the communication costs, though beyond a few fragments there are diminishing returns for utility (at no cost to privacy).

The following theorem states the privacy guarantees of report fragmenting. It analyzes the situation in which an adversary has gained access to $t \leq \tau$ fragments. It demonstrated that the privacy of a respondent degrades gracefully as more fragments are exposed to an adversary.

\begin{thm}
\label{thm:record_frag}
For any $\epslt{f},\epslt{b}>0$, an $\epslt{b}$-DP local randomizer~$\calR_b$, an $\epslt{f}$-DP local randomizer $\calR_f$, an integer $\tau$, and a set of indices $J\subseteq [\tau]$ of size $t$, consider the algorithm $\calM_J$ that for $(y_1,\ldots,y_\tau) = \rfrag(\calR_b, \calR_f, \epslt{b}, \epslt{f}, \tau)$ outputs $y_J = (y_i)_{i\in J}$. Then
$\calM_J$ is an $\eps$-DP local randomizer for $\eps = \ln\left(\frac{e^{\epslt{b}+t\epslt{f}}+1}{e^{\epslt{b}}+e^{t\epslt{f}}}\right) \leq \min\{\epslt{b},t\epslt{f}\}$.
\end{thm}
We stated Theorem \ref{thm:record_frag} for the standard replacement DP. If $\calR_b$ satisfies only removal $\epslt{b}$-DP then $\calM_J$ has the same $\epslt{b}$ for removal DP.
The proof is based on a general result showing how DP guarantees
are amplified when each data element is preprocessed by a local randomizer.
(Details %
in Appendix \ref{app:record_fragmenting}.)

\mypar{Report fragmenting for histograms} Here we instantiate report fragmenting in the context of histograms. Recall, for the histogram computation problem described in Section \ref{sec:attribFrag}, each data sample is $\boldx=\left(x^{(1)},\cdots,x^{(k)}\right)$ is a one-hot vector in $k$ dimensions. In report fragmenting with privacy backstop, we do the following: For each $i\in[k]$, we run an instance of Algorithm \ref{Alg:ReportFrag} independently, with $x^{(i)}$ as respondent data. One can view the set of report fragments generated by all the execution of Algorithm \ref{Alg:ReportFrag} as a matrix: $M(\boldx)=[m_{i,j}]_{\tau\times k}$, where $m_{i,j}$ refers to the $i$-th report generated for the $j$-th domain element.
To be most effective, report fragments should be sent according to respondent's trust in shuffler instances.

For the report fragmenting above, we obtain the following accuracy/privacy tradeoff (proof in Appendix \ref{app:record_fragmenting}).

\begin{thm}[Utility/privacy tradeoff]
For a per-report local privacy budget of $\epslt{f} > 1$, backstop privacy budget of $\epslt{b}$, and number of reports $\tau$, Algorithm $\rfrag(\afrag(\calR_{k\text{-RAPPOR}}), \tau)$ satisfies removal  $\ln\left(\frac{e^{\epslt{b}+\tau\epslt{f}}+1}{e^{\epslt{b}}+e^{\tau\epslt{f}}}\right)$-local differential privacy and $(\epsc,\delta)$-central DP where, for any $\delta < 1/2$:
\[
\epsc = \min\left\{\sqrt{\frac{8\tau\epslt{f} \log^2(\tau\epslt{f}/\delta)}{n}}, \sqrt\frac{64{e^{\epslt{b}}\log(4/\delta)}}{n}
\right\},
\]
has accuracy $\alpha$ with probability at least $1-\beta$ with:

\begin{equation*}
\alpha=O\left(\sqrt{\frac{\log(k/\beta)}{n\tau\epslt{f}}}+\sqrt\frac{\log(k/\beta)}{ne^{\epslt{b}}}\right).
\end{equation*}

\label{cor:recFrag}
\end{thm}

\section{Crowds and crowd IDs}
\label{sec:crowd}

Foundational to this work
is the concept of a crowd:
a sufficiently large set of LDP reports
gathered from a large enough set of distinct respondents,
such that each LDP report can become ``lost in the crowd'' and thereby anonymous.
As discussed in Section~\ref{Sec:prelims},
the shuffler intermediary
must ensure, independently,
that a sufficiently large crowd
is present
on every one of the shuffler's reporting channels.
Channels are
equivalent to (but more efficient than)
a distinct shuffler 
with its own public identity,
and channels are only hosted on a single shuffler 
for efficiency.
As such,
the identity of the channel
that a report is sent on
must be assumed to be public.

As an alternative,
the ESA architecture
described how respondents
could send LDP reports annotated by
a ``Crowd~ID''
that
could be hidden
by cryptographic means
from both network attackers
and the shuffler intermediaries
(using blinded shuffling).
In ESA, the reports for each Crowd~ID
were grouped together, shuffled separately, and
only output if their cardinality was sufficient;
furthermore, this cardinality threshold
was randomized for privacy.
Revisiting this alternative,
we find that annotating LDP reports by IDs
can be helpful, in those cases where
respondents have an existing reason to
publicly self-identify as belonging to a data partition---e.g., 
because
they are unable to hide
their use of certain computer hardware or software,
or do not want to hide
their coarse-grained location, nationality, or language preferences.
On the other hand,
given the strength of the recent privacy amplification results based on anonymity,
we find
little to no value remaining in
the use of
ESA's Crowd~IDs as
a distinct reporting channel
(i.e., reporting some data
via an LDP report and some data via that report's ID annotation).

We can formally define ESA's Crowd~IDs
as being the set of indices $\{1,\ldots,m\}$
for any partitioning
of an
underlying dataset of LDP records
$D=\{x_1,\dots,x_n\}\in\calD^n$
into disjoint subsets
$D=D_1\cup\dots\cup D_m$.
For tasks like those in Section~\ref{sec:attribFrag},
separately analyzing each subset $D_i$
can significantly improve utility
whenever reports that carry the same signal
are partitioned
into the same subset---i.e.,
if reports about the same values
are associated with the same ID.
The expected (un-normalized) $\ell_\infty$-norm estimation error
for each partition $D_i$
will be
$\sqrt{{|D_i|}/{e^{\epsl}}}$,
if the records in the dataset have an $\epsl$ privacy guarantee,
compared to
$\sqrt{{|D|}/{e^{\epsl}}}$
for the whole dataset.
Therefore,
for equal-size Crowd~ID partitions,
the utility of monitoring can be improved
by a factor of $\sqrt{m}$,
and, if partition sizes vary a lot,
the estimation error 
may be improved much more
for the smaller partitions.

However,
the utility improvement of Crowd~IDs must come at a cost to privacy.
After all,
Crowd~IDs are visible to the analyzer
and can be considered 
as the first component of a report pair, along with their associated LDP report.
As such,
their total privacy cost
can only be bounded by
$\epsl + \widehat{\epsl}$:
the sum of each LDP report's $\epsl$ bound 
and any bound $\widehat{\epsl}$ that holds for its associated Crowd~ID
(and this $\widehat{\epsl}$ may be $\infty$).

Even without a bound on the Crowd~ID privacy loss,
respondents may want to send ID-annotated LDP reports.
In particular, this may be because
partitioning is based on aspects of data that raise few privacy concerns,
or are seen as being public already (e.g., the rough geographic origin of
a mobile phone's data connection).
Alternatively, this may be because
respondents see a direct benefit from
sending reports in a manner that improves the utility of monitoring.

For example,
respondents may desire to receive improved services
by sending reports whose IDs
depend on 
the version of their software, the type of their hardware device,
and their general location (e.g., metropolitan area).
Or,
to help build better predictive keyboards,
respondents may
send LDP reports about the
words they type
annotated by their software's
preferred-language settings
(e.g., EN-US, EN-IN, CHS, or CHT);
such partitioned LDP reporting is realistic
and has been
deployed in practice~\cite{appledp,dplangmodels}.
For lack of a better term,
we can refer to
such partitioning as \emph{natural Crowd~IDs}.

However,
even when Crowd~IDs are derived from public data,
the cardinality of 
each partition may be a privacy concern---at least for small partitions---if
Crowd~IDs are derived without randomization.
The shuffler intermediary
can address this privacy concern
by applying randomized thresholding,
as outlined in the original ESA paper~\cite{prochlo}.
For a more complete description,
Algorithm~\ref{Alg:CrowdThresh}
shows how the shuffler can drop reports
before applying a fixed threshold
in order to make each partition's cardinality
differentially private;
furthermore,
formal privacy and utility guarantee is given
in Theorem~\ref{thm:privCrowds} and Theorem~\ref{thm:utilCrowdThresh}
and Appendix~\ref{app:crowd} includes proofs.

    \begin{algorithm}
        \caption{Randomized Report Deletion.}
        \begin{algorithmic}[1]
            \REQUIRE reports partitioned by Crowd ID: $\{R_i\}_{[P]}$, \\ \hspace*{3em} privacy parameters: $\left(\eps^{\sf cr},\delta^{\sf cr}\right)$.
            \FOR{$i\in[P]$}
                \STATE $n_i\leftarrow |R_i|$
                \STATE $\hat n_i\leftarrow \max\{n_i +{\sf Laplace}\left(\frac{2}{\eps^{\sf cr}}\right) - \frac{2}{\eps^{\sf cr}}\log\left(\frac{2}{\delta^{\sf cr}}\right),0\}$
                \IF{$\hat n_i \leq n_i$}
                    \STATE \small{$R'_i \leftarrow R_i \backslash \left\{(n_i-\hat n_i) \text{ uniformly chosen records} \right\}$}
                \ELSE
                    \STATE Abort
                \ENDIF
            \ENDFOR
            \RETURN The new partitioning by Crowd ID: $(R'_1, \ldots, R'_P)$
        \end{algorithmic}
        \label{Alg:CrowdThresh}
    \end{algorithm}
    
    \begin{thm}[Privacy guarantee]
    Algorithm \ref{Alg:CrowdThresh} satisfies $\left(\eps^{\sf cr},\delta^{\sf cr}\right)$-central differential privacy on the counts of records in each crowd.
    \label{thm:privCrowds}
    \end{thm}
    
    \begin{thm}[Utility guarantee]
    Algorithm \ref{Alg:CrowdThresh} ensures that for all crowds $i$, $\left|R_i \setminus R'_i\right| \leq \frac{4}{\eps^{\sf cr}}\log\left(\frac{2P}{\delta^{\sf cr}}\right)$ with prob.~$\geq 1-\delta^{\sf cr}$. 
    \label{thm:utilCrowdThresh}
    \end{thm}

\mypar{Data-derived Crowds~IDs}
In addition to natural Crowd~IDs,
ESA proposed that LDP reports could be partitioned
in a purely data-dependent manner---e.g.,
by deriving Crowd~IDs by using deterministic hash functions
on the data being reported---and
reported on the utility of such partitioning
in experiments~\cite{prochlo}.
While such \emph{data-derived Crowd~IDs}
can improve utility,
their privacy/utility tradoffs cannot compete
with those offered by recent privacy amplification results based on anonymity.
The following simple example
serves to illustrate
how amplification-by-shuffling
have made data-derived Crowd~IDs obsolete.

Let's assume LDP records are partitioned by a hash function $h\colon\calD\to[m]$,
for $m=2$,
with the output of $h$ defining a binary data-derived Crowd~ID.
For worst-case analysis,
we must assume a degenerate $h$ that maps any particular $z\in\calD$ to $0$
and all other values in $\calD$ to $1$.
Therefore, the Crowd~ID
must be treated as holding the same information
as any value $z$
contained in an LDP report with an $\epsl$ privacy guarantee;
this entails that 
the Crowd~ID must be randomized
to establish for it a privacy bound $\widehat{\epsl}$,
if the privacy loss for any value $z$ is to be limited.
As a result,
ID-annotated LDP reports
have a combined privacy bound of 
$\epsl + \widehat{\epsl}$,
and any fixed privacy budget must be
split between those two parameters.

ESA proposed that data-derived
Crowd~IDs could be subjected to little randomization
(i.e., that $\widehat{\epsl} \gg \epsl$).
Thereby,
ESA implicitly discounted the privacy loss of data-derived Crowd~IDs,
with the justification
that they were only revealed when 
the cardinality of report subsets was above a randomized, large threshold.
In certain special cases---e.g., when $\epsl=0$---such
discounting may be appropriate,
since randomized aggregate cardinality counts can limit the risk
due to circumstances like that of the degenerate hash function $h$ above.
However,
in general,
accurately accounting for the privacy loss
bounded by $\epsl + \widehat{\epsl}$
reveals that it is best to not utilize data-derived Crowd~IDs at all.
The best privacy/utility tradeoff is achieved
by setting $\widehat{\epsl}=0$ and not splitting the privacy budget at all
(cf.\ Table~\ref{fig:horse_1_report} and Table~\ref{fig:horse_1_sketch_report}),
while
amplification-by-shuffling with attribute fragmenting
can be used to establish meaningful central privacy guarantees.

\begin{algorithm}[th]
    \caption{LDP-SGD; client-side}
    \label{alg:localG}
    \begin{algorithmic}[1]
        \REQUIRE Local privacy parameter: $\epsle$, current model: $\theta_t\in\mathbb{R}^d$, $\ell_2$-clipping norm: $L$.
        \STATE Compute clipped gradient \[\boldx\leftarrow \nabla \ell(\theta_t;d)\cdot \min\left\{1,\frac{L}{\|\nabla \ell(\theta_t;d)\|_2}\right\}.\]
        \STATE $\boldz_i\leftarrow
                \begin{cases}
                    L\cdot \frac{\boldx}{\|\boldx\|_2} & \text{w.p. } \frac{1}{2} + \frac{\|\boldx\|_2}{2L}, \\
                    -L\cdot \frac{\boldx}{\|\boldx\|_2} & \text{otherwise}.
                \end{cases}$ \vspace{5pt}
            \STATE Sample $\boldv\sim_u \bfS^d$, the unit sphere in $d$ dimensions.
            \STATE $\hat{\boldz}\leftarrow
            \begin{cases}
                \text{sgn}(\langle \boldz,\boldv\rangle)\cdot \boldv  & \text{w.p. } \frac{e^{\epsle}}{1+e^{\epsle}}.\\
                -\text{sgn}(\langle \boldz,\boldv\rangle)\cdot \boldv  & \text{otherwise}.\\
            \end{cases}$
        \vspace{3pt}
        \RETURN $\hat{\boldz}$.
    \end{algorithmic}
\end{algorithm}

\section{Machine Learning in the ESA framework}
\label{sec:mlESA}

In this section we demonstrate that ESA framework is suitable for training machine learning models with strong local and central differential privacy guarantees. We show both theoretically (for convex models), and empirically (in general) that one can have \emph{strong per epoch} local differential privacy (denoted by $\epsle$), and good central differential privacy overall, while achieving nearly state-of-the-art (for differentially private models) accuracy on benchmark data sets (e.g., MNIST and CIFAR-10).

Per-epoch local differential privacy refers to the LDP guarantee for a respondent over a single pass over the dataset. Here we assume that each epoch is executed on a separate shuffler, and the adversary can observe the traffic onto \emph{only one} of those shufflers. However, it is worth mentioning that the central differential privacy guarantee we provide is \emph{over the complete execution} of the model training algorithm.

Formally, we show the following:
\begin{enumerate}
\item For convex Empirical Risk Minimization problems (ERMs), with local differential privacy guarantees per report on the data sample, and amplification via shuffling in the ESA framework, we achieve \emph{optimal} privacy/utility tradeoffs w.r.t.\ excess empirical risk and the corresponding central differential privacy guarantee.
\item Empirically, we show that one can achieve accuracies of $95\%$ on MNIST, $70\%$ on CIFAR-10, and $78\%$ on Fashion-MNIST, with per epoch $\epsle\approx 1.9$.
\end{enumerate}

\begin{algorithm}[thb]
    \caption{LDP-SGD; server-side}
    \label{alg:servrG}
    \begin{algorithmic}[1]
        \REQUIRE Local privacy budget per epoch: $\epsle$, number of epochs: $T$, parameter set: $\calC$.
        \STATE $\theta_0\leftarrow\{0\}^d$.
        \FOR{$t\in[T]$}
            {\STATE Send $\theta_t$ to all clients.}
            {\STATE Collect shuffled responses $(\hat{\boldz}_i)_{i\in [n]}$. \label{step:shuff}}
            \STATE Noisy gradient: \begin{small}$\boldg_t\leftarrow\frac{L\sqrt \pi}{2}\cdot\frac{\Gamma\left(\frac{d-1}{2}+1\right)}{\Gamma\left(\frac{d}{2}+1\right)}\cdot \frac{e^{\epsle}+1}{e^{\epsle}-1}\left(\frac{1}{n}\sum\limits_{i\in [n]} \hat{\boldz}_i\right)$\end{small}.
            \STATE Update: $\theta_{t+1}\leftarrow \Pi_{\calC}\left(\theta_t - \eta_t\cdot \boldg_t\right)$, where $\Pi_\calC(\cdot)$ is the $\ell_2$-projection onto set $\calC$, and $\eta_t=\frac{\|\calC\|_2\sqrt{n}}{L\sqrt d}\cdot\frac{e^{\epsle} - 1}{e^{\epsle} + 1}$.
        \ENDFOR
        \RETURN $\theta_{\sf priv}\leftarrow\theta_T$.
    \end{algorithmic}
\end{algorithm}

In the rest of this section, we state the algorithm, privacy analysis, and the utility analysis for convex losses. We defer the empirical evaluation to Section \ref{sec:mlESAexpt}.

\mypar{Empirical Risk Minimization (ERM)} Consider a dataset $D=(x_1,\ldots,x_n)\in\calD^n$, a set of models $\calC\subseteq\mathbb{R}^d$ which is not necessarily convex, and a loss function $\ell\colon\calC\times\calD\to\mathbb{R}$. The problem of ERM is to estimate a model $\hat\theta\in\calC$ such that: \[R(\hat\theta):=\frac{1}{n}\sum\limits_{i=1}^n\ell(\hat\theta;x_i)-\min\limits_{\theta\in\calC}\frac{1}{n}\sum\limits_{i=1}^n\ell(\theta;x_i)\] is small. In this work we revisit the locally differentially private SGD algorithm of Duchi et al.~\cite{duchi2018minimax}, denoted LDP-SGD (Algorithms \ref{alg:localG} and \ref{alg:servrG}), to estimate a $\theta_{\sf priv}\in\calC$ s.t. i) $R(\theta_{\sf priv})$ is small, and ii) the computation of $R(\theta_{\sf priv})$ satisfies per-epoch local differential privacy of $\epsle$, and overall central differential privacy of $(\epsc,\delta)$ (Theorem \ref{thm:learning}). We remark that, by adapting the analysis from \cite{BassilyFTT19}, one can similarly address the problem of stochastic convex optimization in which the goal is to minimize the expected population loss on a dataset drawn i.i.d.~from some distribution.
At a high level, LDP-SGD follows the following template of noisy stochastic gradient descent~\cite{williams-mcsherry,song2013stochastic,BST14}.
\begin{enumerate}
    \item \textbf{Encode:} Given a current state $\theta_t$, apply $\epsle$-DP randomizer from \cite{duchi2018minimax} to the gradient at $\theta_t$ on all (or a subset of) the data samples in $D$.
    \item {\bf Shuffle:} Shuffle all the gradients received.
    \item {\bf Analyze:} Average these gradients, and call it $\boldg_t$.  Update the current model as $\theta_{t+1}\leftarrow \theta_t-\eta_t\cdot \boldg_t$, where $\eta$ is the learning rate.
    \item Perform steps (1)--(3) for $T$ iterations.
\end{enumerate}

In Theorem \ref{thm:learning}, we state the privacy guarantees for LDP-SGD. Furthermore, we show that under central differential guarantee achieved via shuffling, in the case of convex ERM (i.e., when the the loss function $\ell$ is convex in its first parameter), we are able to recover the \emph{optimal privacy/utility tradeoffs (up to logarithmic factors in $n$)} w.r.t.\ the central differential privacy stated in \cite{BST14}. %
(proof in
Appendix \ref{app:learning}).
\begin{thm}[Privacy/utility tradeoff]
Let per-epoch local differential privacy budget be $\epsle\leq(\log n)/4$.
\begin{enumerate}
    \item {\bf Privacy guarantee; applicable generally:} Over $T$ iterations, in the shuffle model, LDP-SGD satisfies $\left(\epsc,\delta\right)$-central differential privacy where:
    \[
        \epsc = O\left(\frac{e^{\epsle}-1}{\sqrt{n}}\cdot \sqrt{T\log^2(T/\delta)}\right).
    \]
    \item {\bf Utility guarantee; applicable with convexity:} If we set $T=n/\log^2 n$, and the loss function $\ell(\cdot;\cdot)$ is convex in its first parameter and $L$-Lipschitz w.r.t.\ $\ell_2$-norm, the expected excess empirical loss satisfies
        \alequn{&\E\left[\frac{1}{n}\sum\limits_{i=1}^n\ell(\theta_{\sf priv};x_i)\right]-\min\limits_{\theta\in\calC}\frac{1}{n}\sum\limits_{i=1}^n\ell(\theta;x_i)
    \\&
    =O\left(\frac{L\|\calC\|_2\sqrt{d}\log^2 n}{n}\cdot\frac{e^{\epsle} + 1}{e^{\epsle} -1 }\right).}
    Here $\|\calC\|_2$ is the $\ell_2$-diameter of the set $\calC$.
\end{enumerate}
\label{thm:learning}
\end{thm}

\mypar{Reducing communication cost using PRGs}
LDP-SGD is designed to operate in a distributed setting and it is useful to design techniques to minimize the overall communication from devices to a server. Observe that in the client-side algorithm (Algorithm \ref{alg:localG}) the only object that depends on data is the sign of the inner product in the computation of $\hat \boldz$. By agreeing with the server on a common sampling procedure $\mathsf{Samp}\colon\zo^\mathsf{len} \rightarrow \mathbf{S}^d$ taking $\mathsf{len}$ uniform bits and producing a uniform sample in $\mathbf{S}^d$, clients can communicate $\text{sgn}(\langle \boldz,\mathsf{Samp}(r)\rangle)$ and randomness $r$ instead of~$\hat \boldz$. This can be further minimized by replacing randomness $r$ of length $\mathsf{len}$ with the seed $s$ of length 128 bits and producing $r \leftarrow \mathsf{PRG}(s)$ where $\mathsf{PRG}$ is a pseudorandom generator stretching uniform short seeds to potentially much longer pseudorandom sequences. Thus, communication can be reduced to~129 bits by sending $(\text{sgn},s)$ and the server reconstructing $\hat \boldz = (\text{sgn}) \mathsf{Samp}(\mathsf{PRG}(s))$.

Note that only the utility of this scheme is affected by the quality of the pseudorandom generator (i.e., the uniform randomness of the PRG).
Revealing the PRG seed $s$ is equivalent to publishing $\boldv$, which is independent of the user's input $\boldz$;
therefore, reducing communication through the use of a PRG with suitable security properties
does not affect the privacy guarantees of the resulting mechanism.

\newif\ifsmother\smothertrue
\newif\ifhorseRev\horseRevtrue
\newif\ifmap\maptrue
\newif\ifgirl\girltrue
\newif\ifreal\realtrue
\section{Experimental Evaluation}
\label{sec:expt}

\newcommand{\bitsonehot}[0]{\textrm{\#bits\textsuperscript{1-hot}}}
\newcommand{\bitssketch}[0]{\textrm{\#bits\textsuperscript{sketch}}}

\begin{table*}[p]
  \caption{Results of experiments reconstructing the heavy-hitters dataset whose distribution is given in Table~\ref{tab:dataset_statistics}.
    Different utility results from anonymous LDP reports with attribute- and report fragmenting (with $\tau=4$, $16$ and $256$ reports),
    at central privacy $(\epsc, \deltac)$-central DP with $\deltac = 5\times 10^{-10}$.
    We report the expected number of bits set (and, therefore, the messages sent)
    for one-hot encoding and count-sketch with attribute fragmenting, represented by \bitsonehot\ and \bitssketch,
    for sketch-based reports using the parameters of Apple's real-world deployment~\cite{appledp}.
}
\label{fig:real-1-crowd}
\centering
\newcolumntype{D}{ >{\arraybackslash} m{5cm} }
\newcolumntype{N}{ >{\centering\arraybackslash} m{5.4cm} }
\begin{tabular}{ D N N }
\makecell[c]{Privacy Guarantees}
& \makecell{One-hot encoding \\(domain size $1\textrm{,}778\textrm{,}120$)}
& \makecell{Count sketch encoding \\($1\textrm{,}024$ hash functions, sketch size $65\textrm{,}536$)}\\
\toprule

\multicolumn{3}{r}{\includegraphics[scale=.18]{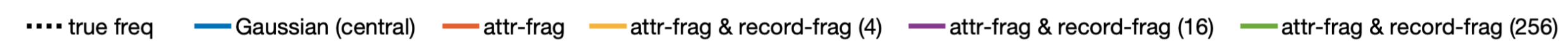}} \\

\vspace*{-11ex}
\begin{tabular}{l}
For one-hot encoding, $\epsc=0.0025$ \\ 
For sketching, $ 0.0025 \leq \epsc \leq \epslt{\infty}$ \\
(from known analyses) \\
$\sigma=1821.02$ \\
For attribute fragmenting, $\epslt{\infty}=1.78$ \\
For $\tau=4$, $\epslt{\infty}=1.45$, $\epslt{1}=0.47$ \\
For $\tau=16$, $\epslt{\infty}=1.50$, $\epslt{1}=0.13$ \\
For $\tau=256$, $\epslt{\infty}=1.52$, $\epslt{1}=0.01$ \\
\bitsonehot$=256589.00$ \\
\bitssketch$=9457.76$
\end{tabular}
&\includegraphics[scale=.25]{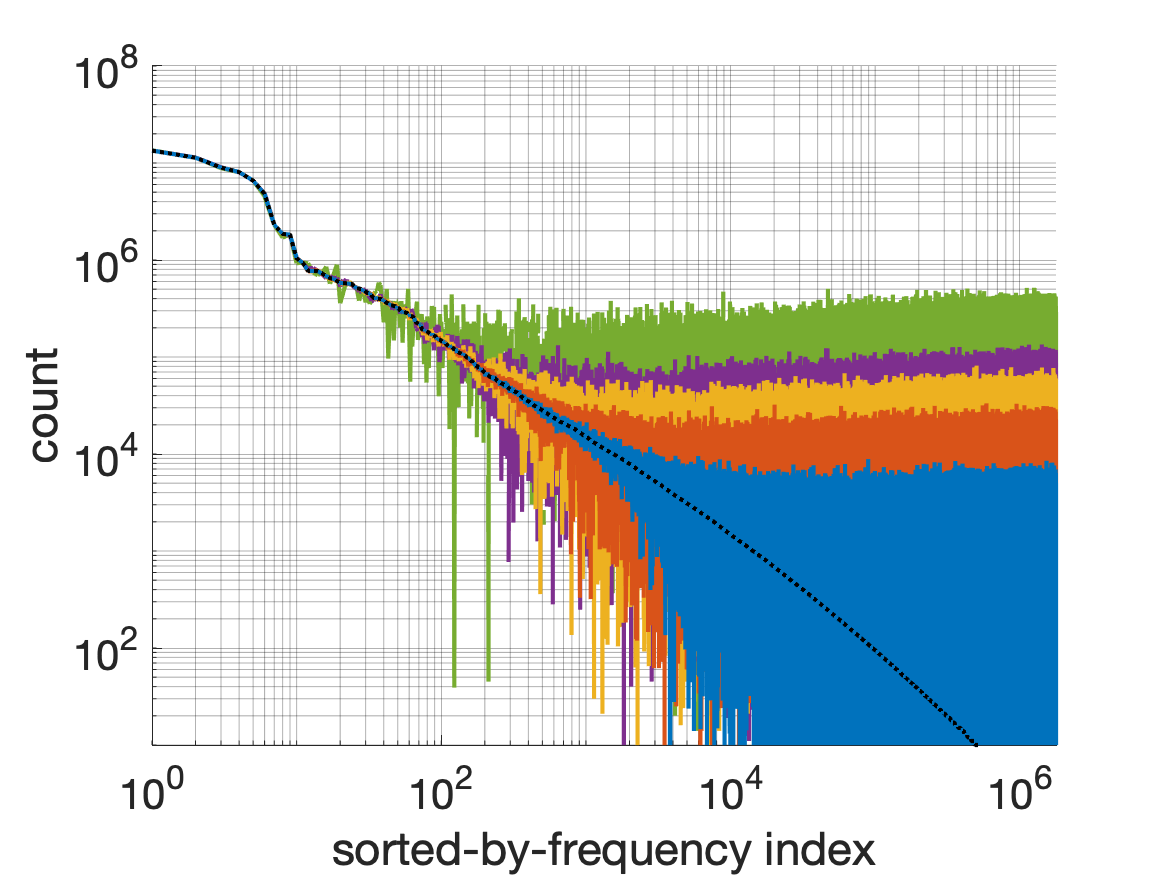}
&\includegraphics[scale=.25]{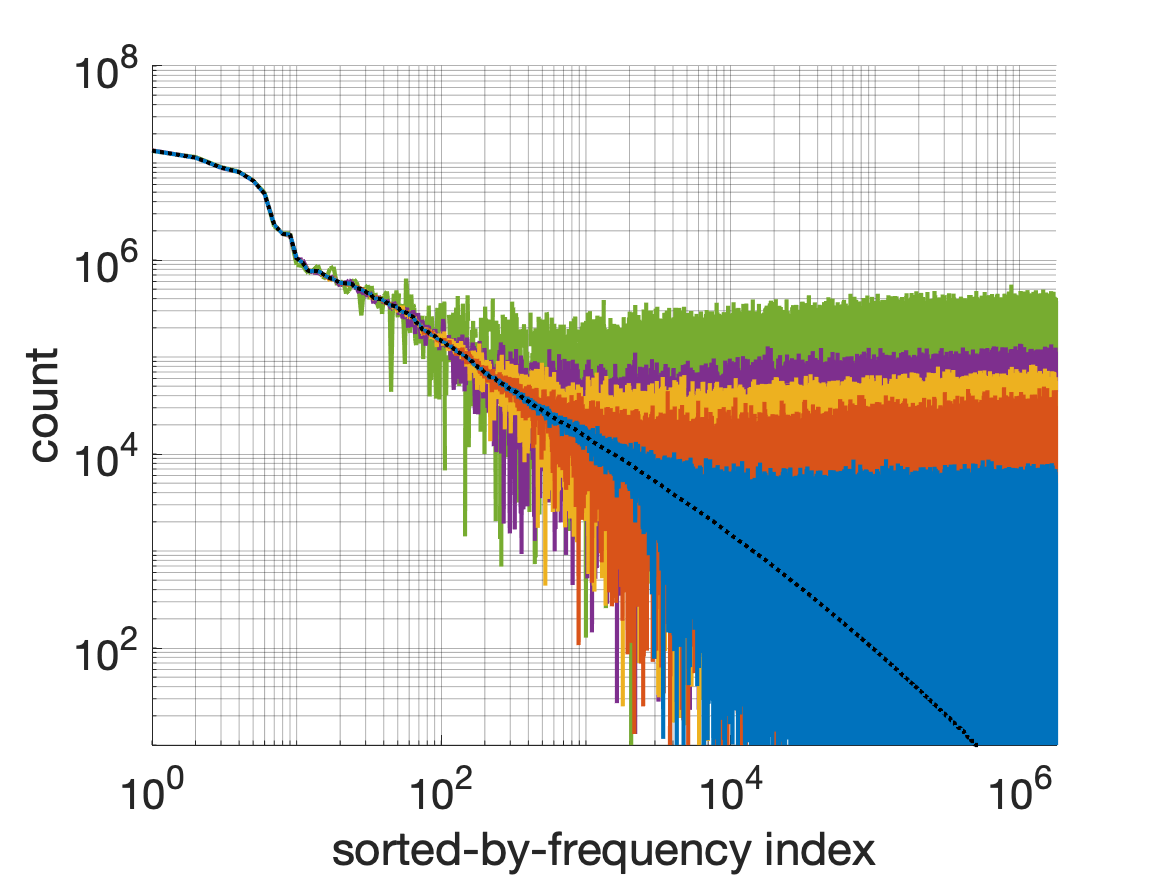}
\\
\midrule

\vspace*{-11ex}
\begin{tabular}{l}
For one-hot encoding, $\epsc=0.01$ \\ 
For sketching, $ 0.01 \leq \epsc \leq \epslt{\infty}$ \\
(from known analyses) \\
$\sigma=455.34$ \\
$\epslt{\infty}=4.07$ \\
For $\tau=4$, $\epslt{1}=2.47$ \\
For $\tau=16$, $\epslt{1}=1.30$ \\
For $\tau=256$, $\epslt{1}=0.11$ \\
\bitsonehot$=29856.75$ \\
\bitssketch$=1101.36$
\end{tabular}
&\includegraphics[scale=.25]{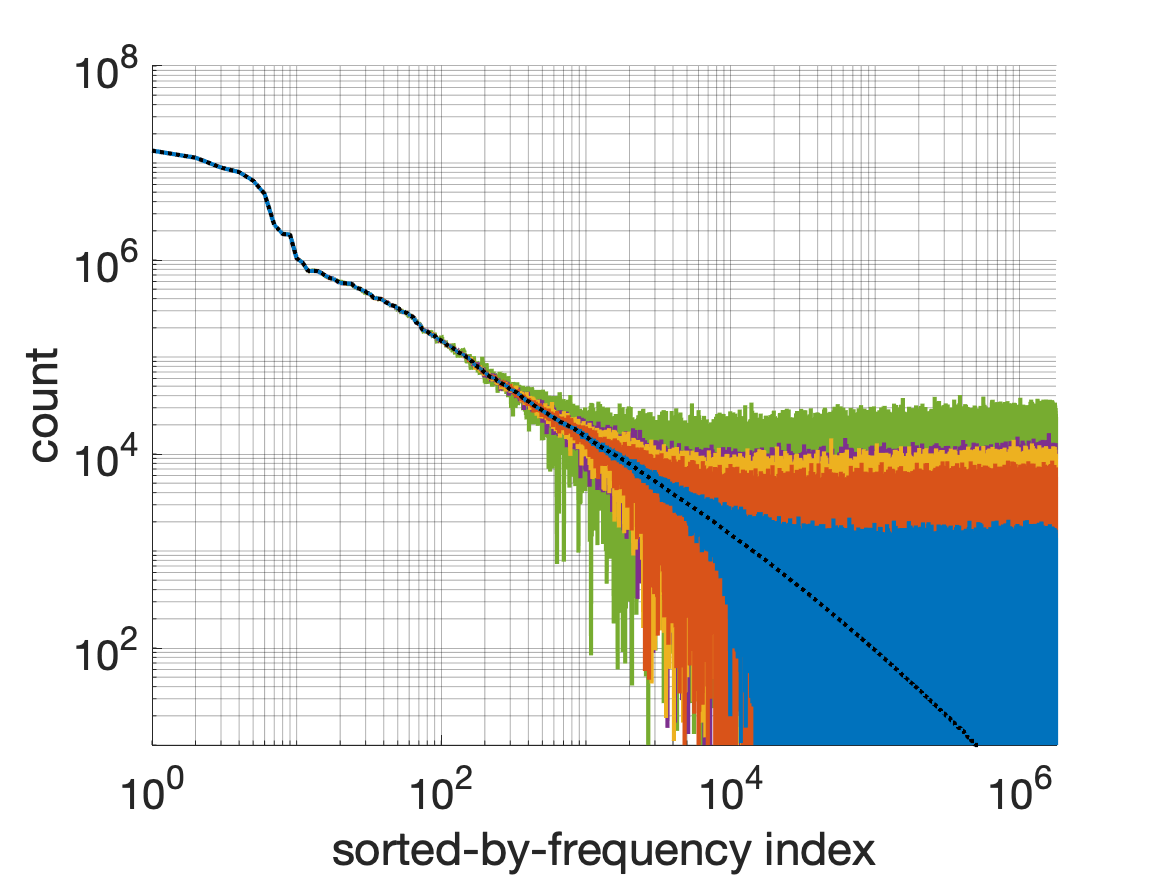}
&\includegraphics[scale=.25]{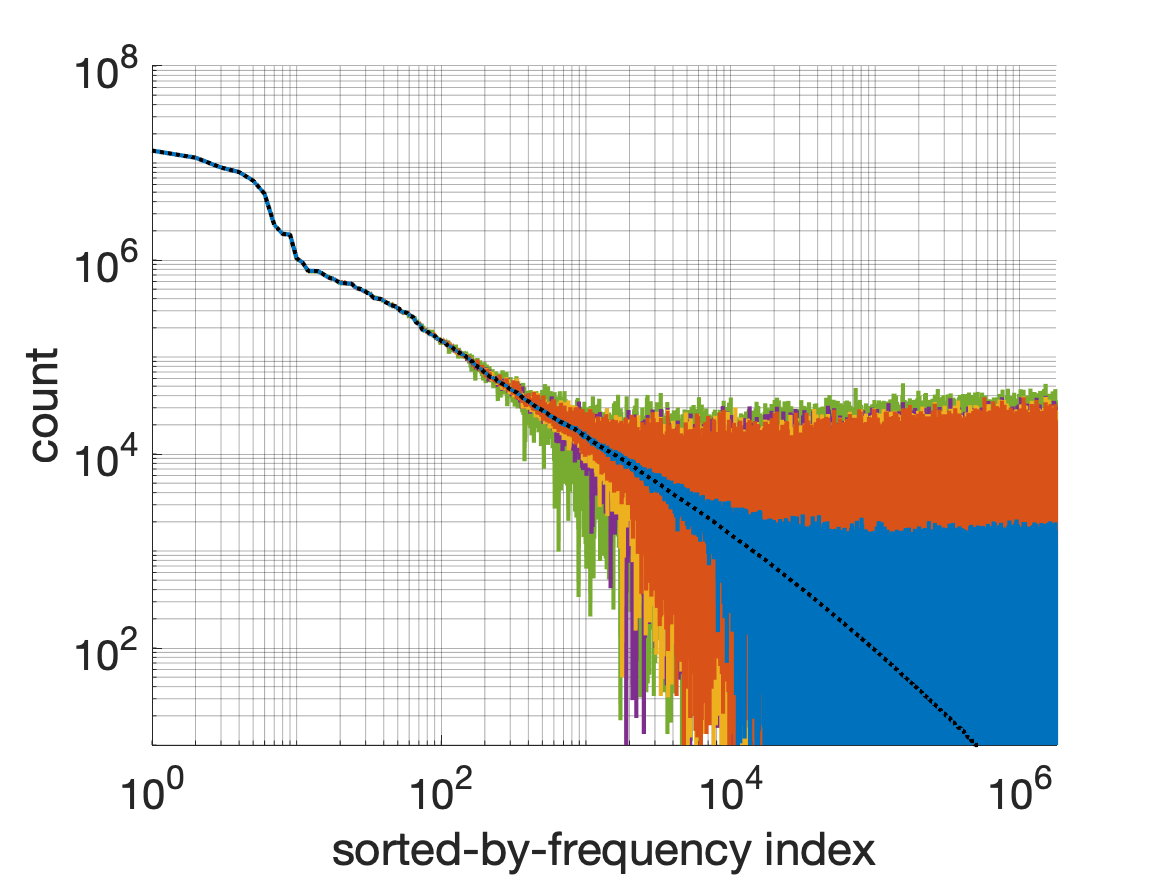}
\\
\midrule

\vspace*{-11ex}
\begin{tabular}{l}
For one-hot encoding, $\epsc=0.05$ \\ 
For sketching, $ 0.05 \leq \epsc \leq \epslt{\infty}$ \\
(from known analyses) \\
$\sigma=91.16$ \\
$\epslt{\infty}=7.235$ \\
For $\tau=4$, $\epslt{1}=5.63$ \\
For $\tau=16$, $\epslt{1}=4.40$ \\
For $\tau=256$, $\epslt{1}=1.72$ \\
\bitsonehot$=1281.93$ \\
\bitssketch$=48.21$
\end{tabular}
&\includegraphics[scale=.25]{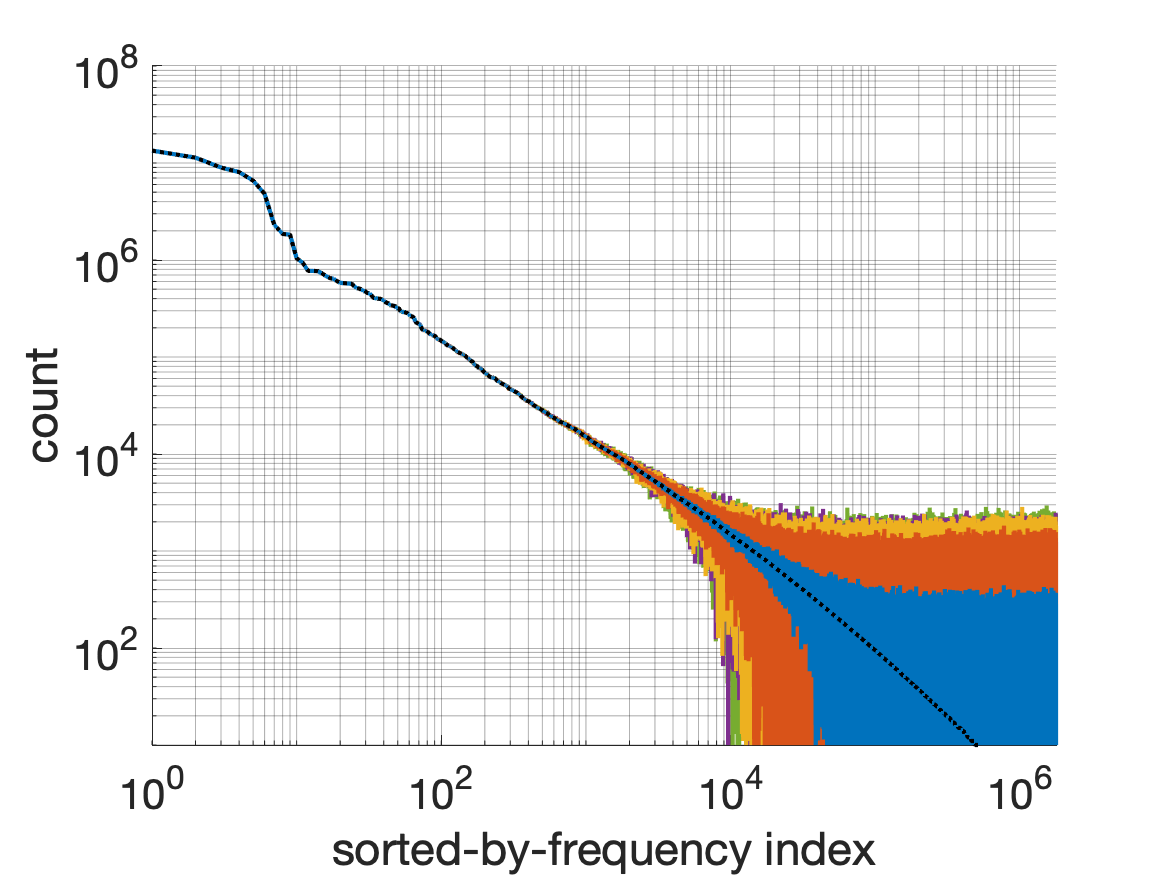}
&\includegraphics[scale=.25]{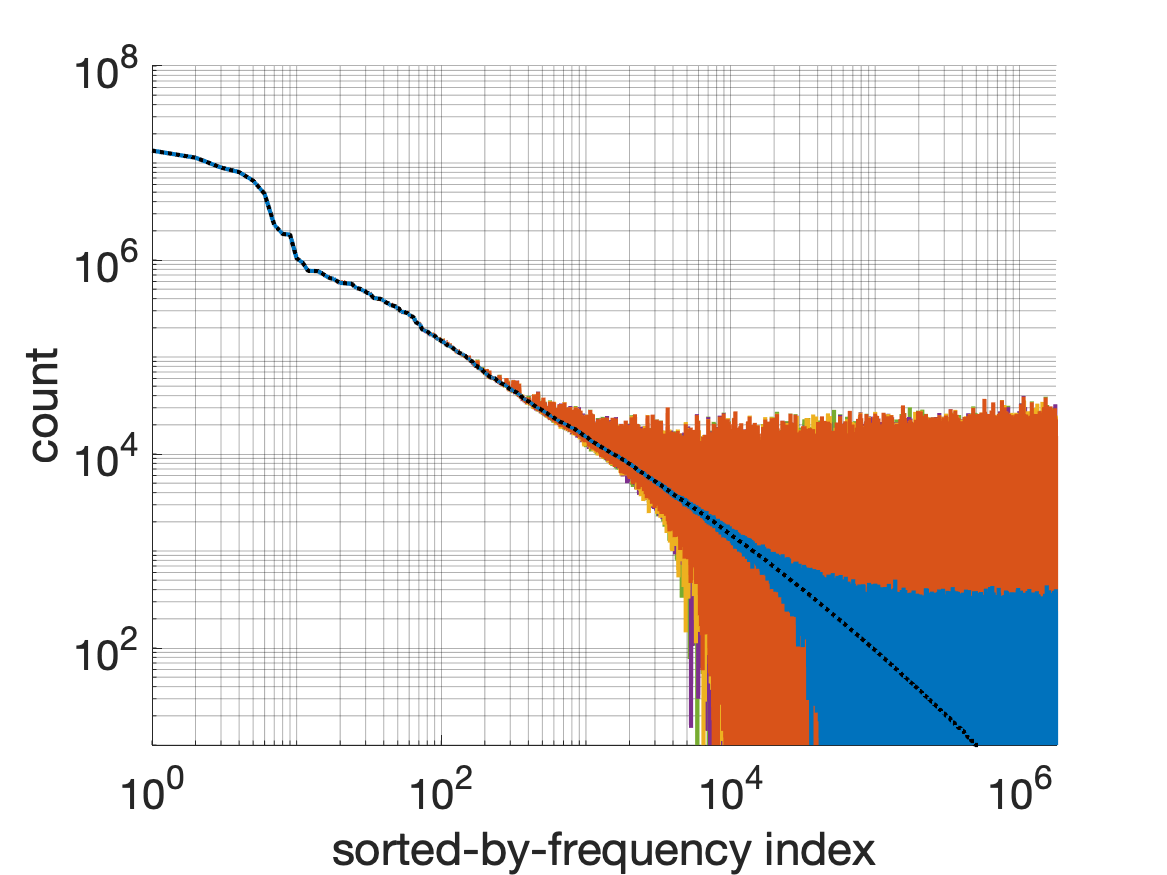}
\\
\midrule

\vspace*{-11ex}
\begin{tabular}{l}
For one-hot encoding, $\epsc=0.25$ \\ 
For sketching, $ 0.25 \leq \epsc \leq \epslt{\infty}$ \\
(from known analyses) \\
$\sigma=18.32$ \\
$\epslt{\infty}=10.40$ \\
For $\tau=4$, $\epslt{1}=8.79$ \\
For $\tau=16$, $\epslt{1}=7.56$ \\
For $\tau=256$, $\epslt{1}=4.85$ \\
\bitsonehot$=55.11$ \\
\bitssketch$=2.99$
\end{tabular}
&\includegraphics[scale=.25]{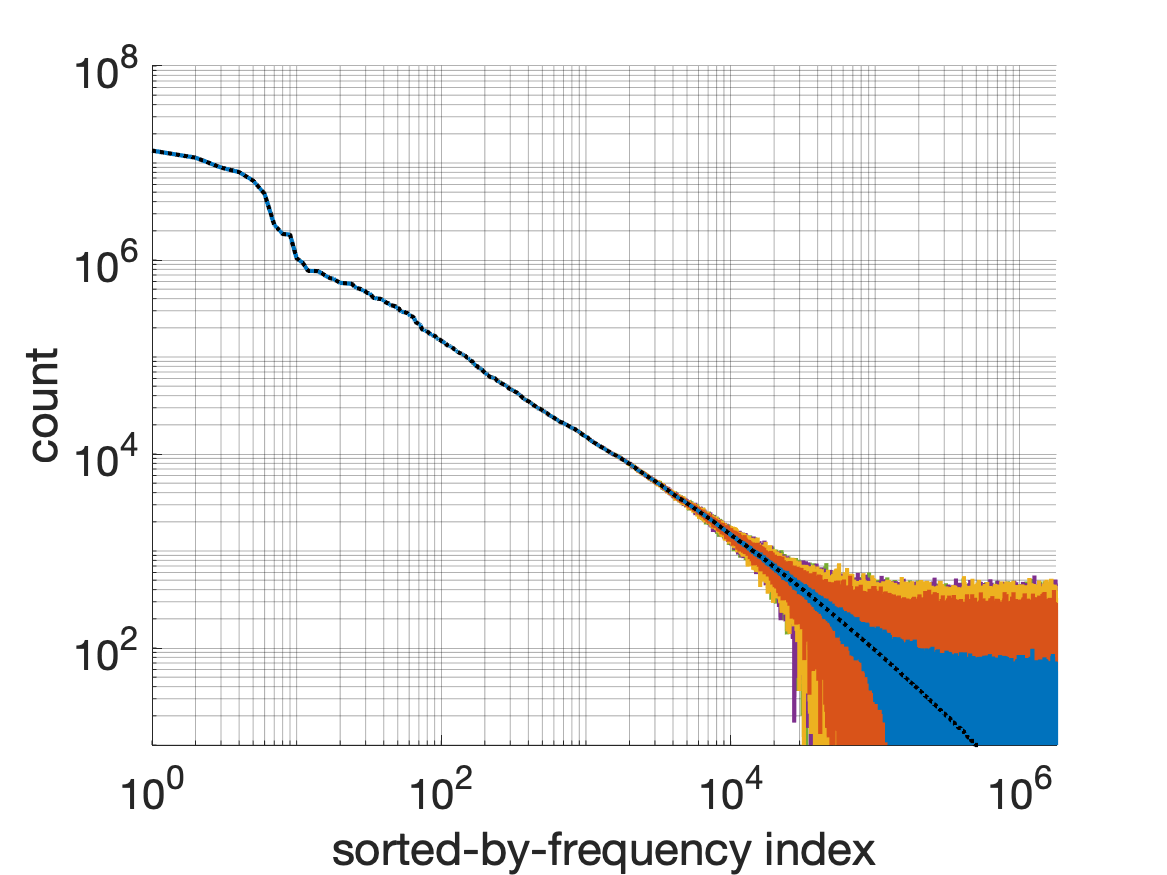}
&\includegraphics[scale=.25]{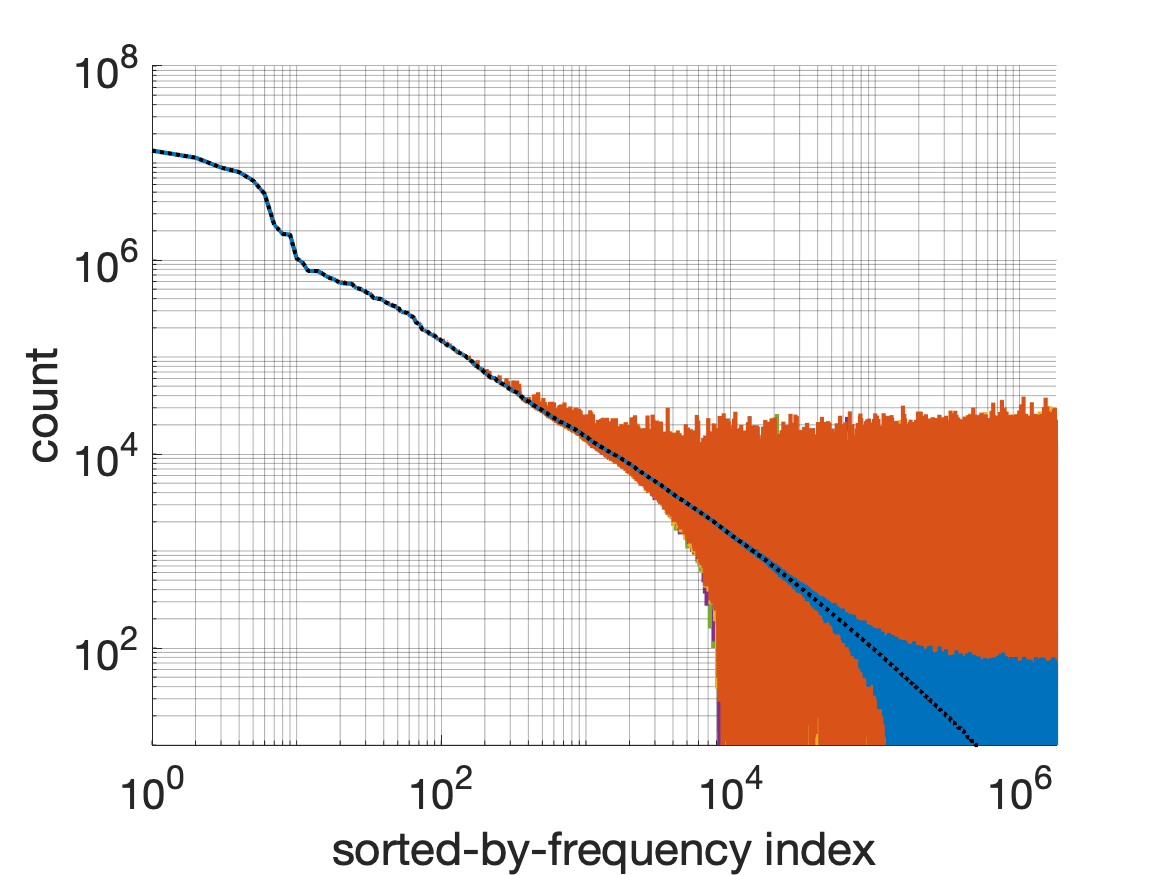}
\\
\midrule

\vspace*{-11ex}
\begin{tabular}{l}
For one-hot encoding, $\epsc=1.0$ \\ 
For sketching, $ 1.0 \leq \epsc \leq \epslt{\infty}$ \\
(from known analyses) \\
$\sigma=4.66$ \\
$\epslt{\infty}=12.99$ \\
For $\tau=4$, $\epslt{1}=11.38$ \\
For $\tau=16$, $\epslt{1}=10.15$ \\
For $\tau=256$, $\epslt{1}=7.44$ \\
\bitsonehot$=5.06$ \\
\bitssketch$=1.15$
\end{tabular}
&\includegraphics[scale=.25]{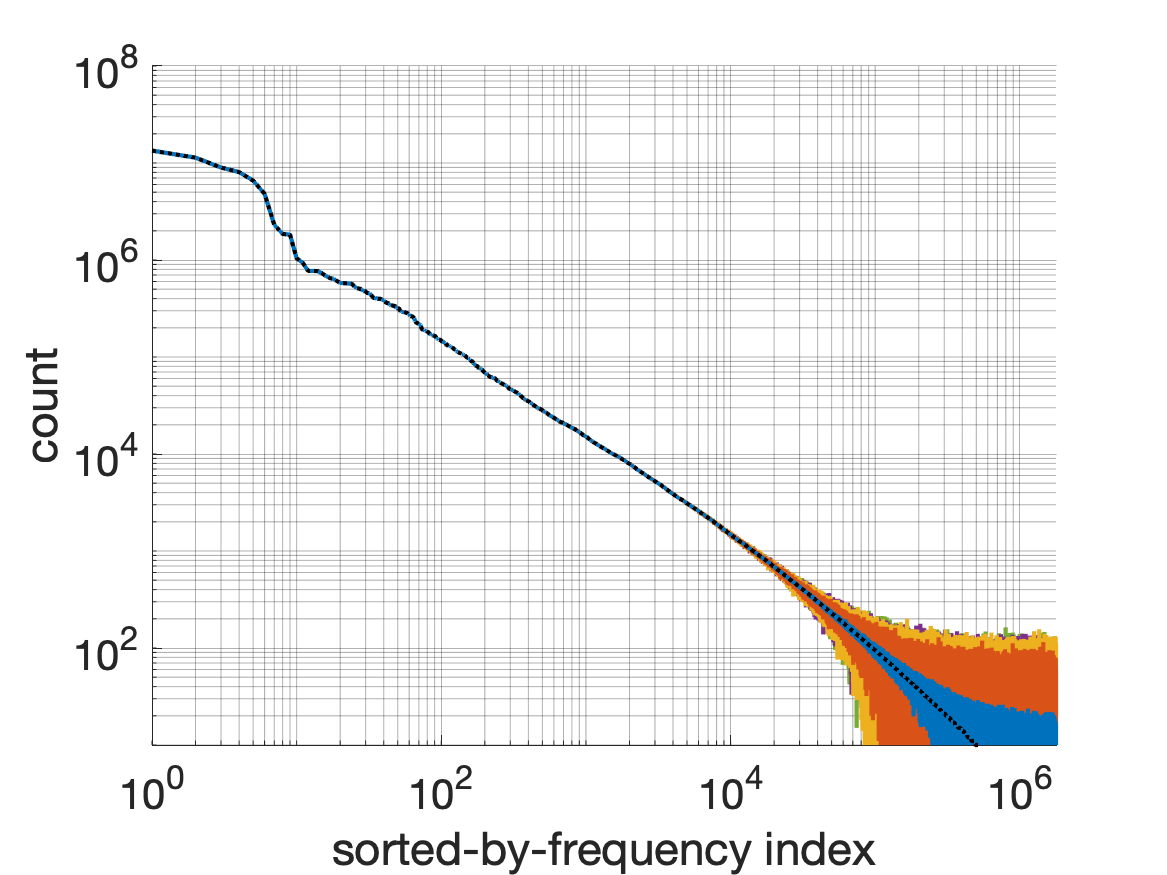}
&\includegraphics[scale=.25]{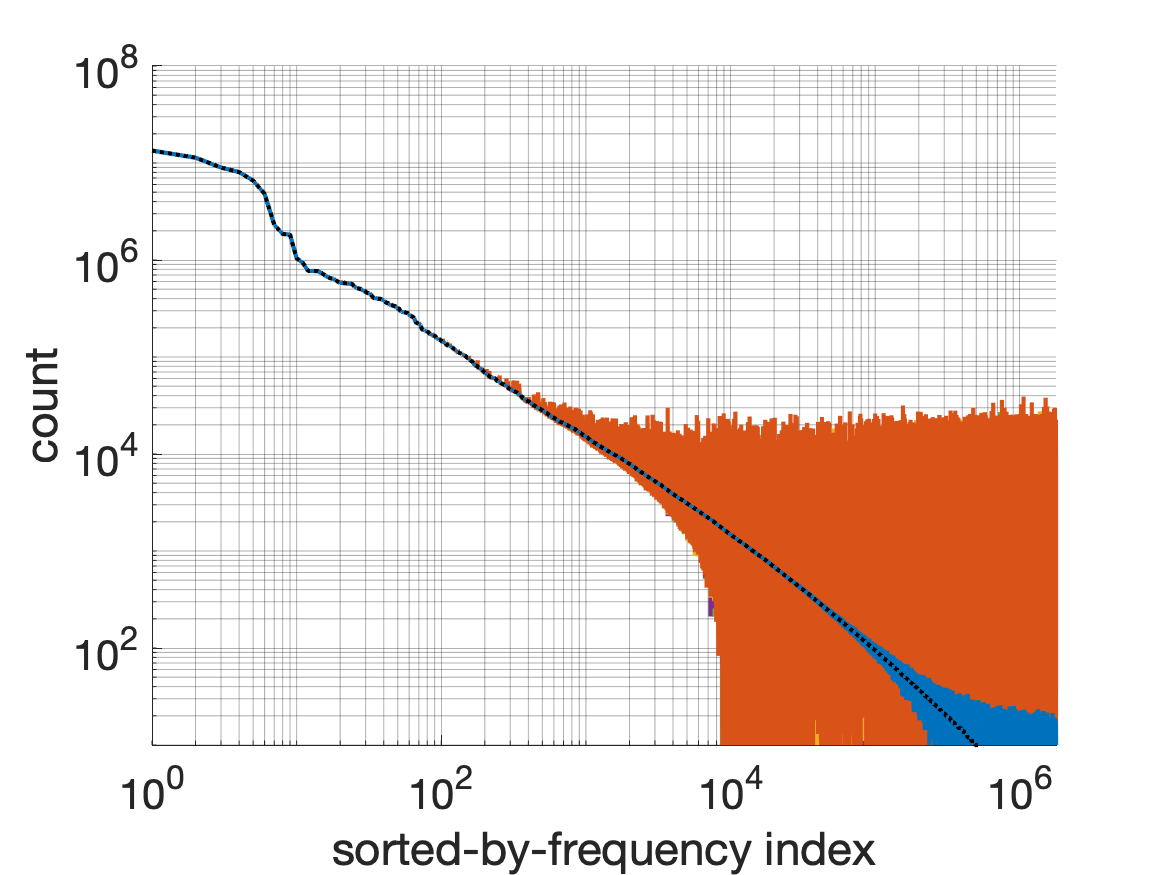}
\\
\end{tabular}
\end{table*}

\begin{table*}[p]
\caption{Statistics of datasets in experiments; in images, we take each unit of luminosity as being one respondent's presence.}
\centering
\newcolumntype{C}{ >{\arraybackslash} m{1.9cm} }
\newcolumntype{D}{ >{\centering\arraybackslash} m{3.8cm} }
\setlength\tabcolsep{2.5pt}
\begin{tabular}{@{}C@{~~~}DDDD@{}}
& \textbf{\map}  & \textbf{\horseRev} & \textbf{\girl} & \textbf{\real}    \\ \toprule
Image size       & $1365\times 2048$   & $274\times 320$ & $721\times 497$ &--\\
Domain size  & \nbr{2795520} & \nbr{87680} & \nbr{358337} & \nbr{1778120} \\
\makecell[l]{Count of\\``respondents''}  & \nbr{236559063} & \nbr{1914589}  & \nbr{50409435} & \nbr{203950512}  \\
\parbox[top][0.1in][c]{1.9cm}{\raggedright{Per-pixel luminosity (i.e., \mbox{``respondent''} count) sorted by magnitude}}
& \includegraphics[scale=.19]{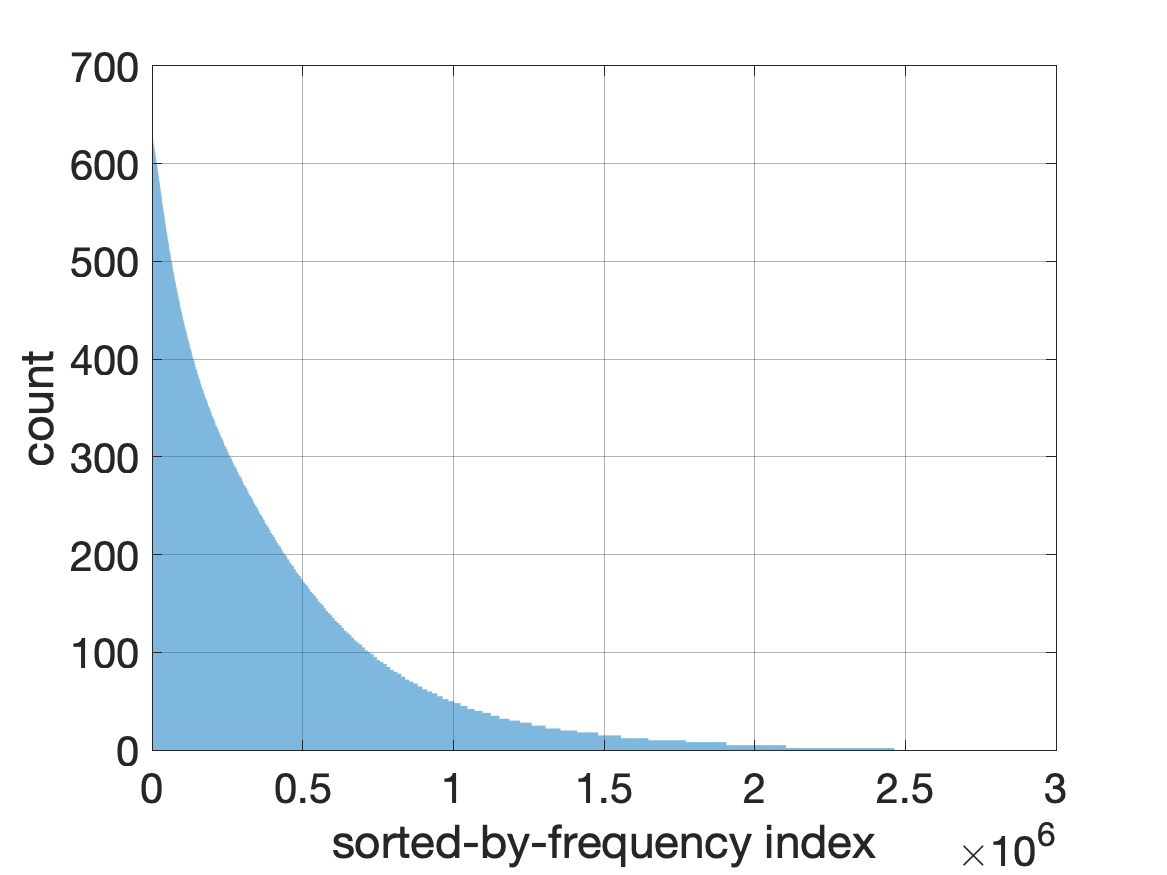}
& \includegraphics[scale=.19]{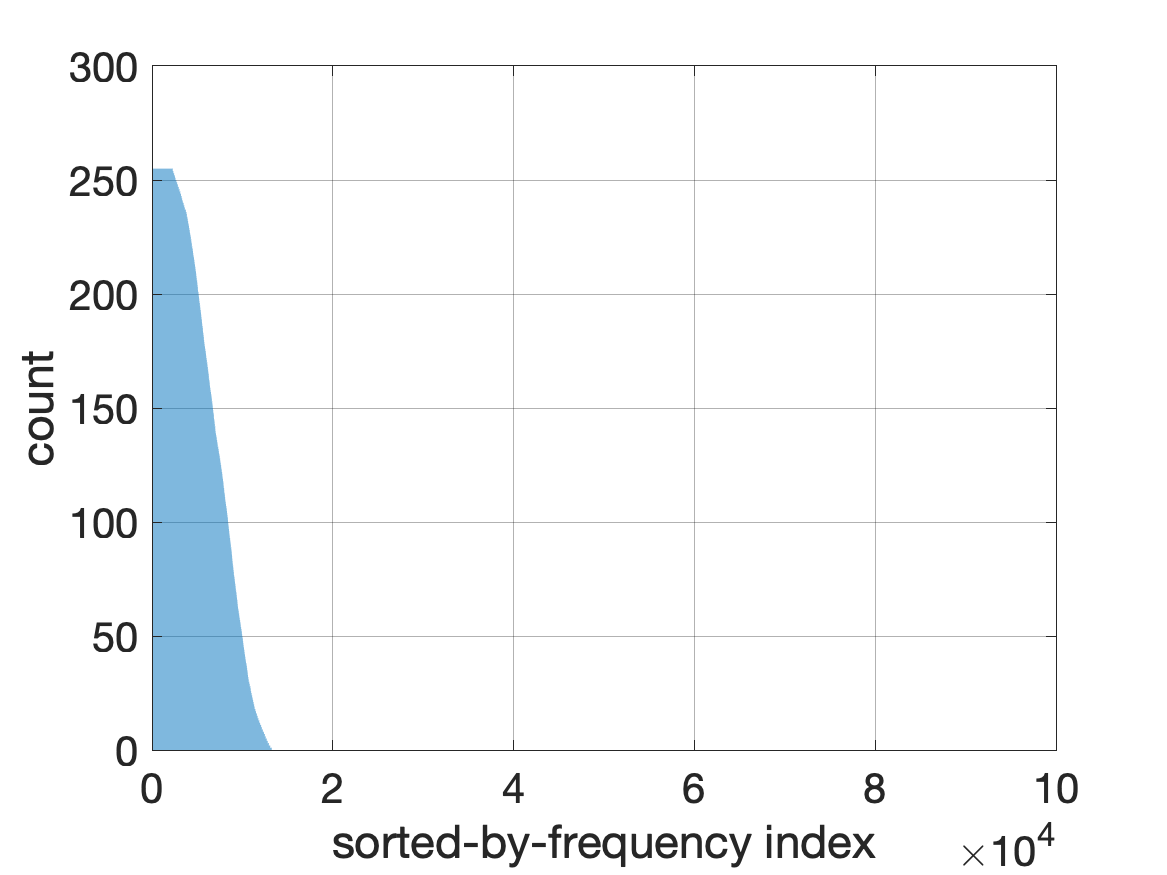}
& \includegraphics[scale=.19]{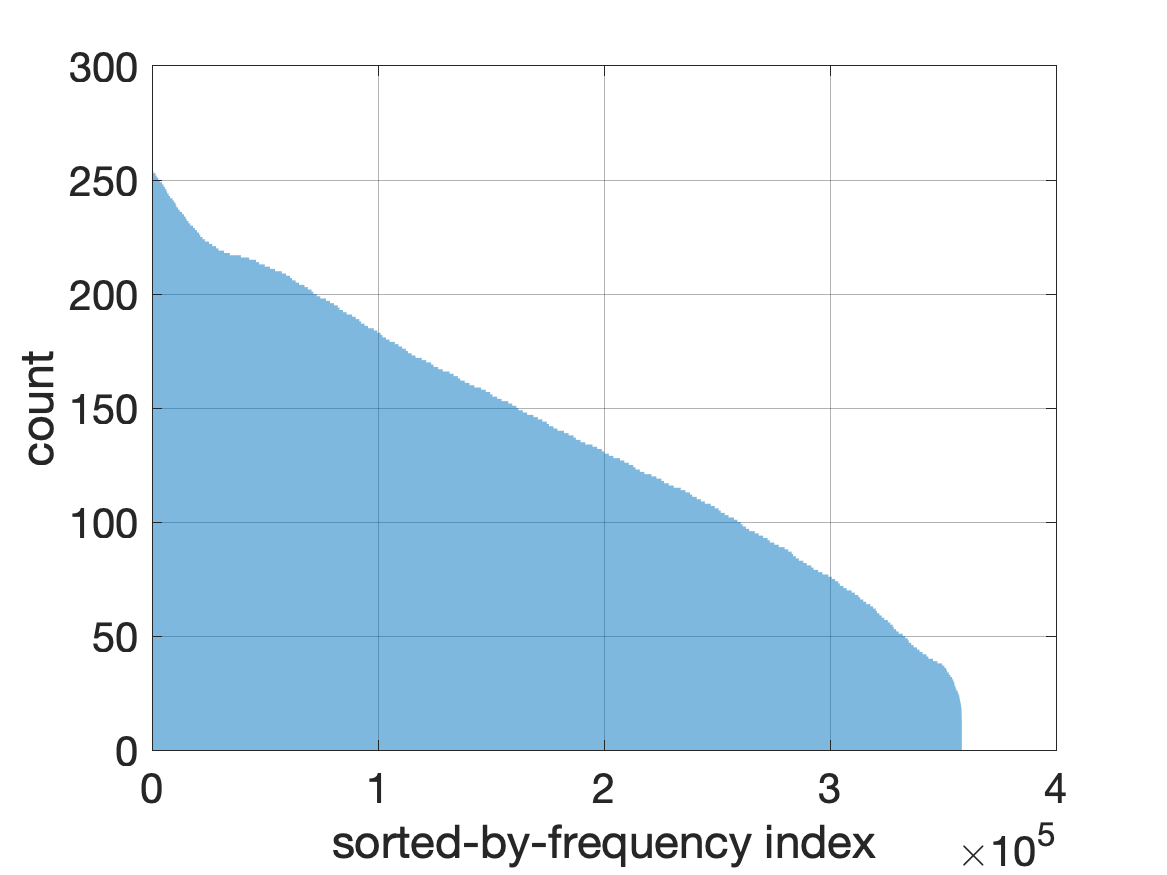}
& \includegraphics[scale=.19]{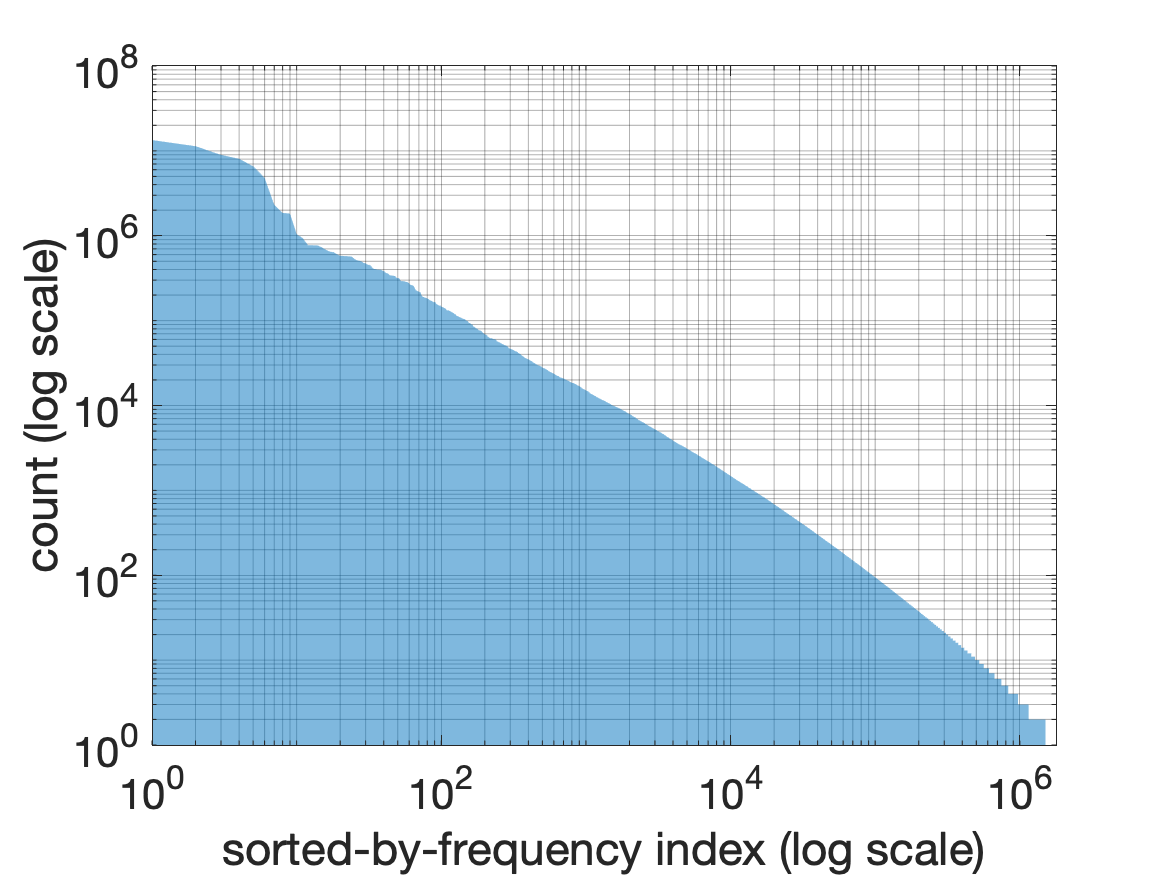}
\end{tabular}
\label{tab:dataset_statistics}

\vspace*{\floatsep}%

\caption{Reconstructions
of the Table~\ref{tab:dataset_statistics} datasets
with an $(\epsc, \deltac)$ central privacy guarantee,
based on reports using removal LDP and attribute fragmenting.
\ifsketch
The initial three rows
show reconstructions from
reports using randomized one-hot encodings
of ``respondent'' data.
The last row is based on
$65\textrm{,}536$-bit-long
Count-Mean-Sketch-encoded reports
using $1\mathrm{,}024$ hash functions,
just like those used in Apple's real-world deployment~\cite{appledp}.
As $\epslt{\infty}$ increases,
the expected number of bits set in the encodings (\#bits)
is greatly reduced,
making it practical
to send
each bit as a
separate, anonymous
report fragment.
\fi
}
\label{fig:unified_attr}
\begin{tabular}{@{}C@{~~~}DDDD@{}}
&\textbf{\map}  & \textbf{\horseRev} & \textbf{\girl} & \textbf{\real} \\
& ($\deltac = 5\times 10^{-10}$)
& ($\deltac = 5\times 10^{-8}$)
& ($\deltac = 5\times 10^{-9}$)
& ($\deltac = 5\times 10^{-10}$)
\\ \toprule
\multirow{4}{*}{
\parbox[top][0.1in][c]{1.9cm}{\raggedright{LDP reports with $\epsl=2.0$ and a varying central epsilon guarantee}}
}
&
\ifbigpdf
\includegraphics[scale=.04]{fig/map2.5/map2.5_rep1_rr01_eps2.0_crowd1_1bitS_res0.png}
\else
\includegraphics[scale=.33]{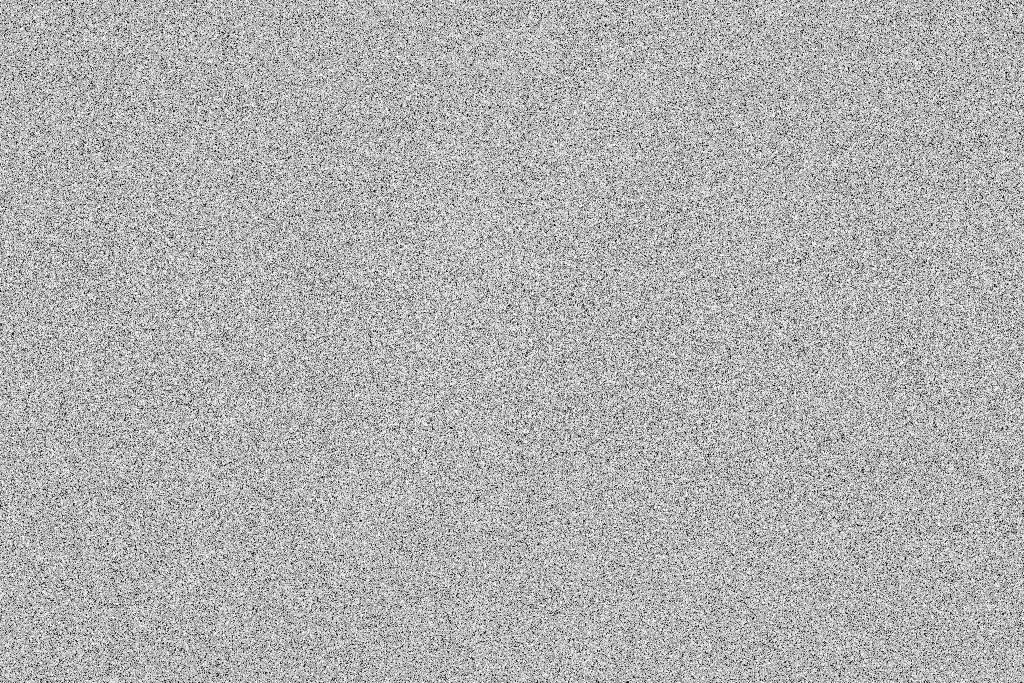}
\fi
& \includegraphics[scale=.2]{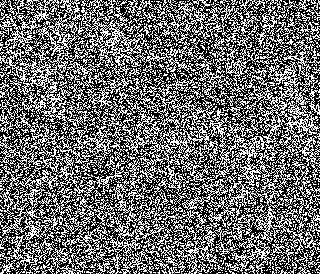}
& \includegraphics[scale=.1]{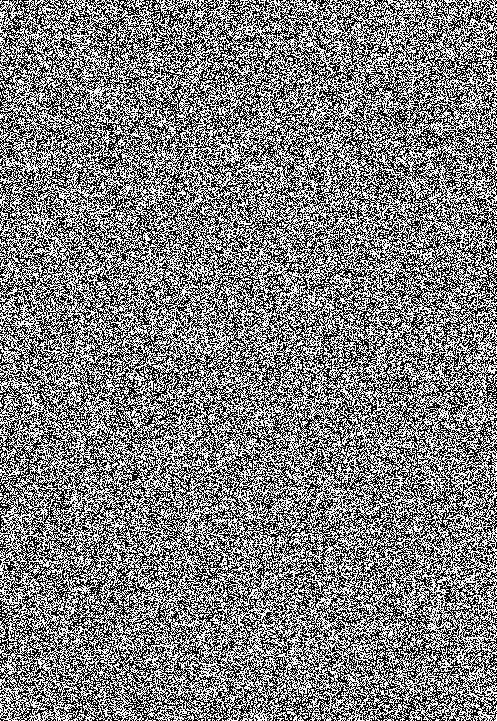}
& \includegraphics[scale=.18]{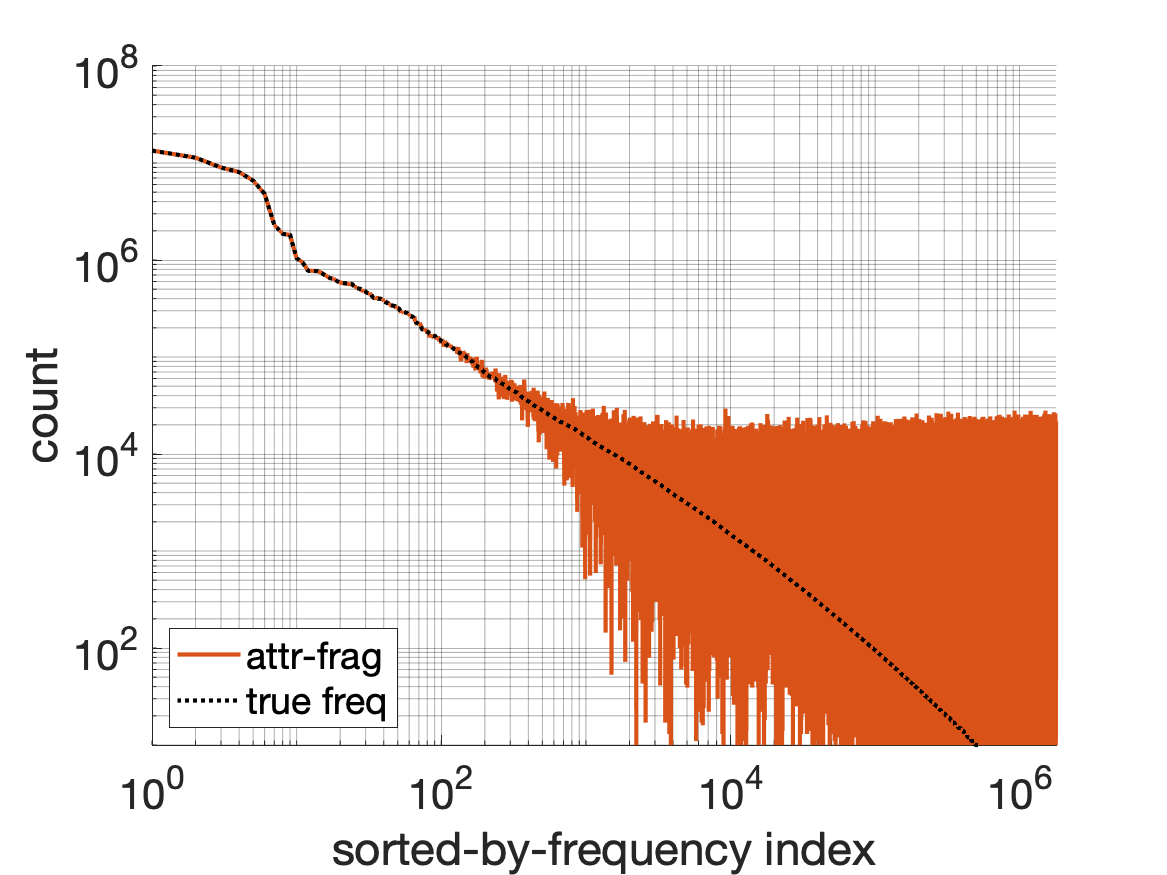}
\\
&$\epsc=0.0011$
&$\epsc=0.0111$
&$\epsc=0.0023$
&$\epsc=0.0012$
\\
&$\rmse=182.17$
&$\rmse=129.13$
&$\rmse=138.06$
&$\rmse=3565.88$
\\
\ifsketch
&$\#\textrm{bits}=333234.91$
&$\#\textrm{bits}=10452.47$
&$\#\textrm{bits}=42715.58$
&$\#\textrm{bits}=211957.86$
\\
\fi
\midrule
\multirow{4}{*}{
\parbox[top][0.1in][c]{1.9cm}{\raggedright{High-epsilon LDP reports with a central guarantee $\epsc=0.05$}}
}
&
\ifbigpdf
\includegraphics[scale=.04]{fig/map2.5/map2.5_rep1_rr01_eps7.385_crowd1_1bitS_res0.png}
\else
\includegraphics[scale=.33]{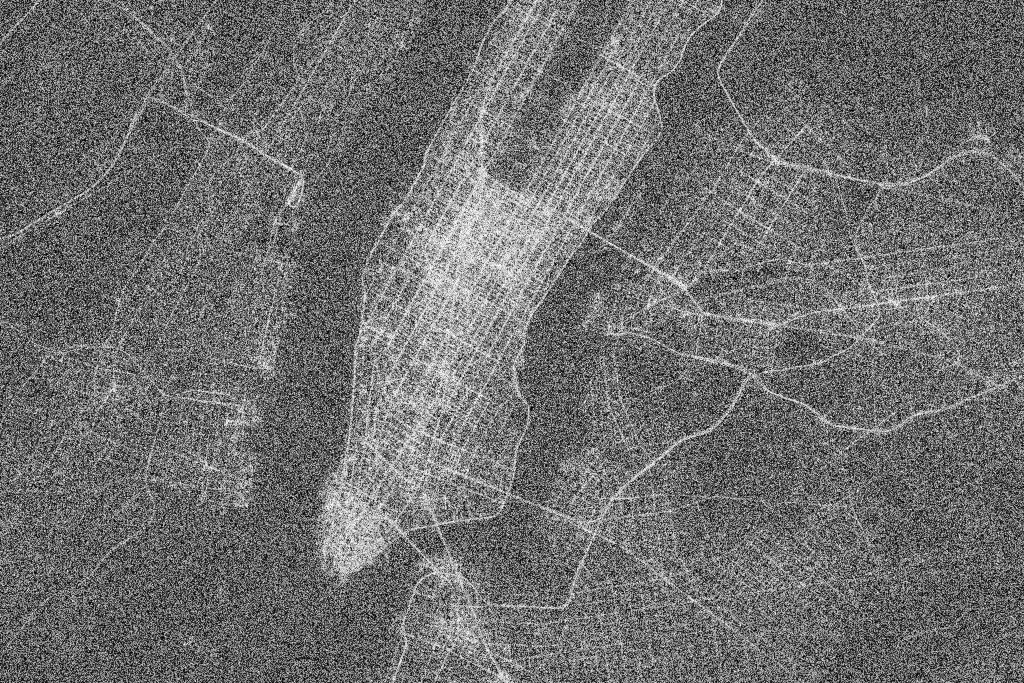}
\fi
& \includegraphics[scale=.2]{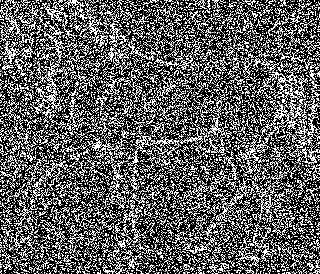}
& \includegraphics[scale=.1]{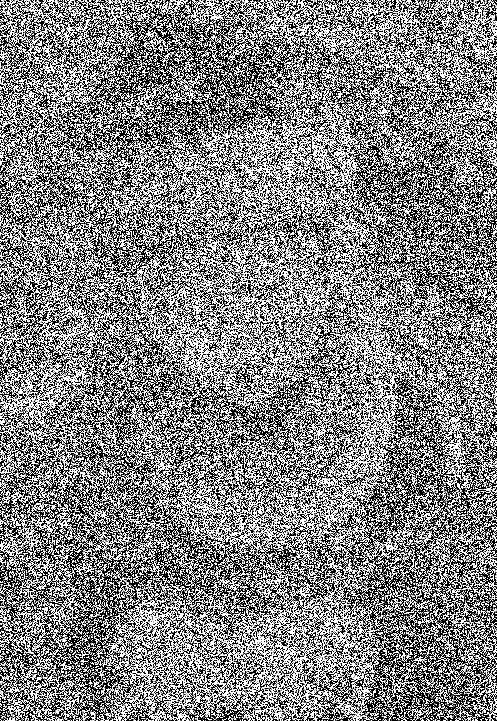}
& \includegraphics[scale=.18]{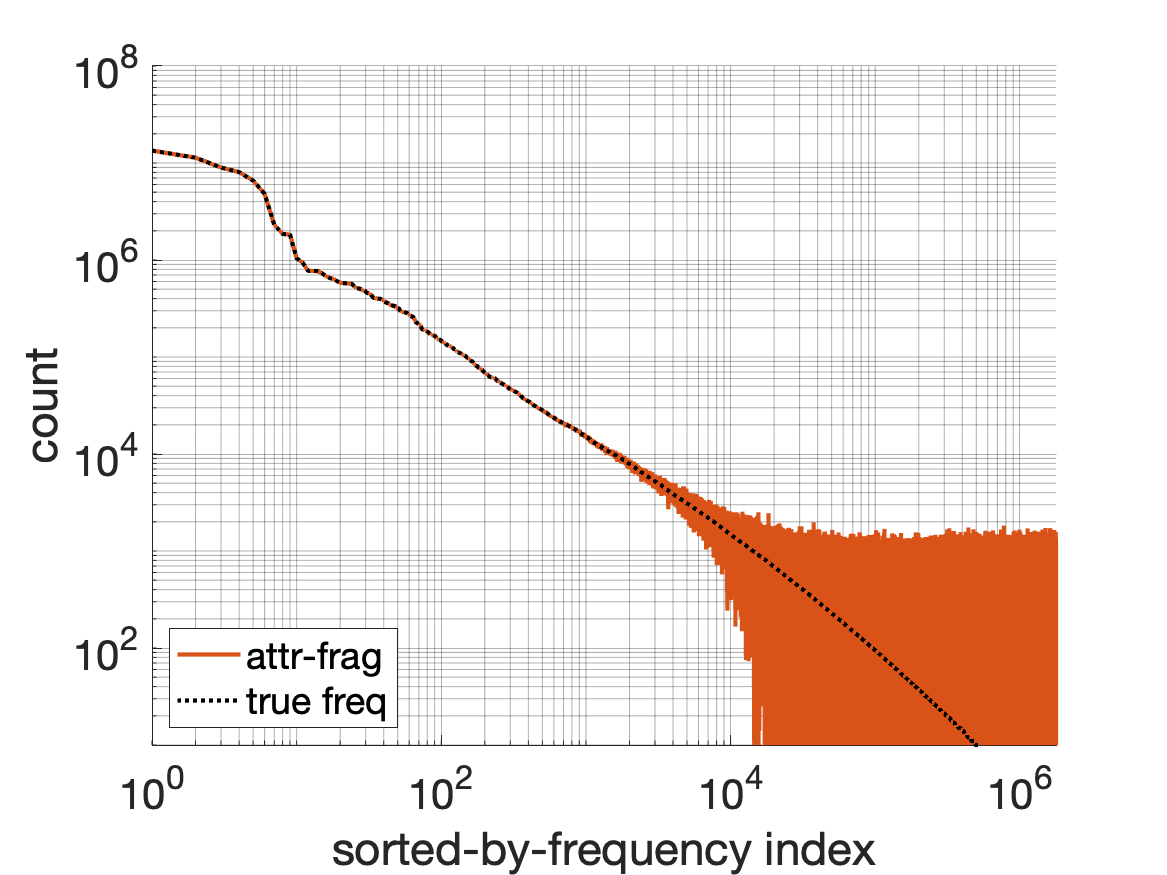}
\\
&$\epslt{\infty}=7.385$
&$\epslt{\infty}=2.94$
&$\epslt{\infty}=5.95$
&$\epslt{\infty}=7.235$
\\
&$\rmse=150.54$
&$\rmse=118.40$
&$\rmse=121.81$
&$\rmse=234.28$
\\
\ifsketch
&$\#\textrm{bits}=1734.52$
&$\#\textrm{bits}=4403.42$
&$\#\textrm{bits}=932.34$
&$\#\textrm{bits}=1281.93$
\\
\fi
\midrule
\multirow{4}{*}{
\parbox[top][0.1in][c]{1.9cm}{\raggedright{High-epsilon LDP reports with a central guarantee $\epsc=1.0$}}
}
&
\ifbigpdf
\includegraphics[scale=.04]{fig/map2.5/map2.5_rep1_rr01_eps13.14_crowd1_1bitS_res0.png}
\else
\includegraphics[scale=.33]{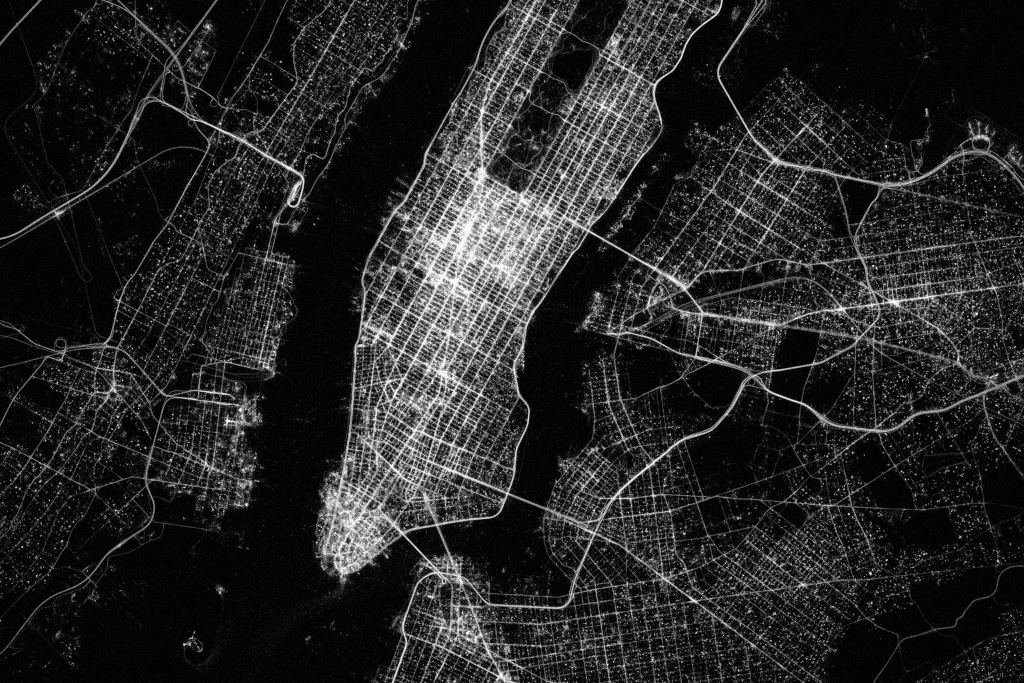}
\fi
& \includegraphics[scale=.2]{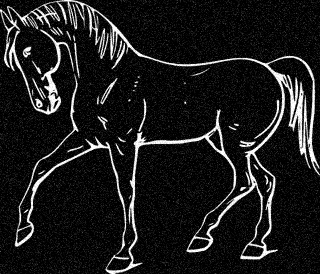}
& \includegraphics[scale=.1]{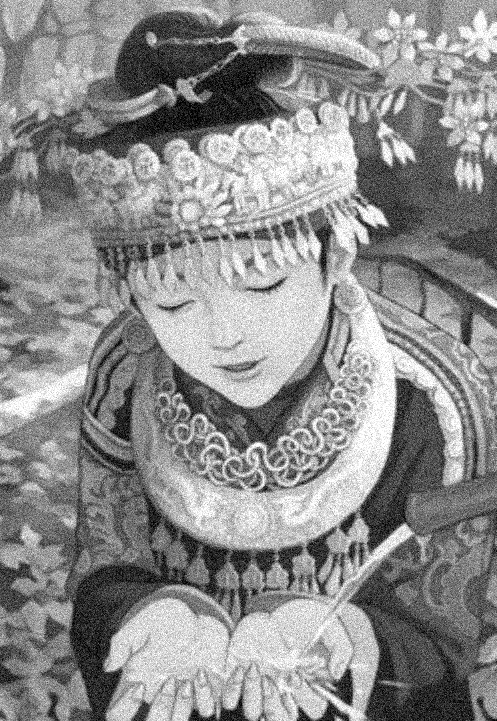}
& \includegraphics[scale=.18]{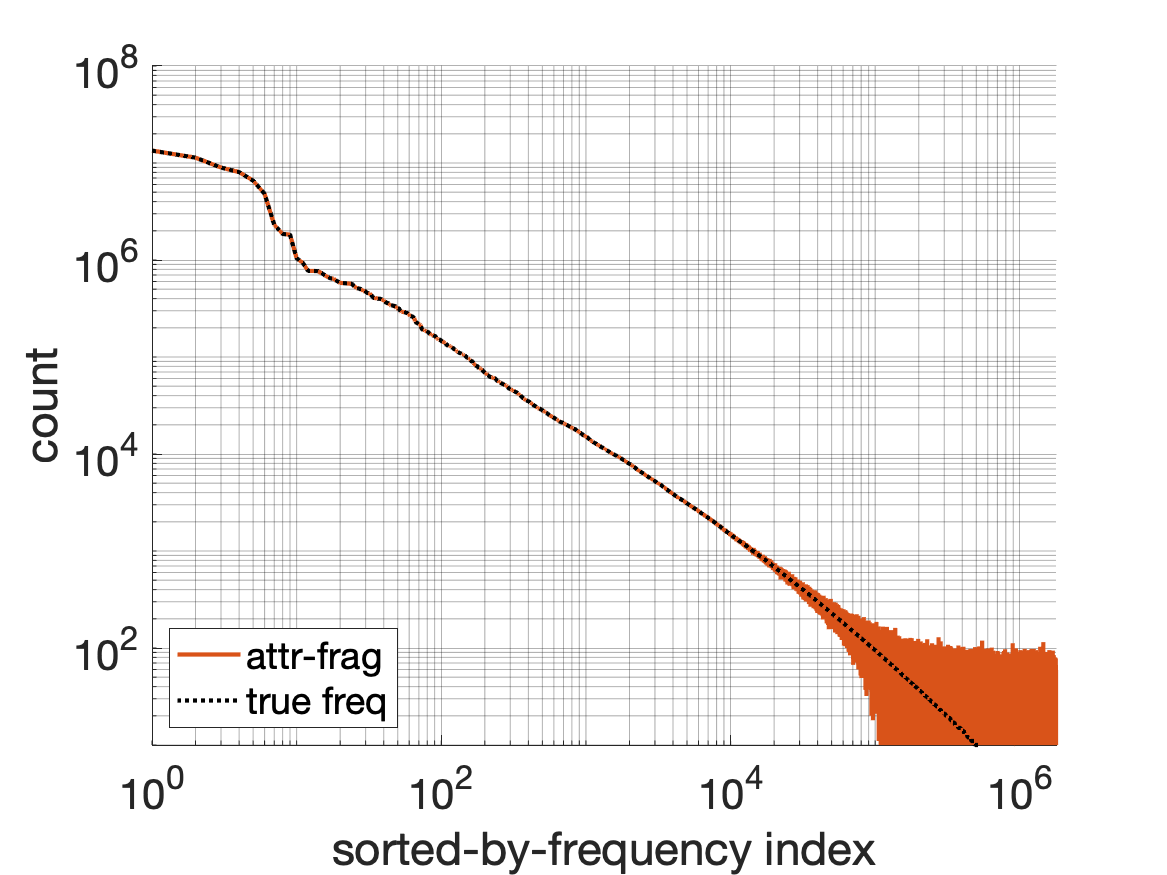}
\\
&$\epslt{\infty}=13.14$
&$\epslt{\infty}=8.55$
&$\epslt{\infty}=11.7$
&$\epslt{\infty}=12.99$
\\
&$\rmse=61.31$
&$\rmse=12.96$
&$\rmse=20.13$
&$\rmse=15.82$
\\
\ifsketch
&$\#\textrm{bits}=6.49$
&$\#\textrm{bits}=17.97$
&$\#\textrm{bits}=3.97$
&$\#\textrm{bits}=5.06$
\\
\midrule
\fi
\ifsketch
\multirow{4}{*}{
\parbox[top][0.1in][c]{1.9cm}{\raggedright{Sketch-based high-epsilon LDP reports. %
Known analyses imply a central guarantee of $1 \leq \epsc  \leq \epslt{\infty}$
}}
}
&
\ifbigpdf
\includegraphics[scale=.04]{fig/map2.5/map2.5_rep1_rr01_eps13.14_hash1024_sketch65536_crowd1_1bitS_res0.png}
\else
\includegraphics[scale=.33]{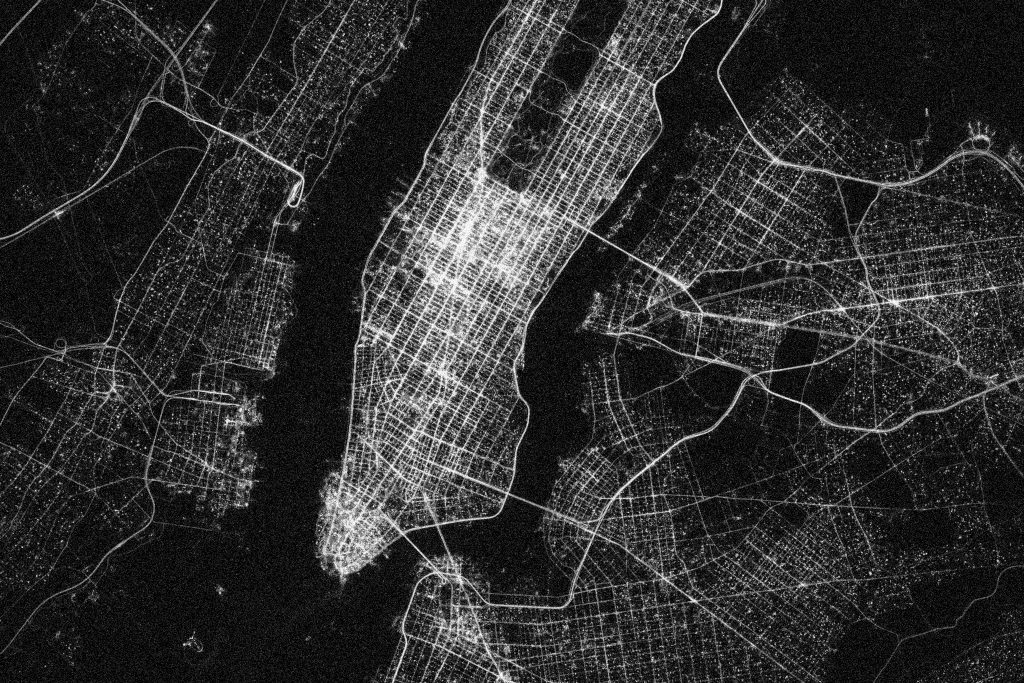}
\fi
& \includegraphics[scale=.2]{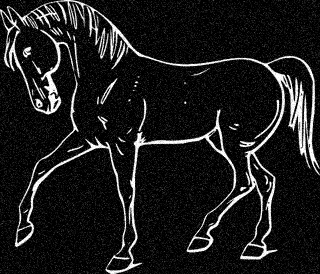}
& \includegraphics[scale=.1]{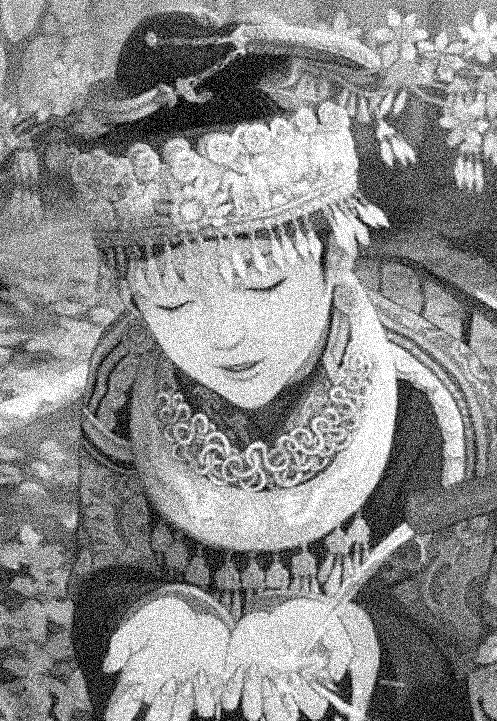}
& \includegraphics[scale=.18]{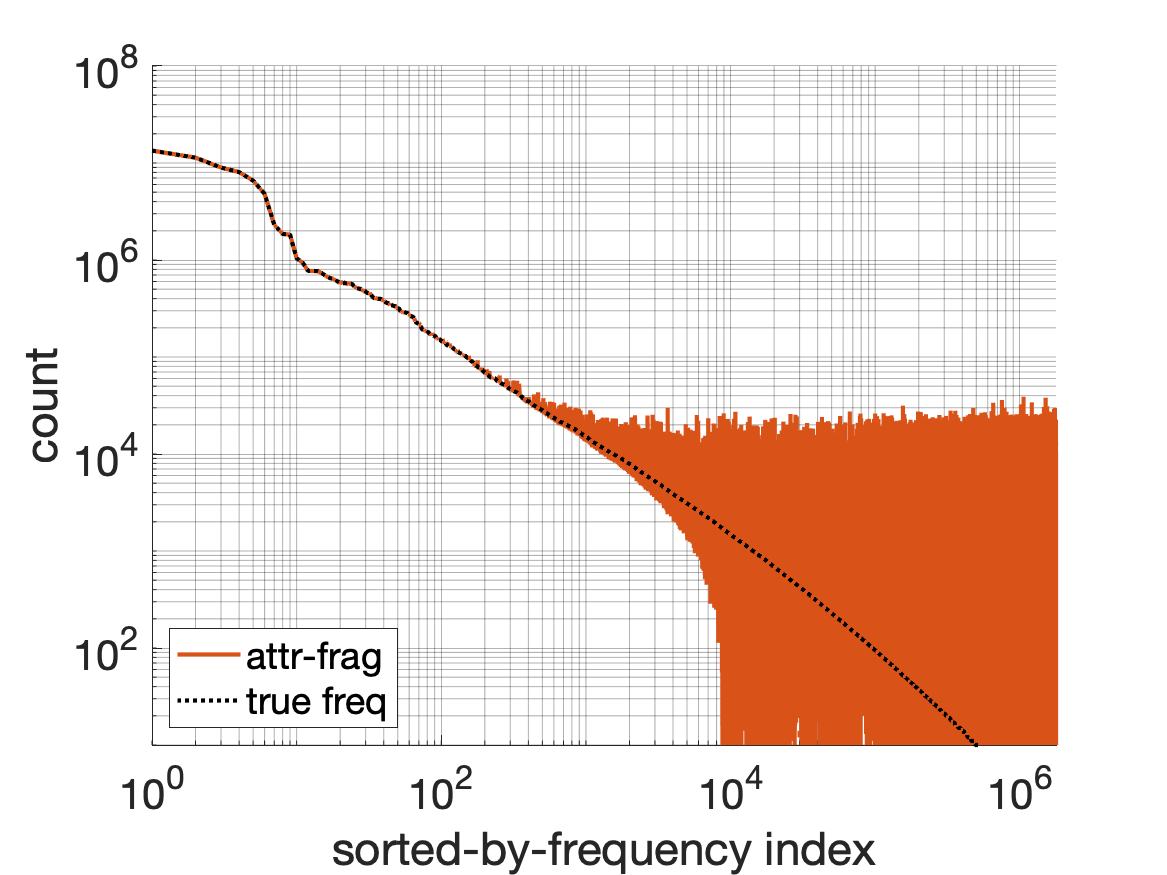}
\\
&$\epslt{\infty}=13.14$
&$\epslt{\infty}=8.55$
&$\epslt{\infty}=11.7$
&$\epslt{\infty}=12.99$
\\
&$\rmse=79.73$
&$\rmse=13.25$
&$\rmse=34.54$
&$\rmse=2583.13$
\\
&$\#\textrm{bits}=1.13$
&$\#\textrm{bits}=13.68$
&$\#\textrm{bits}=1.54$
&$\#\textrm{bits}=1.15$
\fi
\end{tabular}
\end{table*}

\newif\ifonehash\onehashtrue

\ifsmother

\fi

\ifhorseRev
\begin{table*}[t!]
\caption{Experimental results of reconstructing the \horseRev{} 
image dataset
by varying $\epsc$ and evaluating a central DP mechanism, attribute fragmenting, and both attribute and report fragmenting achieving $(\epsc, \deltac)$-central DP with $\deltac = 5\times 10^{-8}$.
}
\label{fig:horse-1-crowd}
\centering
\newcolumntype{D}{ >{\centering\arraybackslash} m{2cm} }
\newcolumntype{N}{ >{\centering\arraybackslash} m{2.6cm} }
\begin{tabular}{ D N N N N N N }

\multirow{2}{*}{\makecell{Central privacy \\ guarantee}}
& \multirow{2}{*}{\makecell{No LDP \\(Gaussian mechanism)}}
& \multirow{2}{*}{\makecell{Attribute-fragmented \\LDP~report}}
& \multicolumn{3}{c}{Attribute- and report-fragmented LDP reports}
\\ \cmidrule(lr){4-6}
& & &$\tau=4$ reports &$\tau=16$ reports &$\tau=256$ reports
\\ \toprule

\ifbigpdf
\multirow{3}{*}{$\epsc=0.05$}
& \includegraphics[scale=.2]{fig/horseRev/horseRev_Gauss0.1.png} 
& \includegraphics[scale=.2]{fig/horseRev/horseRev_rep1_rr01_eps2.94_crowd1_1bitS_res0.png} 
& \includegraphics[scale=.2]{fig/horseRev/horseRev_rep1-4_rr01_eps2.94-1.595_crowd1_1bitS_res0.png}
& \includegraphics[scale=.2]{fig/horseRev/horseRev_rep1-16_rr01_eps2.94-0.56_crowd1_1bitS_res0.png}
& \includegraphics[scale=.2]{fig/horseRev/horseRev_rep1-256_rr01_eps2.94-0.035_crowd1_1bitS_res0.png}
\\
\else
\multirow{2}{*}{$\epsc=0.05$}
\fi
&$\sigma=80.42$, \quad $\epslt{*}=\infty$
&$\epslt{\infty}=2.94$, $\epslt{1}=2.94$ 
&$\epslt{\infty}=2.91$, $\epslt{1}=1.37$%
&$\epslt{\infty}=2.94$, $\epslt{1}=0.50$%
&$\epslt{\infty}=2.94$, $\epslt{1}=0.03$%
\\
&$\rmse=51.09$
&$\rmse=118.40$
&$\rmse=128.05$
&$\rmse=132.46$
&$\rmse=139.47$
\\
\midrule
      
\ifbigpdf
\multirow{3}{*}{$\epsc=0.25$}
& \includegraphics[scale=.2]{fig/horseRev/horseRev_Gauss0.5.png} 
& \includegraphics[scale=.2]{fig/horseRev/horseRev_rep1_rr01_eps5.96_crowd1_1bitS_res0.png} 
& \includegraphics[scale=.2]{fig/horseRev/horseRev_rep1-4_rr01_eps5.96-4.575_crowd1_1bitS_res0.png}
& \includegraphics[scale=.2]{fig/horseRev/horseRev_rep1-16_rr01_eps5.96-3.19_crowd1_1bitS_res0.png}
& \includegraphics[scale=.2]{fig/horseRev/horseRev_rep1-256_rr01_eps5.96-0.7_crowd1_1bitS_res0.png}
\\
\else
\multirow{2}{*}{$\epsc=0.25$}
\fi
&$\sigma=16.18$, \quad $\epslt{*}=\infty$
&$\epslt{\infty}=5.96$, $\epslt{1}=5.96$ 
&$\epslt{\infty}=5.96$, $\epslt{1}=4.35$%
&$\epslt{\infty}=5.96$, $\epslt{1}=3.13$%
&$\epslt{\infty}=5.96$, $\epslt{1}=0.70$%
\\
&$\rmse=10.72$
&$\rmse=45.01$
&$\rmse=63.01$
&$\rmse=63.67$
&$\rmse=82.12$ \\
\midrule
      
\ifbigpdf
\multirow{3}{*}{$\epsc=0.5$}
& \includegraphics[scale=.2]{fig/horseRev/horseRev_Gauss1.png} 
& \includegraphics[scale=.2]{fig/horseRev/horseRev_rep1_rr01_eps7.28_crowd1_1bitS_res0.png} 
& \includegraphics[scale=.2]{fig/horseRev/horseRev_rep1-4_rr01_eps7.28-5.895_crowd1_1bitS_res0.png}
& \includegraphics[scale=.2]{fig/horseRev/horseRev_rep1-16_rr01_eps7.28-4.51_crowd1_1bitS_res0.png}
& \includegraphics[scale=.2]{fig/horseRev/horseRev_rep1-256_rr01_eps7.28-1.765_crowd1_1bitS_res0.png}
\\
\else
\multirow{2}{*}{$\epsc=0.5$}
\fi
&$\sigma=8.15$, \quad $\epslt{*}=\infty$
&$\epslt{\infty}=7.28$, $\epslt{1}=7.28$ 
&$\epslt{\infty}=7.28$, $\epslt{1}=5.67$%
&$\epslt{\infty}=7.28$, $\epslt{1}=4.45$%
&$\epslt{\infty}=7.28$, $\epslt{1}=1.76$%
\\
&$\rmse=5.40$
&$\rmse=23.79$
&$\rmse=33.60$
&$\rmse=33.59$
&$\rmse=36.27$
\\
\midrule
      
\ifbigpdf
\multirow{3}{*}{$\epsc=0.75$}
& \includegraphics[scale=.2]{fig/horseRev/horseRev_Gauss1.5.png} 
& \includegraphics[scale=.2]{fig/horseRev/horseRev_rep1_rr01_eps8.03_crowd1_1bitS_res0.png} 
& \includegraphics[scale=.2]{fig/horseRev/horseRev_rep1-4_rr01_eps8.03-6.645_crowd1_1bitS_res0.png}
& \includegraphics[scale=.2]{fig/horseRev/horseRev_rep1-16_rr01_eps8.03-5.255_crowd1_1bitS_res0.png}
& \includegraphics[scale=.2]{fig/horseRev/horseRev_rep1-256_rr01_eps8.03-2.49_crowd1_1bitS_res0.png}
\\
\else
\multirow{2}{*}{$\epsc=0.75$}
\fi
&$\sigma=5.47$, \quad $\epslt{*}=\infty$
&$\epslt{\infty}=8.03$, $\epslt{1}=8.03$ 
&$\epslt{\infty}=8.03$, $\epslt{1}=6.42$%
&$\epslt{\infty}=8.03$, $\epslt{1}=5.19$%
&$\epslt{\infty}=8.03$, $\epslt{1}=2.49$%
\\
&$\rmse=3.67$
&$\rmse=16.46$
&$\rmse=23.22$
&$\rmse=23.20$
&$\rmse=24.17$
\\
\midrule

\ifbigpdf
\multirow{3}{*}{$\epsc=1.0$}
& \includegraphics[scale=.2]{fig/horseRev/horseRev_Gauss2.png} 
& \includegraphics[scale=.2]{fig/horseRev/horseRev_rep1_rr01_eps8.55_crowd1_1bitS_res0.png} 
& \includegraphics[scale=.2]{fig/horseRev/horseRev_rep1-4_rr01_eps8.55-7.165_crowd1_1bitS_res0.png}
& \includegraphics[scale=.2]{fig/horseRev/horseRev_rep1-16_rr01_eps8.55-5.775_crowd1_1bitS_res0.png}
& \includegraphics[scale=.2]{fig/horseRev/horseRev_rep1-256_rr01_eps8.55-3.005_crowd1_1bitS_res0.png}
\\
\else
\multirow{2}{*}{$\epsc=1.0$}
\fi
&$\sigma=4.14$, \quad $\epslt{*}=\infty$
&$\epslt{\infty}=8.55$, $\epslt{1}=8.55$ 
&$\epslt{\infty}=8.55$, $\epslt{1}=6.94$%
&$\epslt{\infty}=8.55$, $\epslt{1}=5.71$%
&$\epslt{\infty}=8.55$, $\epslt{1}=3.00$%
\\
&$\rmse=2.78$
&$\rmse=12.96$
&$\rmse=17.92$
&$\rmse=18.01$
&$\rmse=18.39$
\\
\midrule

\end{tabular}
\end{table*}
\else
\fi

\ifsketch
\ifhorseRev
\begin{table*}[t!]
\caption{Experimental results for
LDP reports using count-sketch encodings
with $1\mathrm{,}024$ hash functions and sketch size $65\mathrm{,}536$,
following Apple's practical deployment~\cite{appledp}.
Like in Table~\ref{fig:horse-1-crowd}, the task is to reconstruct  the \horseRev{} dataset, for varying $\epsc$, using a central DP mechanism, attribute fragmenting, and both attribute and report fragmenting at $(\epsc, \deltac)$-central DP with $\deltac = 5\times 10^{-8}$.
To demonstrate the estimation error introduced by the sketching algorithm, the last row gives the non-private baseline.
}
\centering
\newcolumntype{D}{ >{\centering\arraybackslash} m{2.3cm} }
\newcolumntype{N}{ >{\centering\arraybackslash} m{2.3cm} }
\begin{tabular}{ D N N N N N N }

\multirow{2}{*}{\makecell{Central privacy \\ guarantee}}
& \multirow{2}{*}{\makecell{Attribute-fragmented \\LDP~report}}
& \multicolumn{3}{c}{Attribute- and report-fragmented LDP reports}
\\ \cmidrule(lr){3-5}
& &$\tau=4$ reports &$\tau=16$ reports &$\tau=256$ reports
\\ \toprule

\ifbigpdf
\multirow{3}{*}{$ 0.05 \leq \epsc \leq \epslt{\infty}$}
& \includegraphics[scale=.2]{fig/horseRev/horseRev_rep1_rr01_eps2.94_hash1024_sketch65536_crowd1_1bitS_res0.png} 
& \includegraphics[scale=.2]{fig/horseRev/horseRev_rep1-4_rr01_eps2.94-1.595_hash1024_sketch65536_crowd1_1bitS_res0.png}
& \includegraphics[scale=.2]{fig/horseRev/horseRev_rep1-16_rr01_eps2.94-0.56_hash1024_sketch65536_crowd1_1bitS_res0.png}
& \includegraphics[scale=.2]{fig/horseRev/horseRev_rep1-256_rr01_eps2.94-0.035_hash1024_sketch65536_crowd1_1bitS_res0.png}
\\
\else
\multirow{2}{*}{$ 0.05 \leq \epsc \leq \epslt{\infty}$}
\fi
&$\epslt{\infty}=2.94$, $\epslt{1}=2.94$ 
&$\epslt{\infty}=2.91$, $\epslt{1}=1.37$
&$\epslt{\infty}=2.94$, $\epslt{1}=0.50$
&$\epslt{\infty}=2.94$, $\epslt{1}=0.03$
\\
&$\rmse=118.57$
&$\rmse=128.34$
&$\rmse=132.52$
&$\rmse=139.47$
\\
\midrule
        
\ifbigpdf
\multirow{3}{*}{$ 0.25 \leq \epsc \leq \epslt{\infty}$}
& \includegraphics[scale=.2]{fig/horseRev/horseRev_rep1_rr01_eps5.96_hash1024_sketch65536_crowd1_1bitS_res0.png} 
& \includegraphics[scale=.2]{fig/horseRev/horseRev_rep1-4_rr01_eps5.96-4.575_hash1024_sketch65536_crowd1_1bitS_res0.png}
& \includegraphics[scale=.2]{fig/horseRev/horseRev_rep1-16_rr01_eps5.96-3.19_hash1024_sketch65536_crowd1_1bitS_res0.png}
& \includegraphics[scale=.2]{fig/horseRev/horseRev_rep1-256_rr01_eps5.96-0.7_hash1024_sketch65536_crowd1_1bitS_res0.png}
\\
\else
\multirow{2}{*}{$ 0.25 \leq \epsc \leq \epslt{\infty}$}
\fi
&$\epslt{\infty}=5.96$, $\epslt{1}=5.96$ 
&$\epslt{\infty}=5.96$, $\epslt{1}=4.35$
&$\epslt{\infty}=5.96$, $\epslt{1}=3.13$
&$\epslt{\infty}=5.96$, $\epslt{1}=0.70$
\\
&$\rmse=45.47$
&$\rmse=62.82$
&$\rmse=63.70$
&$\rmse=81.91$
\\
\midrule
        
\ifbigpdf
\multirow{3}{*}{$ 0.5 \leq \epsc \leq \epslt{\infty}$}
& \includegraphics[scale=.2]{fig/horseRev/horseRev_rep1_rr01_eps7.28_hash1024_sketch65536_crowd1_1bitS_res0.png} 
& \includegraphics[scale=.2]{fig/horseRev/horseRev_rep1-4_rr01_eps7.28-5.895_hash1024_sketch65536_crowd1_1bitS_res0.png}
& \includegraphics[scale=.2]{fig/horseRev/horseRev_rep1-16_rr01_eps7.28-4.51_hash1024_sketch65536_crowd1_1bitS_res0.png}
& \includegraphics[scale=.2]{fig/horseRev/horseRev_rep1-256_rr01_eps7.28-1.765_hash1024_sketch65536_crowd1_1bitS_res0.png}
\\
\else
\multirow{2}{*}{$ 0.5 \leq \epsc \leq \epslt{\infty}$}
\fi
&$\epslt{\infty}=7.28$, $\epslt{1}=7.28$ 
&$\epslt{\infty}=7.28$, $\epslt{1}=5.67$
&$\epslt{\infty}=7.28$, $\epslt{1}=4.45$
&$\epslt{\infty}=7.28$, $\epslt{1}=1.76$
\\
&$\rmse=24.11$
&$\rmse=33.62$
&$\rmse=33.78$
&$\rmse=36.66$
\\
\midrule
        
\ifbigpdf
\multirow{3}{*}{$ 0.75 \leq \epsc \leq \epslt{\infty}$}
& \includegraphics[scale=.2]{fig/horseRev/horseRev_rep1_rr01_eps8.03_hash1024_sketch65536_crowd1_1bitS_res0.png} 
& \includegraphics[scale=.2]{fig/horseRev/horseRev_rep1-4_rr01_eps8.03-6.645_hash1024_sketch65536_crowd1_1bitS_res0.png}
& \includegraphics[scale=.2]{fig/horseRev/horseRev_rep1-16_rr01_eps8.03-5.255_hash1024_sketch65536_crowd1_1bitS_res0.png}
& \includegraphics[scale=.2]{fig/horseRev/horseRev_rep1-256_rr01_eps8.03-2.49_hash1024_sketch65536_crowd1_1bitS_res0.png}
\\
\else
\multirow{2}{*}{$ 0.75 \leq \epsc \leq \epslt{\infty}$}
\fi
&$\epslt{\infty}=8.03$, $\epslt{1}=8.03$ 
&$\epslt{\infty}=8.03$, $\epslt{1}=6.42$
&$\epslt{\infty}=8.03$, $\epslt{1}=5.19$
&$\epslt{\infty}=8.03$, $\epslt{1}=2.49$
\\
&$\rmse=17.05$
&$\rmse=23.60$
&$\rmse=23.40$
&$\rmse=24.48$
\\
\midrule

\ifbigpdf
\multirow{3}{*}{$ 1.0 \leq \epsc \leq \epslt{\infty}$}
& \includegraphics[scale=.2]{fig/horseRev/horseRev_rep1_rr01_eps8.55_hash1024_sketch65536_crowd1_1bitS_res0.png} 
& \includegraphics[scale=.2]{fig/horseRev/horseRev_rep1-4_rr01_eps8.55-7.165_hash1024_sketch65536_crowd1_1bitS_res0.png}
& \includegraphics[scale=.2]{fig/horseRev/horseRev_rep1-16_rr01_eps8.55-5.775_hash1024_sketch65536_crowd1_1bitS_res0.png}
& \includegraphics[scale=.2]{fig/horseRev/horseRev_rep1-256_rr01_eps8.55-3.005_hash1024_sketch65536_crowd1_1bitS_res0.png}
\\
\else
\multirow{2}{*}{$ 1.0 \leq \epsc \leq \epslt{\infty}$}
\fi
&$\epslt{\infty}=8.55$, $\epslt{1}=8.55$ 
&$\epslt{\infty}=8.55$, $\epslt{1}=6.94$
&$\epslt{\infty}=8.55$, $\epslt{1}=5.71$
&$\epslt{\infty}=8.55$, $\epslt{1}=3.00$
\\
&$\rmse=13.25$
&$\rmse=18.32$
&$\rmse=18.43$
&$\rmse=18.74$
\\
\midrule

\ifbigpdf
\multirow{3}{*}{$\epsc=\infty$}
& \includegraphics[scale=.2]{fig/horseRev/horseRev_rep1_nonPriv_hash1024_sketch65536_crowd1_res0.png} 
& \includegraphics[scale=.2]{fig/horseRev/horseRev_rep1-4_nonPriv_hash1024_sketch65536_crowd1_res0.png}
& \includegraphics[scale=.2]{fig/horseRev/horseRev_rep1-16_nonPriv_hash1024_sketch65536_crowd1_res0.png}
& \includegraphics[scale=.2]{fig/horseRev/horseRev_rep1-256_nonPriv_hash1024_sketch65536_crowd1_res0.png}
\\
\else
\multirow{2}{*}{$\epsc=\infty$}
\fi
&$\epslt{*}=\infty$
&$\epslt{*}=\infty$
&$\epslt{*}=\infty$
&$\epslt{*}=\infty$
\\
&$\rmse=4.12$
&$\rmse=4.12$
&$\rmse=4.12$
&$\rmse=4.12$
\\
\midrule
\end{tabular}
\end{table*}
\fi
\fi

\ifhorseRev
\ifonehash
\begin{table*}[t!]
\caption{Experimental results of reconstructing the \horseRev{} 
image dataset
from
a collection of anonymous reports that result from
running the LDP reporting
protocol one, two, or five times
for every single respondent,
using all, half, or a fifth
of each respondent's $\epslt{}$ privacy budget, respectively.
In each case, the LDP reports
also utilize attribute fragmenting.
In all experiments, the best utility is achieved
when the entire  $\epslt{}$ privacy budget
is used to construct LDP reports in a single run.
}
\label{fig:horse_1_report}
\centering
\newcolumntype{D}{ >{\centering\arraybackslash} m{2cm} }
\newcolumntype{N}{ >{\centering\arraybackslash} m{2.5cm} }
\begin{tabular}{ D N N N }
Local privacy guarantee &Single LDP run &Two LDP runs &Five LDP runs
\\ \toprule

\ifbigpdf
\multirow{2}{*}{Total $\epslt=2.94$}
&\includegraphics[scale=.2]{fig/horseRev/horseRev_rep1_rr01_eps2.94_crowd1_1bitS_res0.png}
&\includegraphics[scale=.2]{fig/horseRev/horseRev_rep1_rr01_eps2.94_rhash2_crowd1_1bitS_res0.png}
&\includegraphics[scale=.2]{fig/horseRev/horseRev_rep1_rr01_eps2.94_rhash5_crowd1_1bitS_res0.png}
\\
\else
{Total $\epslt=2.94$}
\fi
&$\rmse=118.40$
&$\rmse=157.76$
&$\rmse=166.50$
\\
\midrule

\ifbigpdf
\multirow{2}{*}{Total $\epslt=5.96$}
&\includegraphics[scale=.2]{fig/horseRev/horseRev_rep1_rr01_eps5.96_crowd1_1bitS_res0.png}
&\includegraphics[scale=.2]{fig/horseRev/horseRev_rep1_rr01_eps5.96_rhash2_crowd1_1bitS_res0.png}
&\includegraphics[scale=.2]{fig/horseRev/horseRev_rep1_rr01_eps5.96_rhash5_crowd1_1bitS_res0.png}
\\
\else
{Total $\epslt=5.96$}
\fi
&$\rmse=45.01$
&$\rmse=127.07$
&$\rmse=154.87$
\\
\midrule

\ifbigpdf
\multirow{2}{*}{Total $\epslt=7.28$}
&\includegraphics[scale=.2]{fig/horseRev/horseRev_rep1_rr01_eps7.28_crowd1_1bitS_res0.png}
&\includegraphics[scale=.2]{fig/horseRev/horseRev_rep1_rr01_eps7.28_rhash2_crowd1_1bitS_res0.png}
&\includegraphics[scale=.2]{fig/horseRev/horseRev_rep1_rr01_eps7.28_rhash5_crowd1_1bitS_res0.png}
\\
\else
{Total $\epslt=7.28$}
\fi
&$\rmse=23.79$
&$\rmse=107.85$
&$\rmse=148.45$
\\
\midrule

\ifbigpdf
\multirow{2}{*}{Total $\epslt=8.03$}
&\includegraphics[scale=.2]{fig/horseRev/horseRev_rep1_rr01_eps8.03_crowd1_1bitS_res0.png}
&\includegraphics[scale=.2]{fig/horseRev/horseRev_rep1_rr01_eps8.03_rhash2_crowd1_1bitS_res0.png}
&\includegraphics[scale=.2]{fig/horseRev/horseRev_rep1_rr01_eps8.03_rhash5_crowd1_1bitS_res0.png}
\\
\else
{Total $\epslt=8.03$}
\fi
&$\rmse=16.46$
&$\rmse=96.02$
&$\rmse=145.09$
\\
\midrule

\ifbigpdf
\multirow{2}{*}{Total $\epslt=8.55$}
&\includegraphics[scale=.2]{fig/horseRev/horseRev_rep1_rr01_eps8.55_crowd1_1bitS_res0.png}
&\includegraphics[scale=.2]{fig/horseRev/horseRev_rep1_rr01_eps8.55_rhash2_crowd1_1bitS_res0.png}
&\includegraphics[scale=.2]{fig/horseRev/horseRev_rep1_rr01_eps8.55_rhash5_crowd1_1bitS_res0.png}
\\
\else
{Total $\epslt=8.55$}
\fi
&$\rmse=12.96$
&$\rmse=88.49$
&$\rmse=141.44$
\\
\midrule

\end{tabular}
\end{table*}
\fi
\fi

\ifonehash
\ifhorseRev
\begin{table*}[t!]
\caption{Experimental results of reconstructing the \horseRev{} 
image dataset
from LDP reports about the results of one, two, or five sketching hash functions,
based on count sketching with $1\textrm{,}024$ hash functions and sketch size $65\textrm{,}536$.
(In each case, the LDP reports
also utilize attribute fragmenting.)
In all experiments, the best utility is achieved
when LDP reports
use the entire $\epslt{}$ privacy budget
to report on the result of a single hash function.
}
\label{fig:horse_1_sketch_report}
\centering
\newcolumntype{D}{ >{\centering\arraybackslash} m{2cm} }
\newcolumntype{N}{ >{\centering\arraybackslash} m{2.5cm} }
\begin{tabular}{ D N N N }
Local privacy guarantee &One hash function & Two hash functions & Five hash functions
\\ \toprule

\ifbigpdf
\multirow{2}{*}{Total $\epslt=2.94$}
&\includegraphics[scale=.2]{fig/horseRev/horseRev_rep1_rr01_eps2.94_hash1024_sketch65536_crowd1_1bitS_res0.png}
&\includegraphics[scale=.2]{fig/horseRev/horseRev_rep1_rr01_eps2.94_hash1024_sketch65536_rhash2_crowd1_1bitS_res0.png}
&\includegraphics[scale=.2]{fig/horseRev/horseRev_rep1_rr01_eps2.94_hash1024_sketch65536_rhash5_crowd1_1bitS_res0.png}
\\
\else
{Total $\epslt=2.94$}
\fi
&$\rmse=118.57$
&$\rmse=129.86$
&$\rmse=135.17$
\\
\midrule

\ifbigpdf
\multirow{2}{*}{Total $\epslt=5.96$}
&\includegraphics[scale=.2]{fig/horseRev/horseRev_rep1_rr01_eps5.96_hash1024_sketch65536_crowd1_1bitS_res0.png}
&\includegraphics[scale=.2]{fig/horseRev/horseRev_rep1_rr01_eps5.96_hash1024_sketch65536_rhash2_crowd1_1bitS_res0.png}
&\includegraphics[scale=.2]{fig/horseRev/horseRev_rep1_rr01_eps5.96_hash1024_sketch65536_rhash5_crowd1_1bitS_res0.png}
\\
\else
{Total $\epslt=5.96$}
\fi
&$\rmse=45.47$
&$\rmse=105.85$
&$\rmse=126.16$
\\
\midrule

\ifbigpdf
\multirow{2}{*}{Total $\epslt=7.28$}
&\includegraphics[scale=.2]{fig/horseRev/horseRev_rep1_rr01_eps7.28_hash1024_sketch65536_crowd1_1bitS_res0.png}
&\includegraphics[scale=.2]{fig/horseRev/horseRev_rep1_rr01_eps7.28_hash1024_sketch65536_rhash2_crowd1_1bitS_res0.png}
&\includegraphics[scale=.2]{fig/horseRev/horseRev_rep1_rr01_eps7.28_hash1024_sketch65536_rhash5_crowd1_1bitS_res0.png}
\\
\else
{Total $\epslt=7.28$}
\fi
&$\rmse=24.11$
&$\rmse=89.41$
&$\rmse=122.00$
\\
\midrule

\ifbigpdf
\multirow{2}{*}{Total $\epslt=8.03$}
&\includegraphics[scale=.2]{fig/horseRev/horseRev_rep1_rr01_eps8.03_hash1024_sketch65536_crowd1_1bitS_res0.png}
&\includegraphics[scale=.2]{fig/horseRev/horseRev_rep1_rr01_eps8.03_hash1024_sketch65536_rhash2_crowd1_1bitS_res0.png}
&\includegraphics[scale=.2]{fig/horseRev/horseRev_rep1_rr01_eps8.03_hash1024_sketch65536_rhash5_crowd1_1bitS_res0.png}
\\
\else
{Total $\epslt=8.03$}
\fi
&$\rmse=17.05$
&$\rmse=78.79$
&$\rmse=118.56$
\\
\midrule

\ifbigpdf
\multirow{2}{*}{Total $\epslt=8.55$}
&\includegraphics[scale=.2]{fig/horseRev/horseRev_rep1_rr01_eps8.55_hash1024_sketch65536_crowd1_1bitS_res0.png}
&\includegraphics[scale=.2]{fig/horseRev/horseRev_rep1_rr01_eps8.55_hash1024_sketch65536_rhash2_crowd1_1bitS_res0.png}
&\includegraphics[scale=.2]{fig/horseRev/horseRev_rep1_rr01_eps8.55_hash1024_sketch65536_rhash5_crowd1_1bitS_res0.png}
\\
\else
{Total $\epslt=8.55$}
\fi
&$\rmse=13.25$
&$\rmse=71.53$
&$\rmse=117.06$
\\
\midrule

\end{tabular}\vspace*{5ex}
\end{table*}
\fi
\fi

\ifmap
\begin{table*}[t!]
\centering
\caption{Results of experiments reconstructing the phone-location \map{} dataset by varying $\epsc$ and evaluating a central DP mechanism, attribute fragmenting, and both attribute and report fragmenting achieving $(\epsc, \deltac)$-central DP with $\deltac = 5\times 10^{-10}$.}
\label{fig:map-1-crowd}
\newcolumntype{D}{ >{\centering\arraybackslash} m{1.6cm} }
\newcolumntype{N}{ >{\centering\arraybackslash} m{3.2cm} }
\setlength\tabcolsep{0.5pt} %
\begin{tabular}{D N N N N N N}

\multirow{2}{*}{\makecell{Central privacy \\ guarantee}}
& \multirow{2}{*}{\makecell{No LDP \\(Gaussian mechanism)}}
& \multirow{2}{*}{\makecell{Attribute-fragmented \\LDP~report}}
& \multicolumn{3}{c}{Attribute- and report-fragmented LDP reports}
\\ \cmidrule(lr){4-6}
& & &$\tau=4$ reports &$\tau=16$ reports &$\tau=256$ reports
\\ \toprule

\ifbigpdf
\multirow{3}{*}{$\epsc=0.05$}
&\includegraphics[scale=.04]{fig/map2.5/map2.5_Gauss0.1.png} 
&\includegraphics[scale=.04]{fig/map2.5/map2.5_rep1_rr01_eps7.385_crowd1_1bitS_res0.png}
&\includegraphics[scale=.04]{fig/map2.5/map2.5_rep1-4_rr01_eps7.385-6.0_crowd1_1bitS_res0.png}
&\includegraphics[scale=.04]{fig/map2.5/map2.5_rep1-16_rr01_eps7.385-4.615_crowd1_1bitS_res0.png}
&\includegraphics[scale=.04]{fig/map2.5/map2.5_rep1-256_rr01_eps7.385-1.865_crowd1_1bitS_res0.png}
\\
\else
\multirow{2}{*}{$\epsc=0.05$}
\fi
&$\sigma=91.16$, \hspace{100pt} $\epslt{1}=\infty$
&$\epslt{\infty}=7.39$, \hspace{100pt} $\epslt{1}=7.39$ 
&$\epslt{\infty}=7.39$, \hspace{100pt} $\epslt{1}=5.78$%
&$\epslt{\infty}=7.39$, \hspace{100pt} $\epslt{1}=4.55$%
&$\epslt{\infty}=7.39$, \hspace{100pt} $\epslt{1}=1.86$%
\\
&$\rmse=62.08$
&$\rmse=150.54$
&$\rmse=160.02$
&$\rmse=160.19$
&$\rmse=161.85$
\\
\midrule

\ifbigpdf
\multirow{3}{*}{$\epsc=0.25$}
&\includegraphics[scale=.04]{fig/map2.5/map2.5_Gauss0.5.png} 
&\includegraphics[scale=.04]{fig/map2.5/map2.5_rep1_rr01_eps10.555_crowd1_1bitS_res0.png}
&\includegraphics[scale=.04]{fig/map2.5/map2.5_rep1-4_rr01_eps10.555-9.17_crowd1_1bitS_res0.png}
&\includegraphics[scale=.04]{fig/map2.5/map2.5_rep1-16_rr01_eps10.555-7.78_crowd1_1bitS_res0.png}
&\includegraphics[scale=.04]{fig/map2.5/map2.5_rep1-256_rr01_eps10.555-5.01_crowd1_1bitS_res0.png}
\\
\else
\multirow{2}{*}{$\epsc=0.25$}
\fi
&$\sigma=18.32$, \hspace{100pt} $\epslt{1}=\infty$
&$\epslt{\infty}=10.56$, \hspace{100pt} $\epslt{1}=10.56$ 
&$\epslt{\infty}=10.56$, \hspace{100pt} $\epslt{1}=8.95$%
&$\epslt{\infty}=10.56$, \hspace{100pt} $\epslt{1}=7.72$%
&$\epslt{\infty}=10.56$, \hspace{100pt} $\epslt{1}=5.01$%
\\
&$\rmse=14.80$
&$\rmse=83.01$
&$\rmse=97.15$
&$\rmse=97.24$
&$\rmse=97.28$
\\
\midrule

\ifbigpdf
\multirow{3}{*}{$\epsc=0.5$}
&\includegraphics[scale=.04]{fig/map2.5/map2.5_Gauss1.png} 
&\includegraphics[scale=.04]{fig/map2.5/map2.5_rep1_rr01_eps11.88_crowd1_1bitS_res0.png}
&\includegraphics[scale=.04]{fig/map2.5/map2.5_rep1-4_rr01_eps11.88-10.495_crowd1_1bitS_res0.png}
&\includegraphics[scale=.04]{fig/map2.5/map2.5_rep1-16_rr01_eps11.88-9.105_crowd1_1bitS_res0.png}
&\includegraphics[scale=.04]{fig/map2.5/map2.5_rep1-256_rr01_eps11.88-6.335_crowd1_1bitS_res0.png}
\\
\else
\multirow{2}{*}{$\epsc=0.5$}
\fi
&$\sigma=9.21$, \hspace{100pt} $\epslt{1}=\infty$
&$\epslt{\infty}=11.88$, \hspace{100pt} $\epslt{1}=11.88$ 
&$\epslt{\infty}=11.88$, \hspace{100pt} $\epslt{1}=10.27$%
&$\epslt{\infty}=11.88$, \hspace{100pt} $\epslt{1}=9.04$%
&$\epslt{\infty}=11.88$, \hspace{100pt} $\epslt{1}=6.33$%
\\
&$\rmse=7.83$
&$\rmse=67.31$
&$\rmse=73.74$
&$\rmse=73.76$
&$\rmse=73.75$
\\
\midrule

\ifbigpdf
\multirow{3}{*}{$\epsc=0.75$}
&\includegraphics[scale=.04]{fig/map2.5/map2.5_Gauss1.5.png} 
&\includegraphics[scale=.04]{fig/map2.5/map2.5_rep1_rr01_eps12.63_crowd1_1bitS_res0.png}
&\includegraphics[scale=.04]{fig/map2.5/map2.5_rep1-4_rr01_eps12.63-11.245_crowd1_1bitS_res0.png}
&\includegraphics[scale=.04]{fig/map2.5/map2.5_rep1-16_rr01_eps12.63-9.855_crowd1_1bitS_res0.png}
&\includegraphics[scale=.04]{fig/map2.5/map2.5_rep1-256_rr01_eps12.63-7.085_crowd1_1bitS_res0.png}
\\
\else
\multirow{2}{*}{$\epsc=0.75$}
\fi
&$\sigma=6.18$, \hspace{100pt} $\epslt{1}=\infty$
&$\epslt{\infty}=12.63$, \hspace{100pt} $\epslt{1}=12.63$ 
&$\epslt{\infty}=12.63$, \hspace{100pt} $\epslt{1}=11.02$%
&$\epslt{\infty}=12.63$, \hspace{100pt} $\epslt{1}=9.79$%
&$\epslt{\infty}=12.63$, \hspace{100pt} $\epslt{1}=7.08$%
\\
&$\rmse=5.39$
&$\rmse=63.00$
&$\rmse=66.79$
&$\rmse=66.79$
&$\rmse=66.80$
\\
\midrule

\ifbigpdf
\multirow{3}{*}{$\epsc=1.0$}
&\includegraphics[scale=.04]{fig/map2.5/map2.5_Gauss1.5.png} 
&\includegraphics[scale=.04]{fig/map2.5/map2.5_rep1_rr01_eps13.14_crowd1_1bitS_res0.png}
&\includegraphics[scale=.04]{fig/map2.5/map2.5_rep1-4_rr01_eps13.14-11.755_crowd1_1bitS_res0.png}
&\includegraphics[scale=.04]{fig/map2.5/map2.5_rep1-16_rr01_eps13.14-10.365_crowd1_1bitS_res0.png}
&\includegraphics[scale=.04]{fig/map2.5/map2.5_rep1-256_rr01_eps13.14-7.595_crowd1_1bitS_res0.png}
\\
\else
\multirow{2}{*}{$\epsc=1.0$}
\fi
&$\sigma=4.66$, \hspace{100pt} $\epslt{1}=\infty$
&$\epslt{\infty}=13.14$, \hspace{100pt} $\epslt{1}=13.14$ %
&$\epslt{\infty}=13.14$, \hspace{100pt} $\epslt{1}=11.53$%
&$\epslt{\infty}=13.14$, \hspace{100pt} $\epslt{1}=10.30$%
&$\epslt{\infty}=13.14$, \hspace{100pt} $\epslt{1}=7.59$%
\\
&$\rmse=4.13$
&$\rmse=61.31$
&$\rmse=63.80$
&$\rmse=63.85$
&$\rmse=63.83$
\\
\midrule

\end{tabular}\vspace*{5ex}
\end{table*}
\else
\fi

\ifgirl
\begin{table*}[t!]
\caption{Results of experiments reconstructing the \girl{} image dataset by varying $\epsc$ and evaluating a central DP mechanism, attribute fragmenting, and both attribute and report fragmenting achieving $(\epsc, \deltac)$-central DP with $\deltac = 5\times 10^{-9}$.}
\label{fig:girl-1-crowd}
\centering
\newcolumntype{D}{ >{\centering\arraybackslash} m{2.0cm} }
\newcolumntype{N}{ >{\centering\arraybackslash} m{2.6cm} }
\begin{tabular}{ D N N N N N N }
\multirow{2}{*}{\makecell{Central privacy \\ guarantee}}
& \multirow{2}{*}{\makecell{No LDP\\(Gaussian mechanism)}}
& \multirow{2}{*}{\makecell{Attribute-fragmented \\LDP~report}}
& \multicolumn{3}{c}{Attribute- and report-fragmented LDP reports}
\\ \cmidrule(lr){4-6}
& & &$\tau=4$ reports &$\tau=16$ reports &$\tau=256$ reports
\\ \toprule

\ifbigpdf
\multirow{3}{*}{$\epsc=0.05$}
&\includegraphics[scale=.1]{fig/girl/girl_Gauss0.1.png} 
&\includegraphics[scale=.1]{fig/girl/girl_rep1_rr01_eps5.95_crowd1_1bitS_res0.png} 
&\includegraphics[scale=.1]{fig/girl/girl_rep1-4_rr01_eps5.95-4.565_crowd1_1bitS_res0.png}
&\includegraphics[scale=.1]{fig/girl/girl_rep1-16_rr01_eps5.95-3.18_crowd1_1bitS_res0.png}
&\includegraphics[scale=.1]{fig/girl/girl_rep1-256_rr01_eps5.95-0.695_crowd1_1bitS_res0.png}
\\
\else
\multirow{2}{*}{$\epsc=0.05$}
\fi
&$\sigma=85.95$, \qquad $\epslt{1}=\infty$
&$\epslt{\infty}=5.95$, $\epslt{1}=5.95$ 
&$\epslt{\infty}=5.95$, $\epslt{1}=4.34$%
&$\epslt{\infty}=5.95$, $\epslt{1}=3.12$%
&$\epslt{\infty}=5.95$, $\epslt{1}=0.69$%
\\
&$\rmse=70.68$
&$\rmse=121.81$
&$\rmse=127.40$
&$\rmse=127.50$
&$\rmse=131.22$
\\
\midrule

\ifbigpdf
\multirow{3}{*}{$\epsc=0.25$}
&\includegraphics[scale=.1]{fig/girl/girl_Gauss0.5.png} 
&\includegraphics[scale=.1]{fig/girl/girl_rep1_rr01_eps9.11_crowd1_1bitS_res0.png} 
&\includegraphics[scale=.1]{fig/girl/girl_rep1-4_rr01_eps9.11-7.725_crowd1_1bitS_res0.png}
&\includegraphics[scale=.1]{fig/girl/girl_rep1-16_rr01_eps9.11-6.335_crowd1_1bitS_res0.png}
&\includegraphics[scale=.1]{fig/girl/girl_rep1-256_rr01_eps9.11-3.565_crowd1_1bitS_res0.png}
\\
\else
\multirow{2}{*}{$\epsc=0.25$}
\fi
&$\sigma=17.28$, \qquad $\epslt{1}=\infty$
&$\epslt{\infty}=9.11$, $\epslt{1}=9.11$ 
&$\epslt{\infty}=9.11$, $\epslt{1}=7.50$%
&$\epslt{\infty}=9.11$, $\epslt{1}=6.27$%
&$\epslt{\infty}=9.11$, $\epslt{1}=3.56$%
\\
&$\rmse=17.06$
&$\rmse=63.84$
&$\rmse=80.65$
&$\rmse=80.46$
&$\rmse=81.27$
\\
\midrule

\ifbigpdf
\multirow{3}{*}{$\epsc=0.5$}
&\includegraphics[scale=.1]{fig/girl/girl_Gauss1.png} 
&\includegraphics[scale=.1]{fig/girl/girl_rep1_rr01_eps10.435_crowd1_1bitS_res0.png} 
&\includegraphics[scale=.1]{fig/girl/girl_rep1-4_rr01_eps10.435-9.05_crowd1_1bitS_res0.png}
&\includegraphics[scale=.1]{fig/girl/girl_rep1-16_rr01_eps10.435-7.66_crowd1_1bitS_res0.png}
&\includegraphics[scale=.1]{fig/girl/girl_rep1-256_rr01_eps10.435-4.89_crowd1_1bitS_res0.png}
\\
\else
\multirow{2}{*}{$\epsc=0.5$}
\fi
&$\sigma=8.70$,\qquad $\epslt{1}=\infty$
&$\epslt{\infty}=10.435$, $\epslt{1}=10.435$ 
&$\epslt{\infty}=10.435$, $\epslt{1}=8.83$%
&$\epslt{\infty}=10.435$, $\epslt{1}=7.60$%
&$\epslt{\infty}=10.435$, $\epslt{1}=4.89$%
\\
&$\rmse=8.68$
&$\rmse=36.56$
&$\rmse=49.72$
&$\rmse=49.69$
&$\rmse=49.80$
\\
\midrule

\ifbigpdf
\multirow{3}{*}{$\epsc=0.75$}
&\includegraphics[scale=.1]{fig/girl/girl_Gauss1.5.png} 
&\includegraphics[scale=.1]{fig/girl/girl_rep1_rr01_eps11.18_crowd1_1bitS_res0.png} 
&\includegraphics[scale=.1]{fig/girl/girl_rep1-4_rr01_eps11.18-9.795_crowd1_1bitS_res0.png}
&\includegraphics[scale=.1]{fig/girl/girl_rep1-16_rr01_eps11.18-8.405_crowd1_1bitS_res0.png}
&\includegraphics[scale=.1]{fig/girl/girl_rep1-256_rr01_eps11.18-5.635_crowd1_1bitS_res0.png}
\\
\else
\multirow{2}{*}{$\epsc=0.75$}
\fi
&$\sigma=5.84$,\qquad $\epslt{1}=\infty$
&$\epslt{\infty}=11.18$, $\epslt{1}=11.18$ 
&$\epslt{\infty}=11.18$, $\epslt{1}=9.57$%
&$\epslt{\infty}=11.18$, $\epslt{1}=8.34$%
&$\epslt{\infty}=11.18$, $\epslt{1}=5.63$%
\\
&$\rmse=5.84$
&$\rmse=25.83$
&$\rmse=35.67$
&$\rmse=35.81$
&$\rmse=35.77$
\\
\midrule

\ifbigpdf
\multirow{3}{*}{$\epsc=1.0$}
&\includegraphics[scale=.1]{fig/girl/girl_Gauss2.png} 
&\includegraphics[scale=.1]{fig/girl/girl_rep1_rr01_eps11.7_crowd1_1bitS_res0.png} 
&\includegraphics[scale=.1]{fig/girl/girl_rep1-4_rr01_eps11.7-10.315_crowd1_1bitS_res0.png}
&\includegraphics[scale=.1]{fig/girl/girl_rep1-16_rr01_eps11.7-8.925_crowd1_1bitS_res0.png}
&\includegraphics[scale=.1]{fig/girl/girl_rep1-256_rr01_eps11.7-6.155_crowd1_1bitS_res0.png}
\\
\else
\multirow{2}{*}{$\epsc=1.0$}
\fi
&$\sigma=4.41$,\qquad $\epslt{1}=\infty$
&$\epslt{\infty}=11.7$, $\epslt{1}=11.7$ %
&$\epslt{\infty}=11.7$, $\epslt{1}=10.09$%
&$\epslt{\infty}=11.7$, $\epslt{1}=8.86$%
&$\epslt{\infty}=11.7$, $\epslt{1}=6.15$%
\\
&$\rmse=4.40$
&$\rmse=20.13$
&$\rmse=28.03$
&$\rmse=28.07$
&$\rmse=28.17$
\\
\midrule

\end{tabular}
\end{table*}
\else
\fi

\ifsmother

\fi

This section covers the experimental evaluation of the ideas described in Sections~\ref{sec:attribFrag}--\ref{sec:mlESA}. We consider three scenarios. In the first set of experiments, we consider a typical power law distribution for discovering heavy hitters~\cite{BST17} that is derived from real data collected on a popular browser platform. The second, inspired by increasing uses of differential privacy for hiding potentially sensitive \emph{location} data, considers histogram estimation over \emph{flat-tailed} distributions, where a small number of respondents contribute to a great many number of categories. In order to visualize the privacy/utility tradeoffs, as is natural in these distributions over locations, we select three distributions that correspond to pixel values in three images. The third set of experiments apply ideas in Section~\ref{sec:mlESA} to train models to within state-of-the-art guarantees on standard benchmark datasets.

\subsection{A Dataset with a Heavy-Hitter Powerlaw Distribution}
\label{sec:lthist}
We consider the
``Heavy-hitter'' distribution shown in Table~\ref{tab:dataset_statistics},
as it is representative of on-line behavioral patterns. It comprises 200 million reports collected over a period of one week from a 1.7-million-value domain. The distribution is a mixture of about a hundred heavy hitters and a power law distribution with the probability density function $p(x) \propto x^{-1.35}$. 

Our experiments target different central DP $\epsc$ values to demonstrate the utility of the techniques described in previous sections.
Specifically, we experiment with a few central DP guarantees. For each given $\epsc$, we consider attribute fragmenting with the corresponding $\epsl$ computed using Theorem~\ref{thm:att-frag-privacy}, and report fragmenting with $4$, $16$ and $256$ reports. The fragmenting parameters $\epslt{b}$ and $\epslt{f}$ are selected so that the central DP is $\epsc$ and the variance introduced in the report fragmenting step is roughly the same as that of the backstop step.
We compare the results with a baseline method---the Gaussian mechanism that guarantees only central DP.

\ifsketch
We enforce local differential privacy by randomizing the one-hot encoding of the item, as well as using the private count-sketch algorithm~\cite{BST17,appledp}, which has been demonstrated to work well over distributions with a very large support.
When using private count-sketch, as in~\cite{BST17,appledp}, we use the protocol where each respondent sends one report of their data to \emph{one} randomly sampled hash function. This setting is different from the original non-private count-sketch algorithm, where each respondent sends their data to \emph{all} hash functions. This is because we need to take into consideration the noise used to guarantee local differential privacy. In fact, for the count-sketch algorithm we use, it can be shown~\cite{BST17, bun2018heavy} that under the same local DP budget used in the experiments, the utility is always the best when each respondent sends their data only to one hash function. 
\fi

Table~\ref{fig:real-1-crowd}
shows our experimental results.
In each experiment, we report $\epslt{\infty}$---the LDP guarantee when the adversary observes \emph{all} reports from the respondent, corresponding to Theorem~\ref{thm:record_frag} with $t=\tau$, and (when using report fragmenting) $\epslt{1}$---the LDP guarantee when the adversary observes only \emph{one} report from the respondent, corresponding to Theorem~\ref{thm:record_frag} with $t=1$. For the Gaussian mechanism, we report $\sigma$---the standard deviation of the zero mean Gaussian noise used to achieve the desired level of central privacy.

To measure the utility of the algorithms, we compare the true and estimated frequencies. 
\ifsketch
We also report the expected communication cost for one-hot encoding and count-sketch, as discussed in Section \ref{sec:encodefragment}. The specific sketching algorithm we consider is the one described in~\cite{appledp}.

\fi
\ifsketch
Our experimental results demonstrate that:
\begin{itemize}
\item With attribute fragmenting and report fragmenting with various number of reports, we achieve close to optimal privacy-utility tradeoffs and recover the top 10{,}000 frequent items of the total probability mass with good central differential privacy $\epsc \leq 1$.
\item It is harder to bound the central privacy of count-sketch LDP reports;
  using off-the-shelf parameters~\cite{BST17,appledp}
  results in slightly less communication cost,
  but this can come at a very high cost to utility.
  As we discuss in Section~\ref{sec:attribFrag}, and elsewhere,
  one-hot encodings may be preferable in 
  in the high-epsilon regime,
  at least until stronger results exist for sketch-based encodings.
\end{itemize}
\else
The experimental results demonstrate that with attribute fragmenting and report fragmenting with various number of reports, we achieve close to optimal privacy-utility tradeoffs and recover the top 10{,}000 frequent items of the total probability mass with good central differential privacy $\epsc \leq 1$.
\fi

\begin{table}[th]
\caption{
Alternative 
central differential-privacy bounds for LDP reports like those in the first three rows of Table~\ref{fig:unified_attr},
computed without the use of attribute fragmenting
as the minimum of the LDP guarantee and the central bound from~\cite{privacy-blanket}.}
\label{tab:central_no_attr}
\centering
\begin{tabular}{llll}
&\multicolumn{1}{c}{Row~1}
&\multicolumn{1}{c}{Row~2}
&\multicolumn{1}{c}{Row~3}
\\
&\multicolumn{1}{c}{($\epsl=2.0$)}
&\multicolumn{1}{c}{(With attr.-frag.}
&\multicolumn{1}{c}{(With attr.-frag.}
\\
&\multicolumn{1}{c}{}
&\multicolumn{1}{c}{$\epsc=0.05$)}
&\multicolumn{1}{c}{$\epsc=1.0$)}
\\
\toprule
\textbf{\map}
&$\epsc=0.0729$ 
&$\epsc=7.385 $ 
&$\epsc=13.14 $
\\
\textbf{\horseRev}
&$\epsc=1.0766$
&$\epsc=2.940 $
&$\epsc=\,\;8.55  $
\\
\textbf{\girl}
&$\epsc=0.1517$
&$\epsc=5.950 $
&$\epsc=11.70 $
\\
\textbf{\real}
&$\epsc=0.0788$
&$\epsc=7.235 $
&$\epsc=12.99 $
\end{tabular}
\end{table}

\subsection{Datasets with Low-amplitude and Flat-tailed Distributions}
\label{sec:fthist}
We consider three datasets described below.%

\mypar{Phone Location Dataset} We consider a real-world dataset created by Richard Harris, a graphics editor on The Times's Investigations team showing $235$ million points gathered from $1.2$ million smartphones~\cite{nytimesdata}.\footnote{Direct link to image: https://static01.nyt.com/images/2018/12/14/business\allowbreak/10location-insider/10location-promo-superJumbo-v2.jpg.} The resulting dataset is constructed by taking $2.5\times$ the luminosity values (ranging from $0$ to $255$) of the image to scale up the number of datapoints such that the total number of reports is around $235$ million, with each person reporting coordinates in a $1365 \times 2048$ grid.

\mypar{Horse Image Dataset}
As in the phone location dataset, we consider the dataset corresponding to the image of a sketch of a horse with contours highlighted in white. Due to the majority black nature of this image, it serves as a good test-case for the scenario where the tail is flat, but somewhat sparse.

\mypar{\girl\ Image Dataset}
We use this drawing of a child originally used by Ledig et al.~\cite{image_girl} (converted to a grayscale) to represent a dense distribution with an average luminosity of roughly $140$ and no black pixels. A dense, flat tail distribution is one of the more challenging scenarios for accurately estimating differentially private histograms.

Table~\ref{tab:dataset_statistics} shows the distributions
and statistics of each of these datasets.
As stated before, in our experiments we assume that for every $(x,y)$ with luminosity $L \in [0,255]$, there are $L$ respondents (for the phone location dataset, this count is scaled) each holding a message $(x,y)$. Each $(x,y)$ is converted into a one-hot-encoded LDP report sent using attribute and report fragmenting for improved central privacy.

In Tables~\ref{fig:unified_attr}--\ref{fig:girl-1-crowd}
we report for each dataset
on the results of experiments
similar to those
we performed for the heavy-hitters dataset (shown in Table~\ref{fig:real-1-crowd}).
At various central privacy levels,
we show the measured utility 
of anonymous LDP reporting with attribute and report fragmenting
compared to the utility of analysis without any local privacy guarantee
(the Gaussian mechanism applied to the original data).
To measure utility,
we report the Root Mean Square Error (RMSE) of the resulting histogram estimate.

The essence of our results can be seen in
Table~\ref{fig:unified_attr}, and its companion Table~\ref{tab:central_no_attr}.
At relatively low LDP report privacy of $\epsl=2.0$,
none of the three datasets
can be reconstructed, at all,
whereas at higher $\epsl$
reconstruction becomes feasible;
at $\epsc=1.0$, reconstruction is very good,
and the number of LDP report messages
sent per respondent 
is very low.
As shown in Table~\ref{tab:central_no_attr},
such high utility at a strong central privacy
is only made feasible
by the application of both
amplification-by-shuffling and
attribute fragmenting.

For each of these three datasets,
Tables~\ref{fig:horse-1-crowd}--\ref{fig:girl-1-crowd}
give detailed results of further experiments.\footnote{The reconstructed images missing in these tables are included in ancillary files at \texttt{https://arxiv.org/abs/XXXX.YYYY}.}
Most of these
follow the pattern set by Table~\ref{fig:unified_attr},
while giving more details.
The exceptions are
Table~\ref{fig:horse_1_report} and Table~\ref{fig:horse_1_sketch_report}),
which empirically demonstrate
how each respondent's LDP budget
is best spent on sending a single LDP report
(while appropriately applying attribute or report fragmentation
to that single report).

In our experiments we show:
\begin{enumerate}
\item attribute fragmenting helps us achieve nearly optimal central privacy/accuracy tradeoff,
\item report fragmenting helps us achieve reasonable central privacy with strong per-report local privacy under various number of reports.
\end{enumerate}

\begin{table}[th]
\centering
\caption{Estimating privacy lower bounds via membership inference attacks. }
\label{tab:membership_inference}
\begin{subtable}[h]{1\linewidth}
\caption{TPR$-$FPR. Mean and standard deviation over $10$ runs.}
\begin{tabular}{l c c c c}
TPR$-$FPR   &MNIST                  &Fashion-MNSIT            &CIFAR-10        \\ \toprule
ESA         &$0.0017\pm 0.0014$   &$0.0130\pm 0.0024$     &$0.0097\pm 0.0014$     \\
DPSGD       &$0.0017\pm 0.0016$   &$0.0122\pm 0.0012$     &$0.0095\pm 0.0005$     \\
\end{tabular}
\end{subtable}

\vspace{10pt}
\begin{subtable}[h]{1\linewidth}
\caption{Upper bound of privacy loss as $\epsc$, and lower bound from membership inference attack using the averaged TPR$-$FPR over $10$ runs.}
\begin{tabular}{l c c c c}
Upper / Lower bd &MNIST              &Fashion-MNSIT      &CIFAR-10    \\ \toprule
ESA                 &$27$ / $0.00171$   &$27$ / $0.01306$   &$71.4$ / $0.00970$   \\
DPSGD               &$9.5$ / $0.00166$  &$9.5$ / $0.01228$  &$9$ / $0.00957$\\
\end{tabular}
\end{subtable}
\end{table}

\mypar{Attribute fragmenting}
Each of Tables~\ref{fig:horse-1-crowd},~\ref{fig:map-1-crowd}, and~\ref{fig:girl-1-crowd}
demonstrate how attribute fragmenting
achieves close to optimal privacy/utility tradeoffs comparable to central DP algorithms.
The improvements on reconstructing the histogram as $\epsc$ values go up demonstrate that the optimality results hold asymptotically and bounds arguing the guarantees of privacy amplification could be tightened.

\mypar{Report \& Attribute fragmenting}
Tables~\ref{fig:horse-1-crowd}--\ref{fig:girl-1-crowd} demonstrate
that by combining report and attribute fragmenting, in a variety of scenarios, we can achieve reasonable accuracy while guaranteeing local and central privacy guarantees and never producing highly-identifying individual reports (per-report privacy $\epslt{1}$'s are small).

\subsection{Machine Learning in the ESA Framework}
\label{sec:mlESAexpt}

\begin{table*}[htb]
\centering
\begin{tabular}{ccccccc}

\multirow{2}{*}{Data set} &\multirow{2}{*}{\# examples} &\multirow{2}{*}{\makecell{LDP bound\\per iteration}} &\multirow{2}{*}{Effective batch size} &\multicolumn{3}{c}{Accuracy in \% (at central privacy bound)} \\
\cmidrule(lr){5-7}
& & & & $\epsc=$5 & $\epsc=$10 & $\epsc=$18 \\
\toprule

& &  & \rule{0pt}{3ex}5000 (Rep. frag$=$1) & 58.6  ($\pm$ 1.9) & 61.2  ($\pm$ 1.3)& 62.6  ($\pm$ 0.7)\\[1.0ex]
\cline{4-7}
{CIFAR-10} & {50000} & {$\epsle=$1.8}  &  \rule{0pt}{3ex}10000 (Rep. frag$=$2)& 59.8  ($\pm$ 1.2)& 63.9  ($\pm$ 0.5)& 65.6  ($\pm$ 0.3)\\[1.0ex]
\cline{4-7}
& & &  \rule{0pt}{3ex}25000 (Rep. frag$=$5)&58.1  ($\pm$ 0.8) & 64  ($\pm$ 0.7)&{\bf 66.6  ($\pm$ 0.4)} \\[1.0ex]
\midrule
& & &  \rule{0pt}{3ex}2000 (Rep. frag$=$1) & 84.2 ($\pm$1.7) & 88.9 ($\pm$1.3) & 89.1 ($\pm$1)\\[1.0ex]
\cline{4-7}
{MNIST} & {60000} & {$\epsle=$1.9} &  \rule{0pt}{3ex}4000 (Rep. frag$=$2)& 85.8 ($\pm$1.8) & 92 ($\pm$0.8) & 93 ($\pm$0.4)\\[1.0ex]
\cline{4-7}
& & &  \rule{0pt}{3ex}10000 (Rep. frag$=$5)&80.5  ($\pm$ 2.2) & 91.2  ($\pm$ 0.7)& {\bf 93.9  ($\pm$ 0.4)}\\[1.0ex]
\midrule
& & & \rule{0pt}{3ex}2000 (Rep. frag$=$1) & 71.1 ($\pm$0.7)& 73.3 ($\pm$0.6)&74.5 ($\pm$0.4) \\[1.0ex]
\cline{4-7}
{Fashion-MNIST} & {60000} & {$\epsle=$1.9}  &  \rule{0pt}{3ex}4000 (Rep. frag$=$2)&70.3 ($\pm$0.8) &74.3 ($\pm$0.4) & {\bf 76.4 ($\pm$0.4)}\\[1.0ex]
\cline{4-7}
& & &  \rule{0pt}{3ex}10000 (Rep. frag$=$5)&67  ($\pm$ 1.8) & 73.3  ($\pm$ 0.6)& {\bf 76.1  ($\pm$ 0.5)}\\[1.0ex]
\midrule
\end{tabular}
\caption{Privacy/utility tradeoff for various data sets. Here $\epsle$ refers to LDP per report fragment, effective batch size corresponds to the number of samples/batch $\times$ number of report fragments. The best known accuracy differentially private training of CIFAR-10 models, with $\epsc=8$ (and $\epsle=\infty$) is 73\% \cite{DP-DL}, for MNIST with $\epsc=3$ (and $\epsle=\infty$) is 98\%~\cite{Shoe}, and for Fashion-MNIST with $\epsc=3$ (and $\epsle=\infty$) is 86\%~\cite{Shoe}. All results are averaged over at least 10 runs.}
\label{tbl:res}
\end{table*}

In this section we provide the empirical evidence of the usefulness of the ESA framework in training machine learning model (using variants of LDP-SGD) with \emph{both} local and central DP guarantees. In particular, we show that with per-epoch local DP as small as $\approx 2$, one can can achieve close to state-of-the-art accuracy on benchmark data sets with reasonable central differential privacy guarantees. \emph{We want to emphasize that state-of-the-art results \cite{DP-DL,TFpriv,Shoe} we compare against do not offer any local DP guarantees.} We consider three data sets, MNIST, Fashion-MNIST, and CIFAR-10. We first describe the privacy budget accounting for central differential privacy and then we state the empirical results.

We train our learning models using LDP-SGD, with the modification that we train with randomly sub-sampled mini-batches, rather than full-batch gradient as described in Algorithm \ref{alg:servrG}. The privacy accounting is done as follows. i) Fix the $\epsle$ per mini-batch gradient computation, ii) Amplify the privacy via privacy amplification by shuffling using \cite{BalleBG18}, and iii) Use advanced composition over all the iterations \cite{bun2016concentrated}. Because of the LDP randomness added in Algorithm~\ref{alg:localG} (LDP-SGD; client-side) of Section~\ref{sec:mlESA}, Algorithm~\ref{alg:servrG} (LDP-SGD; server-side) typically requires large mini-batches. Due to engineering considerations, we simulate large batches via \emph{report fragmenting}, as we do not envision the behavior to be significantly different on a real mini-batch of the same size.\footnote{Note that this is only to overcome engineering constraints; we do not need group privacy accounting as this only simulates a larger implementation.} Formally, to simulate a batch size of $m$ with a set of $s$ individual gradients, we report $\tau=m/s$ i.i.d.\ LDP reports of the gradient from each respondent, with  $\epsle$-local differential privacy/report. (To distinguish it from actual batch size, throughout this section we refer to it as \emph{effective batch size}. For privacy amplification by shuffling and sampling, we consider batch size to be $m$.)

\begin{table}[htb]
    \centering
    \begin{tabular}{|c|c|}
    \multicolumn{1}{c}{Layer} &
    \multicolumn{1}{c}{Parameters}  \\
    \hline
    \hline
    Convolution & 16 filters of 8x8, strides 2 \\
    Max-Pooling & 2x2 \\
    Convolution & 32 filters of 4x4, strides 2  \\
    Max-Pooling & 2x2 \\
    Fully connected & 32 units  \\
    Softmax & 10 units \\
    \hline
    \end{tabular}
    \caption{Architecture for MNIST and Fashion-MNIST.}
    \label{tbl:mnist}
\end{table}

\begin{table}[htb]
    \centering
    \begin{tabular}{|c|c|}
    \multicolumn{1}{c}{Layer} &
    \multicolumn{1}{c}{Parameters} \\
    \hline
    \hline
    Conv $\times$ 2 & 32 filters of 3x3, strides 1 \\
    Max-Pooling & 2x2  \\
    Conv $\times$ 2 & 64 filters of 3x3, strides 1 \\
    Max-Pooling & 2x2  \\
    Conv $\times$ 2 & 128 filters of 3x3, strides 1 \\
    Fully connected & 1024 units  \\
    Softmax & 10 units  \\
    \hline
    \end{tabular}
    \caption{Architecture for CIFAR-10.}
        \label{tbl:architecture_cifar}
\end{table}

\mypar{Implementation framework} To implement LDP-SGD, we modify the DP-SGD algorithm in Tensorflow Privacy~\cite{TFpriv} to include the new client-side noise generation algorithm (Algorithm~\ref{alg:localG}) and the privacy accountant.

\mypar{MNIST and Fashion-MNIST Experiments} We train models whose architecture is described in Table \ref{tbl:mnist}. The results on this dataset is summarized in Table \ref{tbl:res}. The non-private accuracy baselines using this architecture are 99\% and 89\% for MNIST and Fashion-MNIST respectively.

We reiterate that the privacy accounting after Shuffling should be considered to be a loose upper bound. To test how much higher the accuracy might reach without accounting for central DP, we also plot the entire learning curve until it saturates (varying batch sizes) at LDP $\epsle=1.9$ per epoch. The accuracy tops out at 95\% and 78\% respectively.

\mypar{CIFAR-10 Experiments} For the CIFAR-10 dataset, we consider the model architecture in Table \ref{tbl:architecture_cifar} following recent work~\cite{DP-DL,Shoe}. Along the lines of work done in these papers, we first train the model without privacy all but the last layer on CIFAR-100 using the same architecture but replacing the softmax layer with one having 100 units (for the 100 classes). Next, we transfer all but the last layer to a new model and only \emph{re-train} the last layer with differential privacy on CIFAR-10.

\begin{figure*}[htb]
\begin{subfigure}{.33\textwidth}
  \centering
  \includegraphics[width=.8\linewidth]{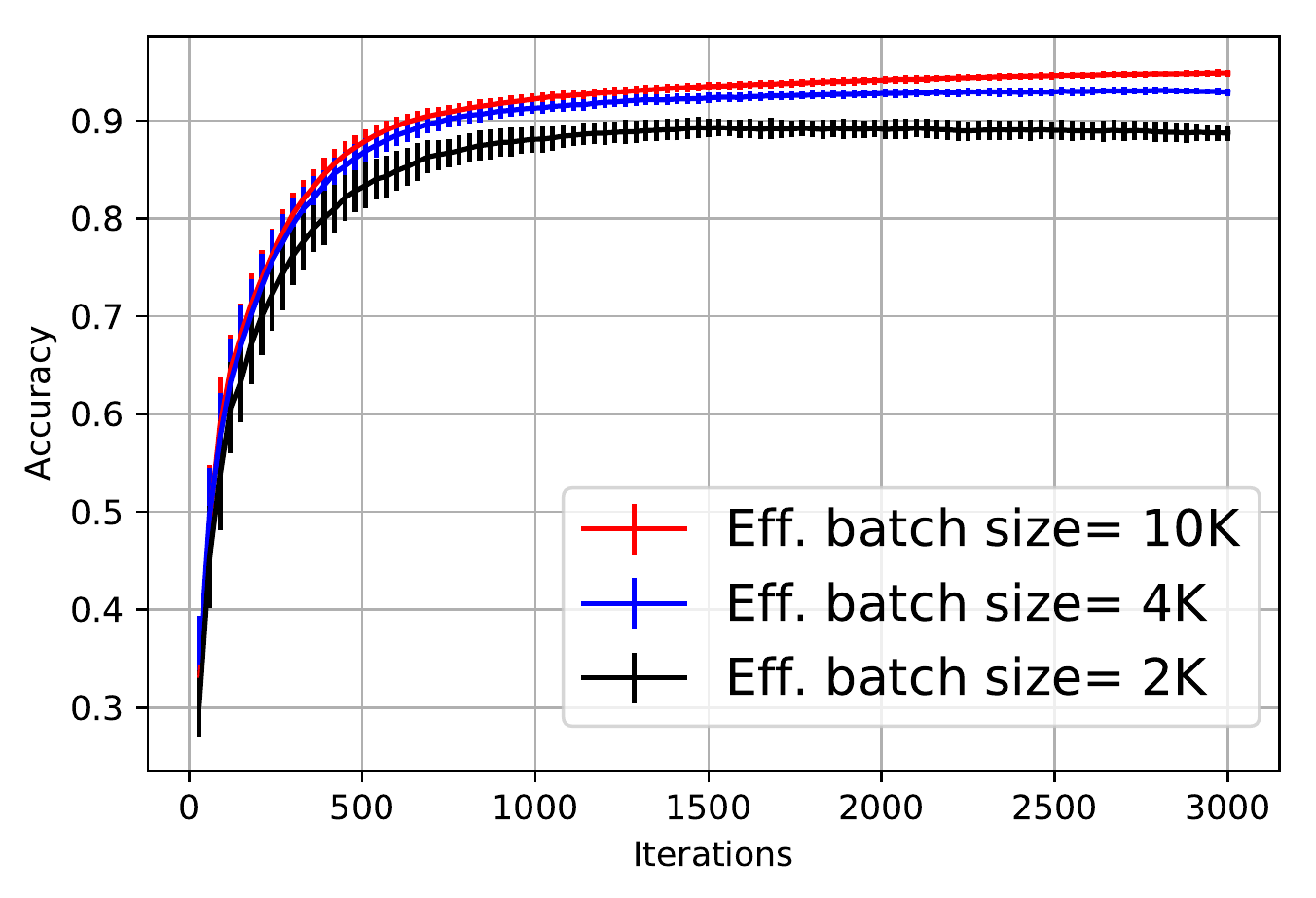}
  \caption{MNIST}
  \label{fig:sub-first}
\end{subfigure}
\begin{subfigure}{.33\textwidth}
  \centering
  \includegraphics[width=.8\linewidth]{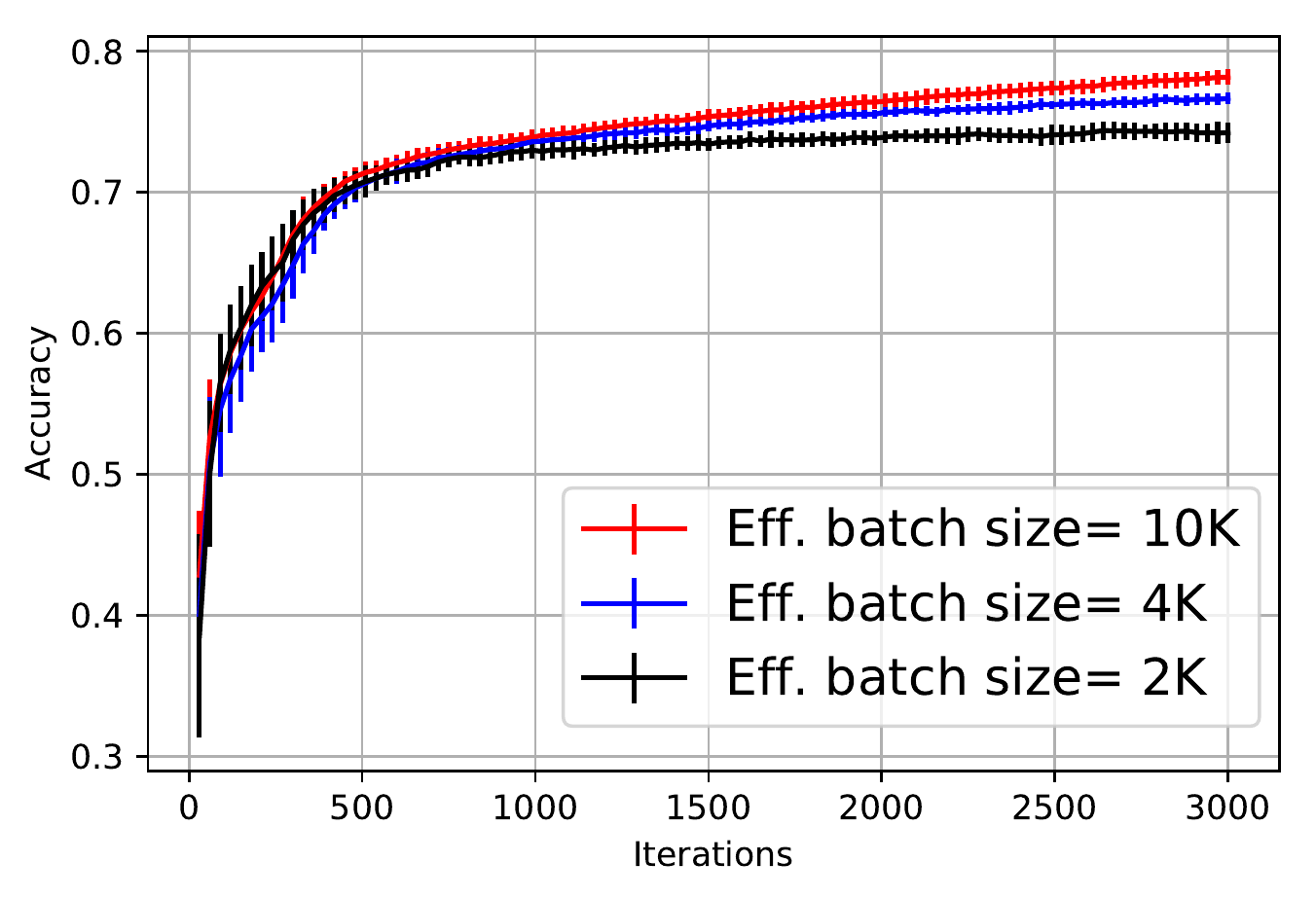}
  \caption{Fashion-MNIST}
  \label{fig:sub-second}
\end{subfigure}
\begin{subfigure}{.33\textwidth}
  \centering
  \includegraphics[width=.8\linewidth]{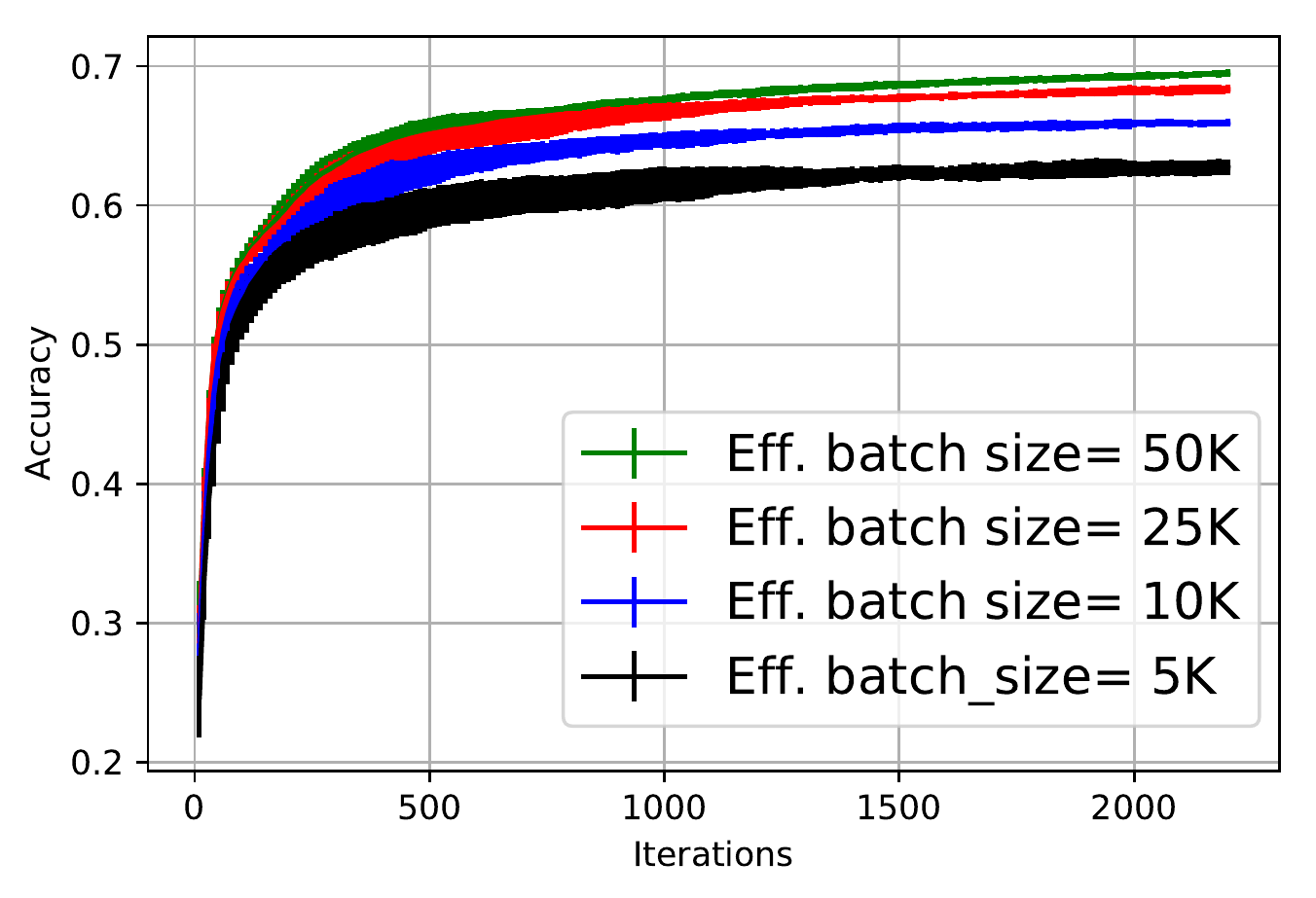}
  \caption{CIFAR-10}
  \label{fig:sub-third}
\end{subfigure}
\caption{Accuracy vs iterations tradeoff on various data sets, with local differential privacy-per-record-per-iteration $\epsle=1.8$ for CIFAR-10, and $\epsle=1.9$ for MNIST and Fashion-MNIST. The plots are over at least ten independent runs.}
\label{fig:MLResl}
\end{figure*}

Our non-private training baseline (of training on all layers) achieves 86\% accuracy. The results of this training method are summarized in Table \ref{tbl:res}. As done with the MNIST experiments, keeping in mind the looseness of the central DP accounting, we also plot in Figure~\ref{fig:sub-third} the complete learning curve up to saturation at LDP $\epsle=1.8$ per epoch. We see that the best achieved accuracy is 70\%.

\mypar{Note on central differential privacy} Since we are translating local differential privacy guarantees to central differential privacy guarantees, our notion of central differential privacy is in the \emph{replacement model}, i.e., two neighboring data sets of the same size but differ by one record. However, the results in \cite{DP-DL,TFpriv,Shoe} are in the \emph{add/removal model}, i.e., two neighboring data sets differ in the presence or absence of one data record. As a blackbox, for any algorithm, $\eps_\mathrm{Add/Remove} \leq \eps_\mathrm{Replace} \leq 2\eps_\mathrm{Add/Remove}$, and for commonly used algorithms, the upper bound is close to tight.

\mypar{Open question} We believe that the current accounting for central differential privacy via advanced composition is \emph{potentially loose}, and one may get stronger guarantees via R\'enyi differential privacy accounting (similar to that in \cite{DP-DL}). %
We leave the problem of tightening the overall central differential privacy guarantee for future work.

\mypar{Estimating lower bounds through membership inference attacks}
We use the membership inference attack to measure the privacy leakage of a model \cite{shokri2017membership, yeometal, jayaraman2019evaluating, erlingsson2019private}. While these measurements yield loose lower bounds on how much information is leaked by a model, it can serve as an effective comparison across models trained with noise subject to different privacy analyses (with their separate upper bounds on differential privacy).

Along the lines of Yeom et al.~\cite{yeometal}, for each model, we measure the average log-loss between true labels and predicted outputs over a set of samples used in training and not in training. One measure of privacy leakage involves the best binary (threshold) classifier based on these loss values to distinguish between in-training and out-training examples. The resulting ROC curve of the classifier across different thresholds can be used to estimate a lower bound on the privacy parameter. Specifically, it is easy to strengthen the results in Yeom et al.~\cite{yeometal} to show that the difference between the true positive rate (TPR) and false positive rate (FPR) at any threshold is bound by $1-e^{-\eps}$ for a model satisfying $\eps$-differential privacy. Thus, the lower bound $\eps \geq -\log\left(\text{max}(\text{TPR}-\text{FPR})\right)$.
The results are shown in Table~\ref{tab:membership_inference} for all models trained.
As can be seen from the results, even though the $\epsc$ upper bound are different for models trained under the ESA framework and those with DPSGD, there is no much difference in the lower bound.

\section{Conclusions}
\label{sec:conclude}

This paper's overall conclusion 
that it is feasible to implement
high-accuracy statistical reporting with
strong central privacy guarantees,
as long as respondent's randomized reports
are anonymized by a trustworthy intermediary.
Sufficient for this are 
a small set of primitives---applied within a relatively simple, abstract attack model---for
both analysis techniques and practical technical mechanisms.
Apart from anonymization itself,
the most critical of these primitives
are those that involve fragmenting of respondents' randomized reports;
first explored in the original ESA paper~\cite{prochlo},
such fragmentation turns out to be critical to achieving
strong central privacy guarantees with high utility,
in our empirical applications on real-world datasets.
As we show here,
those primitives are sufficient to achieve
high utility for difficult tasks
such as iterative training of deep-learning neural networks,
while providing both central and local guarantees
of differential privacy.

In addition, this paper makes it clear that
when it comes to practical applications
of anonymous, differentially-private reporting,
significantly more exploration and empirical evaluation is needed,
along with more refined analysis.
Specifically, this need is made very clear by
the discrepancy this paper finds between the utility and central privacy guarantees
of anonymous one-hot-encoded LDP reports and anonymous sketch-based LDP reports,
witch sketching parameters drawn from those used in real-world deployments.
At the very least, this discrepancy highlights
how practitioners must carefully choose the
mechanisms they use in sketch-based encodings,
and the parameters by which they tune those mechanisms,
in order to achieve good tradeoffs for the dataset and task at hand.
However, the lack of precise
central privacy guarantees for anonymous sketch-based LDP reports
also shows the pressing needs
for better sketch constructions and analyses
that properly account for the anonymity and fragmentation of respondents' reports.
While some recent work has started
to look at better analysis of sketching (e.g.,  asymptotically~\cite{ghazi2019private}),
practitioners should
look towards the excellent tradeoffs
shown here for one-hot-encoded LDP reports,
until further, more practical results are derived  in the large alphabet setting.

\bibliographystyle{IEEEtran}
\bibliography{oakland_esa}

\appendix

\subsection{Missing details from Section \ref{sec:attribFrag}}
\label{app:attribFrag}

\begin{proof}[Proof of~Theorem~\ref{thm:ahd}]
To prove removal LDP we use the reference distribution $\calR_0$ to be randomized response with~$\epsl$ on the $k$-dimensional all-zeros vector $\boldzero$. For any $\boldx \in \calD$ (represented as one-hot binary vector in $k$ dimensions), $\boldx$ and~$\boldzero$ differ in one position and therefore, by standard properties of randomized response Algorithm $\afrag(\calR_{k\text{-RAPPOR}})$ computing $\hat{x}^{(j)} := \calR_j(x^{(j)}, \epsl)$ for $j \in [k]$ satisfies removal $\epsl$-local differential privacy. Furthermore, each $\hat{x}^{(j)}$ by itself is computed with (replacement) $\epsl$-DP. We obtain the central differential privacy guarantee (through amplification via shuffling) by invoking Lemma \ref{lem:abcfd} with $\lambda=\frac{2n}{1+e^{\epsl}}$.

The lower bound of $14\log(4/\delta)$ for $\lambda$ translates (with some simplification) to an upper bound of $\epsl \leq \log n - \log(14\log(4/\delta))$ assumed in the Theorem statement. Furthermore, as $\lambda \geq 14\log(2/\delta) \geq 8\log(2/\delta)$, we have that  $\lambda-\sqrt{2\lambda\log(2/\delta)}$ in  Lemma \ref{lem:abcfd} is at least $\lambda/2$. Simplifying the expression in~\eqref{eq:mixnet-bound}, the central privacy guarantee for each individual bit of any $\hat{\boldx}$ is:
\begin{small}
\begin{align}
\epsc^{\sf bit}& \leq \sqrt{\frac{64\log(4/\delta)}{\lambda}}=\sqrt{\frac{64(1+e^{\epsl})\log(4/\delta)}{2n}}\nonumber\\
&\leq \sqrt{\frac{64\cdot e^{\epsl}\log(4/\delta)}{n}}.
\label{eq:132}
\end{align}
\end{small}
To prove removal central differential privacy for the entire output we define the algorithm $\calM'\colon \calD^n \times 2^{[n]}$ as follows. Given $D=(x_1,\ldots,x_n)$ and a set of indices $I$, $\calM'$ uses the reference distribution $\calR_0$ in place of the local randomizer for each element $x_i$ for which $i\not\in I$. Changing any $\boldx$ to $\boldzero$ for the $i$-th element changes only one input bit. It follows from Eq.~\eqref{eq:132} that the overall $\epsc$ for removal central differential privacy guarantee is $\epsc=\sqrt{\frac{64\cdot e^{\epsl}\log(4/\delta)}{n}}$, which completes the proof.
\end{proof}

\begin{proof}[Proof of~Theorem~\ref{thm:ahds}]
In $\afrag(\calR_{k\text{-RAPPOR}})$ (Algorithm~\ref{Alg:Frag}) consider any $\boldx$ and the corresponding $\hat{\boldx}$, the list of randomized responses $\calR_j(x^{(j)}, \epsl)$. For brevity, consider the random variable $\boldsymbol{\zeta}=\left(\frac{e^{\epsl}+1}{e^{\epsl}-1}\cdot\hat \boldx-\frac{1}{e^{\epsl}-1}\right)$. It follows that  $\mathbb{E}\left[\boldsymbol{\zeta}\right]=\boldx$ and furthermore ${\sf Var}[\boldsymbol{\zeta}]=\frac{e^{\epsl}+1}{e^{\epsl}-1}-1=\Theta\left(1/e^{\epsl}\right)$.
Using standard sub-Gaussian tail bounds, and taking an union bound over the domain $[k]$, one can show that w.p.\ at least $1-\beta$, over all $n$ respondents with data $\boldx_i$, \[\alpha=\left\|\hat{\boldh}-\frac{1}{n}\sum \boldx_i\right\|_\infty=\Theta\left(\sqrt\frac{\log(k/\beta)}{n e^{\epsl}}\right).\] Applying Theorem~\ref{thm:ahd} to compute $\epsc$ in terms of $\epsl$ completes the proof.
\end{proof}

\subsection{Missing details from Section \ref{sec:record_fragmenting}}
\label{app:record_fragmenting}

We start by analyzing the privacy of an arbitrary combination of local DP randomizer followed by an arbitrary differentially private algorithm. To simplify this analysis we show that it suffices to restrict our attention to binary domains.
\begin{lem}
\label{lem:reduce-2-binary}
Assume that for every replacement $(\eps_1,\delta_1)$-DP local randomizer $\calQ_1  \colon \zo \to \zo$ and
every replacement $(\eps_2,\delta_2)$-DP local randomizer $\calQ_2  \colon \zo \to \zo$ we have that $\calQ_2 \circ \calQ_1$ is a replacement $(\eps,\delta)$-DP local randomizer. Then  
for every replacement $(\eps_1,\delta_1)$-DP local randomizer $\calR_1  \colon X \to Y$ and
replacement $(\eps_2,\delta_2)$-DP local randomizer $\calR_2: Y \to Z$ we have that $\calR_2 \circ \calR_1$ is a replacement $(\eps,\delta)$-DP local randomizer.
\end{lem}
\begin{proof}
Let $\calR_1  \colon X \to Y$ be a replacement $(\eps_1,\delta_1)$-local randomizer and $\calR_2\colon Y \to Z$ be a replacement $(\eps_2,\delta_2)$-DP randomizer. Assume for the sake of contradiction that for some $(\eps,\delta)$ there exists
an event $S\subseteq Z$ such that for some $x,x'$:
\[
\Pr[\calR_2(\calR_1(x)) \in S] > e^\eps \Pr[\calR_2(\calR_1(x')) \in S] + \delta .
\]
We will show that then there exist an $(\eps_1,\delta_1)$-DP local randomizer $\calQ_1  \colon \zo \to \zo$ and
$(\eps_2,\delta_2)$-DP local randomizer $\calQ_2  \colon \zo \to \zo$ such that
\[
\Pr[\calQ_2(\calQ_1(0)) = 1] > e^\eps \Pr[\calQ_2(\calQ_1(1)) = 1] + \delta ,
\]
contradicting the conditions of the lemma.

Let $$y_0 := \argmin_{y\in Y}\{\Pr[\calR_2(y)\in S] \} ,$$ $$y_1 := \argmax_{y\in Y}\{\Pr[\calR_2(y)\in S] \} $$
and let $$P_1 := \{y \in Y \ |\ \Pr[\calR_1(x)=y] - e^\eps \Pr[\calR_1(x')=y] > 0 \} .$$
Using this definition and our assumption we get:
\alequn{& \left(\Pr[\calR_1(x) \not\in P_1] - e^\eps \Pr[\calR_1(x') \not\in P_1]\right) \cdot \Pr[\calR_2(y_0)\in S] \\& + \left(\Pr[\calR_1(x) \in P_1] - e^\eps \Pr[\calR_1(x') \in P_1]\right) \cdot  \Pr[\calR_2(y_1)\in S]\\
&\geq  \sum_{y\in Y} \left(\Pr[\calR_1(x)=y] - e^\eps \Pr[\calR_1(x')=y] \right) \cdot  \Pr[\calR_2(y)\in S] \\&
> \delta .
}
We now define 
$\calQ_1(0) := \indi{\calR_1(x) \in P_1}$ and $\calQ_1(1) := \indi{\calR_1(x') \in P_1}$, where $\indi{\cdot}$ denotes the indicator function.
 By this definition $\calQ_1$ is obtained from $\calR_1$ by restricting the set of inputs and postprocessing the output. Thus $\calQ_1$ is a replacement $(\eps_1,\delta_1)$-DP local randomizer.
Next define for $b\in \zo$, $\calQ_2(b) := \indi{\calR_2(y_b) \in S}$. Again, it is easy to see that $\calQ_2$ is a replacement $(\eps_2,\delta_2)$-DP.
We now obtain that 
{\small{\alequn{&\Pr[\calQ_2(\calQ_1(0)) = 1] - e^\eps \Pr[\calQ_2(\calQ_1(1)) = 1] \\
&= \sum_{b\in \zo} \left(\Pr[\calQ_1(0)=b] - e^\eps \Pr[\calQ_1(1)=b] \right) \cdot  \Pr[\calQ_2(b) = 1] \\
 &= \left(\Pr[\calR_1(x) \not\in P_1] - e^\eps \Pr[\calR_1(x') \not\in P_1]\right) \cdot  \Pr[\calR_2(y_0)\in S] \\& + \left(\Pr[\calR_1(x) \in P_1] - e^\eps \Pr[\calR_1(x') \in P_1]\right) \cdot \Pr[\calR_2(y_1)\in S]\\
& > \delta
}}}
as needed for contradiction.
\end{proof}
As an easy corollary of Lemma \ref{lem:reduce-2-binary} we obtain a tight upper bound in the pure differential privacy case.
\begin{cor}\label{cor:reduce-2-binary-pure}
For every replacement $\eps_1$-DP local randomizer $\calR_1  \colon X \to Y$ and
every replacement $\eps_2$-DP local randomizer $\calR_2  \colon Y \to Z$ we have that $\calR_2 \circ \calR_1$ is a replacement $\eps$-DP local randomizer for $\eps = \ln\left(\frac{e^{\eps_1+\eps_2}+1}{e^{\eps_1}+e^{\eps_2}}\right)$.
In addition, if $\calR_1$ is removal $\eps_1$-DP then $\calR_2 \circ \calR_1$ is a removal $\eps$-DP.
\end{cor}
\begin{proof}
By Lemma \ref{lem:reduce-2-binary} it suffices to consider the case where $X=Y=Z=\zo$. Thus it suffices to upper bound the expression:
{\small{$$ \frac{ \Pr[\calR_1(0)=0] \cdot \Pr[\calR_2(0)=1] + \Pr[\calR_1(0)=1] \cdot \Pr[\calR_2(1)=1] }{\Pr[\calR_1(1)=0] \cdot \Pr[\calR_2(0)=1] + \Pr[\calR_1(1)=1] \cdot \Pr[\calR_2(1)=1] } .$$}}
Denoting by $p_0 := \Pr[\calR_1(0)=0]$, $p_1 := \Pr[\calR_1(1)=0]$ and $\alpha = \Pr[\calR_2(0)=1]/\Pr[\calR_2(1)=1]$ the expression becomes:
$$\frac{1 + (\alpha-1)p_0}{1 + (\alpha-1)p_1} .$$
The conditions on $\calR_1$ imply that $\frac{p_0}{p_1}, \frac{1-p_0}{1-p_1} \in [e^{-\eps_1},e^{\eps_1}]$ and $\alpha \in [e^{-\eps_2},e^{\eps_2}]$. Without loss of generality we can assume that $\alpha \geq 1$ and thus the expression is maximized when $\alpha = e^{\eps_2}$ and $p_0>p_1$.
Maximizing the expression under these constraints we obtain that the maximum is $\frac{e^{\eps_1+\eps_2}+1}{e^{\eps_1}+e^{\eps_2}}$ and is achieved when $p_0=1-p_1= e^{\eps_1}/(1+e^{\eps_1})$. In particular, the claimed value of $\eps$ is achieved by the standard binary randomized response with $\eps_1$ and $\eps_2$.

To deal with the case of removal we can simply substitute $\calR_1(x')$ with the reference distribution $\calR_0$ in the analysis to obtain removal DP guarantees for $\calR_2 \circ \calR_1$.
\end{proof}

We remark that it is easy to see that $\frac{e^{\eps_1+\eps_2}+1}{e^{\eps_1}+e^{\eps_2}} \leq \min\{e^{\eps_1},e^{\eps_2}\}$. Also in the regime where $\eps_1,\eps_2 \leq 1$ we obtain that $\eps = O(\eps_1 \eps_2)$, namely the privacy is amplified by applying local randomization.

\begin{proof} [Proof of Theorem \ref{cor:recFrag}]
The proof of local differential privacy is immediate based on Theorem \ref{thm:record_frag}. To obtain the central differential privacy guarantee, we consider each of the terms in the $\min$ expression for $\varepsilon_c$. From the central differential privacy context, each of the shufflers in the execution of Algorithm \ref{Alg:ReportFrag} can be considered to be a \emph{post-processing} of the output of a single shuffler, and the privacy guarantee from this single shuffler should prevail. Each of the individual reports are at most $\varepsilon_b$-locally differentially private, and hence by using the generic privacy amplification by shuffling result from Lemma~\ref{lem:borja}, the second term in the $\varepsilon_c$ follows. To obtain the first term, recall the matrix $M(\boldx)$ in Section \ref{sec:record_fragmenting}. Each row of the matrix satisfies $\varepsilon_0$-local differential privacy, and there are $\tau$~rows in this matrix. Hence, first applying privacy amplification theorem from Lemma \ref{lem:abcfd} on each of the rows independently, and then using advanced composition from Theorem \ref{thm:advc} over the $\tau$~rows, we obtain the first term in $\varepsilon_c$, which completes the proof of the central differential privacy guarantee.

The utility guarantee follows immediately from the utility proof of Theorem \ref{thm:att-frag-accuracy}.
\end{proof}

\subsection{Missing Details from Section \ref{sec:crowd}}
\label{app:crowd}

\begin{proof}[Proof of Theorem \ref{thm:privCrowds}]
The proof follows a similar argument as \cite[Theorem 3.5]{vadhan2017complexity}.
Consider two neighboring data sets $D$ and $D'$, there are only two crowd IDs whose counts get affected. Since the randomization for each of the counts are done independently, we can analyze their privacy independently and then perform standard composition \cite{dwork2014algorithmic}. Consider a crowd $\calD_i$, and the corresponding counts $n_i \neq n'_i$ on data sets $D$ and $D'$ respectively.

Notice that the computation of $\hat{n}_i$ satisfies $\frac{\epsl^{\sf cr}}{2}$-differential privacy by the Laplace mechanism \cite{DMNS}. Now, by the tail probability of Laplace noise, with probability at least $1-\frac{\delta^{\sf cr}}{2}$, the algorithm does not abort on crowd $\calD_i$. In that case, the shuffler can ensure $\hat n_i$ records in $\calD_i$ via dropping records. This would ensure $\left(\frac{\epsl^{\sf cr}}{2},\frac{\delta^{\sf cr}}{2}\right)$-differential privacy.

Therefore, composing the above over the two crowds that are affected by $D$ and $D'$, we complete the proof.
\end{proof}

\begin{proof}[Proof of Theorem \ref{thm:utilCrowdThresh}]
The proof of this theorem follows from standard tail probabilities of the Laplace mechanism. With probability at least $1-\delta^{\sf cr}$, for a given crowd $\calD_i$, the error in the reported count is at most $2T=\frac{4}{\epsl^{\sf cr}}\log\left(\frac{4}{\delta^{\sf cr}}\right)$. Taking an union bound over all the $\xi$ crowds, completes the proof.
\end{proof}

\subsection{Missing Details from Section \ref{sec:mlESA}}
\label{app:learning}

\begin{proof}[Proof of Theorem \ref{thm:learning}]
We will prove the privacy and utility guarantees separately.

\mypar{Privacy guarantee} We will prove this guarantee in  two steps: (i) Amplify the local differential privacy guarantee $\epsl$ per epoch via \cite[Corollary 5.3.1]{privacy-blanket} (see Theorem \ref{thm:borja}), and (ii) Use advanced composition \cite{vadhan2017complexity} to account for the privacy budget. Combination of these two immediately implies the theorem.

\begin{thm}[Corollary 5.1 from \cite{privacy-blanket}]
Let $\calR:\mathbb{X}\to\mathbb{Y}$ be an $\epsle$-local differentially private randomizer, and $\calM$ be the corresponding shuffled mechanism (that shuffles all the locally randomized reports). If $\epsle\leq \log(n)/4$, then $\calM$ satisfies $\left(O\left(\frac{(e^{\epsle}-1)\sqrt{\log(1/\delta)}}{\sqrt n}\right),\delta\right)$-central differential privacy in the shuffled setting.
\label{thm:borja}
\end{thm}

\mypar{Utility guarantee} Here we use the a standard bound on the convergence of Stochastic Gradient Descent (SGD) stated in Theorem \ref{thm:sz}. One can instantiate Theorem \ref{thm:sz} in the context of this paper as follows:
$F(\theta)=\frac{1}{n}\sum\limits_{i=1}^n\ell(\theta;x_i)$, and $\boldg_t$ is the randomized gradient computed in Algorithm \ref{alg:servrG}.

\begin{thm}[Theorem 2 from \cite{shamir2013stochastic}]
Consider a convex function $F:\calC\to\mathbb{R}$ defined over a convex set $\calC\subseteq\mathbb{R}^d$, and consider the following SGD algorithm: $\theta_{t+1}\leftarrow\Pi_{\calC}\left(\theta_t-\frac{c}{\sqrt t}\boldg_t\right)$, where $\Pi_{\calC}\left(\cdot\right)$ is the $\ell_2$-projection operator onto the set $\calC$, $c>0$ is a constant, and $\boldg_t$ has the following properties. i) {\bf[Unbiasedness]} $\E[\boldg_t]=\bigtriangledown F(\theta_t)$, and ii) {\bf[Bounded Variance]} $\E\left[\|\boldg_t\|^2_2\right]=G^2$.  The following is true for any $T>1$.
$$\E[F(\theta_t)]-\min\limits_{\theta\in\calC}F(\theta)\leq \left(\frac{\|\calC\|^2_2}{c}+cG^2\right)\frac{2+\log T}{\sqrt{T}}.$$
\label{thm:sz}
\end{thm}

Following the instantiation above, by the property of the noise distribution, one can easily show that $\E[\boldg_t]=\frac{1}{n}\sum\limits_{i=1}^n\bigtriangledown \ell(\theta_t;x_i)$, and furthermore $\E[\|\boldg_t\|_2]=O\left(\frac{L\sqrt{d}}{\sqrt{n}}\cdot \frac{e^{\epsle}+1}{e^{\epsle}-1}\right)=G$. (See \cite[Appendix I.2]{duchi2018minimax} for the full derivation. Setting $c=\frac{\|\calC\|_2}{G}$, and setting $T=n/\log^2 n$ completes the proof.
\end{proof}

\end{document}